\documentclass{ar-1col}

\jname{Annu. Rev. in Astronomy \& Astrophysics}
\jvol{AA}
\jyear{2016}

\usepackage{natbib}
\newcommand\lsim{\mathrel{\rlap{\lower4pt\hbox{\hskip1pt$\sim$}}
\raise1pt\hbox{$<$}}}
\newcommand\gsim{\mathrel{\rlap{\lower4pt\hbox{\hskip1pt$\sim$}}
\raise1pt\hbox{$>$}}}
\newcommand{\Ham}{\mathcal{H}}
\newcommand{\out}{\rm out}
\newcommand{\inner}{\rm in}

\newcommand{\msun}{M_\odot}
\begin{document}

\markboth{Smadar Naoz}{The EKL Effect and Its Applications}

\title{The Eccentric Kozai-Lidov Effect and Its Applications}

\author{Smadar Naoz
\affil{Department of Physics and Astronomy, University of California, Los Angeles, CA 90095, USA; \\ snaoz@astro.ucla.edu} }

%

\newcommand{\R}{\textfrak{R}}
\newcommand{\M}{M_{\star}}
\newcommand{\Ms}{M_{\odot}}
\newcommand{\Mp}{M_{p}}
\newcommand{\Mj}{M_{J}}
\newcommand{\Mc}{M_{c}}
\newcommand{\Em}{\epsilon_{M}}
\newcommand{\tot}{{\rm tot}} 

%


%

\begin{abstract}
The hierarchical triple body approximation has useful applications to a variety of systems  from planetary and stellar  scales to supermassive black holes. In this approximation, the energy of each orbit is separately conserved  and therefore the two semi-major axes are constants. On timescales much larger than the orbital periods, the orbits  exchange angular momentum which leads to eccentricity and orientation (i.e., inclination) oscillations. The orbits' eccentricity can reach extreme values leading to a nearly radial motion, which can further evolve into short orbit periods and merging binaries. Furthermore, the orbits' mutual inclination may change dramatically from pure prograde to pure retrograde leading to misalignment and a wide range of inclinations.  This dynamical behavior is coined as the ``eccentric Kozai-Lidov" mechanism.
The behavior of such a system is exciting, rich and chaotic in nature. Furthermore, these dynamics are accessible from a large part of the triple body parameter space and can be applied  to diverse range of  astrophysical settings and used to gain insights to many puzzles.
 
\end{abstract}

\maketitle

\tableofcontents

%
%
%



\section{Introduction  }

Triple  systems are common in the Universe. They are found in many different astrophysical settings covering a large range of mass and physical scales, such as triple stars \citep[e.g.,][]{T97,Eggleton+07,Tokovinin14a,Tokovinin14b},  and accreting compact binaries with a companion \citep[such as companions to  X-ray binaries e.g.,][]{1988ApJ...334L..25G,Prodan:2012ey}. In addition, it seems that   supermassive black hole binaries and higher multiples are common and thus, any star in their vicinity forms a triple system \citep[e.g.,][]{Valtonen+96,DiMatteo+05,Khan+12,Kulkarni+12}.  Furthermore, considering the solar system, binaries composed of near earth objects,  asteroids or dwarf planets \citep[for which a substantial fraction seems to reside  in a binary configuration, e.g.,][]{Polishook+06,Nesvorny+11,Margot+15} naturally form a triple system with our Sun. Lastly, it was shown that  Hot Jupiters   are likely to have a far away companion, forming a triple system of star-Hot Jupiter binary with a distant perturber \citep[e.g.,][]{Knutson+14,Ngo+15,Wang+15}.
Stability requirements yield that most of these systems will be hierarchical in scale, with a tight inner binary orbited by a  tertiary on a wider orbit, forming the outer binary. Therefore, in most cases the dynamical behavior of these systems takes place on timescales much longer than the orbital periods.

The study of 
secular perturbations (i.e., long term  phase average evolution over timescales longer than the orbital periods)  in triple systems can be dated back to Lagrange, Laplace and  Poincare. Many years later, the study of secular {\it hierarchical} triple system was addressed by Lidov (1961, where the English translation version was published only in 1962). He studied the orbital evolution of artificial satellites due to gravitational perturbations from an axisymmetric outer potential.   Short time after that,  \citet{Kozai} studied the effects of
Jupiter's gravitational perturbations on an inclined asteroid in our
own solar system. In these settings a relatively  tight inner  binary composed of a primary and a secondary (in these initial studies assumed to be a test particle), is orbited by a far away companion. We denote the inner (outer) orbit  semi-major axis as $a_1$  ($a_2$). 
In this setting the secular approximation  can be utilized. This implies that the energy of each orbit is  conserved separately (as well as the energy of the entire system), thus $a_1$ and $a_2$ are constants during the evolution. The  dynamical behavior  is a result of angular momentum exchange between the two orbits. \citet{Kozai}, for example, expanded the 3-body Hamiltonian in semi-major axis ratio (since the outer orbit is far away, $a_1/a_2$ is a small parameter). He then averaged over the orbits and lastly truncated the expansion to the lowest order, called the quadrupole, which is  proportional to $(a_1/a_2)^2$. 
Both \citet{Kozai} and \citet{Lidov} found that the inner test particle inclination and eccentricity oscillate  on timescales much larger than
its orbital period. In these studies the outer perturber was assumed to carry most of the angular momentum, and thus under the assumption of an axisymmetric outer potential the inner and outer orbits z-component of the angular momenta (along the total angular momentum) are conserved. This led to  large variations between the eccentricity and inclination of the test particle orbit.

While the Kozai-Lidov\footnote{Although Lidov has published his work first, we are using here the alphabetical order for the name of the mechanism.}  mechanism seemed interesting  it was largely ignored for many years. However, about 15-20 years ago, probably correlating with the detection of the  eccentric planet 6 Cyg B, \citep{Cochran+96}, or the close to perpendicular stellar   Algol system \citep{1998EKH,Baron+12}, the Kozai-Lidov mechanism received its deserved attention. However, while the mechanism seemed very promising in addressing these astrophysical phenomena, it was limited to a narrow parts of the parameter space \citep[favoring    close to perpendicular initial orientation  between the two orbits, e.g.,][]{Maechal90,Morbidelli02,3book,Dan} and produced only moderate eccentricity  excitations.  Most of the studies that investigated different astrophysical applications of the Kozai-Lidov mechanism used the   \citet{Kozai} and \citet{Lidov} test particle, axisymmetric outer orbit quadrupole-level approximation. 

This approximation has an analytical solution which  describes (for initially highly inclined orbits $\sim40^\circ-140^\circ$, see below)  the large amplitude oscillations  between the inner orbit's eccentricity and inclination with respect to the outer orbit \citep[e.g.,][]{Kin+99,Morbidelli02}. These oscillations have a well defined maximum and minimum eccentricity and inclination and limits the motion to either prograde ($\leq 90^\circ$) or retrograde ($\geq 90^\circ$) with respect to the outer orbit. The axisymmetric outer orbit quadrupole-level approximation is applicable for an ample amount of systems. For example, this approximation has appropriately described the motion of Earth's artificial satellites under the influence of  gravitational perturbations  from  the moon \citep[e..g][]{Lidov}. Other astrophysical systems for which this approximation is applicable include (but are not limited to) the  effects of the Sun's gravitational perturbation on planetary satellites, since in this case indeed the  satellite mass is negligible compared to the other masses in the system, and the planet's orbit around the Sun is circular.  Indeed it was shown that the axisymmetric outer orbit quadrupole-level approximation can successfully be used to study the 
 inclination distribution of the Jovian irregular satellites \citep[e.g.,][]{Carr+02,Nes+03} or in general the survival of planetary outer satellites \citep[e.g.,][]{Kin+91}, as well as the dynamical evolution of  the orbit of a  Kuiper Belt object satellite due to perturbation form the sun \citep[e.g.,][]{PN09,Naoz10obs}. Indeed this approximation is useful and can be applied in the limit of a circular outer orbit and a test particle inner object.


Recently, \cite{Naoz11,Naoz+11sec} showed that relaxing either one of these assumptions leads to qualitative different  dynamical evolution. Considering systems beyond the test particle approximation, or a circular orbit, requires the next level of approximation, called the octupole--level of approximation \citep[e.g.][]{Har68,Har69,Ford00,Bla+02}. This level of approximation is proportional  to $(a_1/a_2)^3$.
In the octupole--level of approximation, the inner orbit eccentricity can reach extremely  high values, and does not have a well defined value, as the system is chaotic in general \citep{Ford00,Naoz+11sec,Li+13,Li+14Chaos,Tey+13}.  In addition, the inner orbit inclination can flip its orientation from prograde, with respect to the total angular momentum, to retrograde \citep{Naoz11}.  We refer to this process as the {\bf eccentric Kozai--Lidov} (EKL) mechanism. We note that here we follow the literature coined  acronym ``EKL" as oppose to the more chronologically accurate acronym ``ELK."

 As will be discussed below the EKL mechanism taps into larger parts of the parameter space (i.e., beyond the $\sim40^\circ-140^\circ$ range), and results in a richer and far more exciting dynamical evolution. As a consequence this mechanism is applicable to a wide range of systems that allow for eccentric orbits, or three massive bodies, from exoplanetary orbits over  stellar interactions to black hole dynamics.  
   The prospect of forming eccentric or short period planets through three body interactions was the source of many studies \citep[e.g.,][]{Inn+97,Wu+03,Dan,Wu+07,Veras+10,Cor+11,Batygin+11,Naoz11,Naoz+12bin,Petrovich15He,Petrovich15Co}. It also promoted many interesting application for stellar dynamics from stellar mergers \citep[e.g.,][]{PF09,NF,Witzel+14,Stephan+16} to compact binary merger which may prompt supernova explosions for double white dwarf merger \citep[e.g.,][]{Tho10,Katz+12}, or gravitational wave emission for neutron star or black hole binary merger \citep[e.g.,][]{Bla+02,Seto13}.

\section{The hierarchical three body secular approximation}\label{sec:3b}

\begin{figure}
\begin{center}
\includegraphics[width=\linewidth]{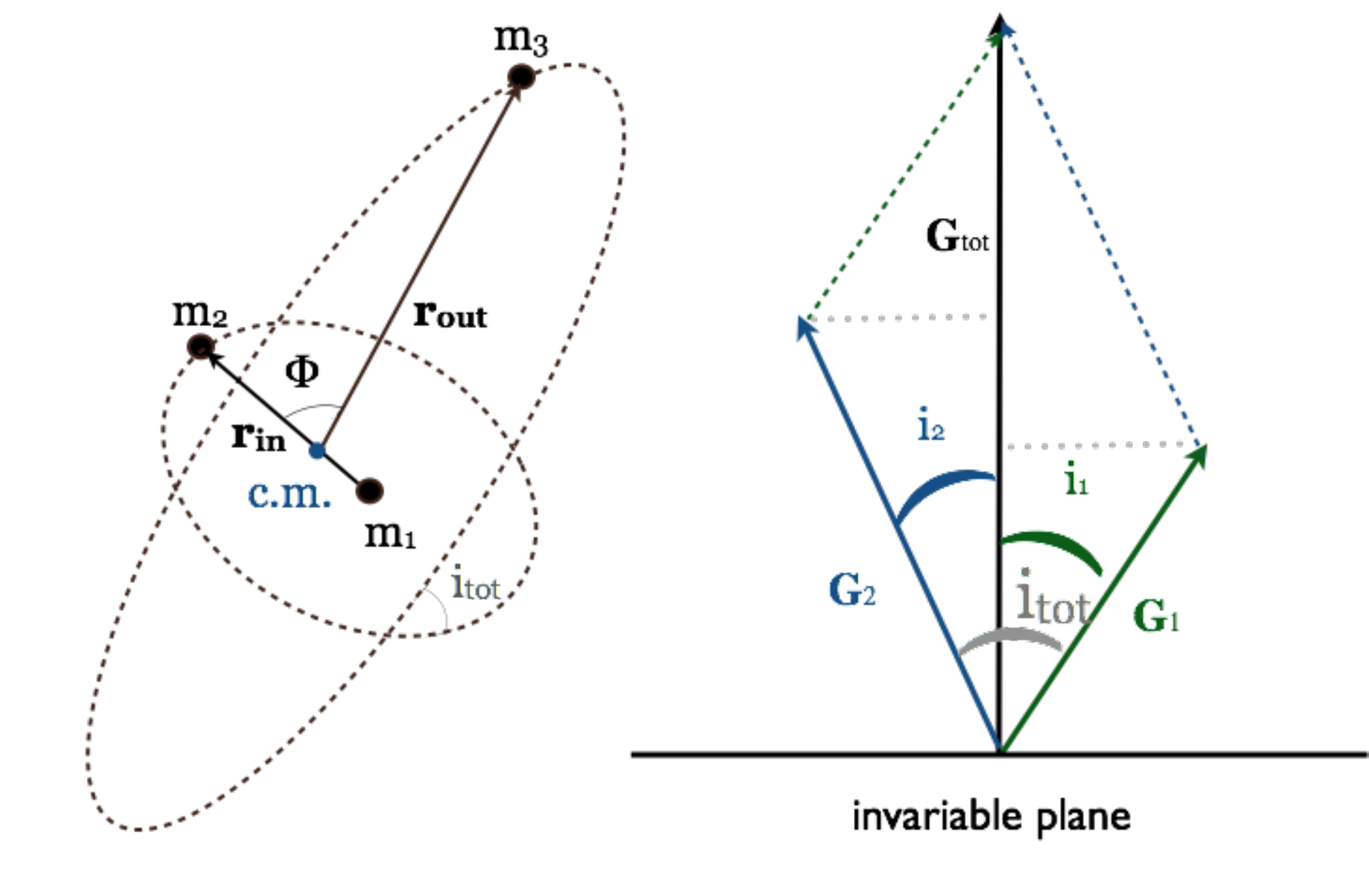}
\caption{\upshape {\bf Schematic description of the coordinate system and the angles used ({\it not to scale}). }
  Left: The three bodies and the relative vectors. 
  Here 'c.m.' denotes the center of mass
  of the inner binary, containing objects of masses $m_1$ and
  $m_2$. The separation vector ${\bf r}_{\in}$ points from $m_1$ to $m_2$;
  ${\bf r}_{\out}$ points from `c.m.' to $m_3$. The angle between the
  vectors ${\bf r}_{\in}$ and ${\bf r}_{\out}$ is $\Phi$. Right: Geometry of the angular momentum vectors and the definition of the relevant inclination angles. We show the total
  angular momentum vector (${\bf G}_\tot$), the angular momentum vector
  of the inner orbit (${\bf G}_1$) with inclination $i_1$ with respect
  to ${\bf G}_\tot$ and the angular momentum vector of the outer orbit
  (${\bf G}_2$) with inclination $i_2$ with respect to
  ${\bf G}_\tot$. The angle between ${\bf G}_1$ and ${\bf G}_2$ defines
  the mutual inclination $i_\tot=i_1+i_2$. The invariable plane is
  perpendicular to ${\bf G}_\tot$, in other words the z axis is parallel to ${\bf G}_\tot$.  } \label{fig:config}
\end{center}
\end{figure}

  In the three-body approximation, dynamical stability requires that either the system has  circular, concentric, coplanar orbits or a hierarchical configuration, in which the {\it inner binary}   is orbited by a third body on a much wider orbit,  the {\it outer binary}  (Figure \ref{fig:config}). In this case the secular approximation (i.e., phase averaged, long term evolution) can be applied, where the interactions between two  non-resonant orbits is equivalent to treating the two orbits as massive wires \citep[e.g.,][]{Maechal90}. Here the line-density is inversely proportional to orbital velocity   and 
 the two orbits torque each other and exchange angular momentum, but
not energy.  Therefore the orbits can change shape and orientation (on
timescales much longer than their orbital periods), but not semi-major
axes of the orbits. The gravitational potential is then expanded in semi-major axis ratio of $a_1/a_2$, where  $a_1$ ($a_2$) is the semi-major axis of the inner (outer) body \citep{Kozai, Lidov}.  This ratio  is a small parameter due to the hierarchical configuration.  
  


 The hierarchical three body system consists of a tight
binary ($m_1$ and $m_2$) and a third body ($m_3$). 
We define  ${\bf r}_{\inner}$ to be the
relative position vector from $m_1$ to $m_2$ and ${\bf r}_{\out}$ the
position vector of $m_3$ relative to the center of mass of the inner
binary (see fig.~\ref{fig:config}).  Using this coordinate system the
dominant motion of the triple can be reduced into two separate
Keplerian orbits: the first describing the relative tight orbit of bodies~1 and~2, and the second describes the wide orbit
of body~3 around the center of mass of bodies~1 and~2. The Hamiltonian
for the system can be decomposed accordingly into two Keplerian
Hamiltonians plus a coupling term that describes the (weak)
interaction between the two orbits.  Let the  semi-major axes (SMAs) of the inner and
outer orbits be $a_1$ and $a_2$, respectively. Then the coupling term
in the complete Hamiltonian can be written as a power series in the
ratio of the semi-major axes $\alpha=a_1/a_2$
\citep[e.g.,][]{Har68}. In a hierarchical system, by definition, this
parameter $\alpha$ is small.

The complete Hamiltonian expanded in orders of $\alpha$ is
\citep[e.g.,][]{Har68},
\begin{equation}
\label{eq:Ham1}
\Ham=\frac{k^2m_1m_2}{2a_1}+\frac{k^2 m_3(m_1+m_2)}{2a_2} + \frac{k^2}{r_2} \sum_{n=2}^{n=\infty} \left(\frac{r_{\inner}}{r_{\out}}\right)^n M_n P_n(\cos{\Phi}) 
\end{equation}
and in terms of the semi-major axises $a_1$ and $a_2$:
\begin{equation}
\label{eq:Ham2}
\Ham= \frac{k^2m_1m_2}{2a_1}+\frac{k^2 m_3(m_1+m_2)}{2a_2} + \frac{k^2}{a_2} \sum_{n=2}^\infty \left(\frac{a_1}{a_2} \right)^n M_n \left(\frac{r_1}{a_1}\right)^n\left(\frac{a_2}{r_2}\right)^{n+1}P_n(\cos{\Phi}) 
\end{equation}
where $k^2$ is the gravitational constant, $P_n$ are Legendre
polynomials, $\Phi$ is the angle between ${\bf r}_{\inner}$ and ${\bf r}_{\out}$
(see Figure \ref{fig:config}) and
\begin{equation}
\label{eq:Mj}
M_n=m_1m_2m_3\frac{m_1^{n-1}-(-m_2)^{n-1}}{(m_1+m_2)^{n}} \ .
\end{equation}
The right term is often called the perturbing function as it describes  the gravitational  perturbations between the two orbits. 
The left two terms in Equation (\ref{eq:Ham2}) are simply the energy of the inner and outer Kepler orbits. Note that the sign convention for this Hamiltonian is positive .

%

The frame of reference chosen throughout this review is the {\it invariable plane} for which the z axis is set along the total angular momentum, which is conserved during the secular evolution of the system (see figure \ref{fig:config}), \citep[e.g.,][]{LZ74}. Another description used in the literature is the vectorial form  \citep[e.g.][]{Katz+11,Boue+14}, which  has been proven to be useful to address different astrophysical setting. Considering the invariable plane it is convenient to 
 adopt the canonical variables known as Delaunay's elements, \citep[e.g.][]{3book}. These describe for each orbit three angles and three conjugate momenta. 
 
 The first set of angles are  the mean anomalies, $M_1$ and $M_2$ (also often denote in the literature as $l_1$ and $l_2$), which describes the position of the object in their orbit. Their 
 conjugate momenta  are:
\begin{eqnarray}
L_1&=&\frac{m_1 m_2}{m_1+m_2}\sqrt{k^2(m_1+m_2)a_1} \ , \label{eq:L1}\\ \nonumber
L_2&=&\frac{m_3(m_1+ m_2)}{m_1+m_2+m_3}\sqrt{k^2(m_1+m_2+m_3)a_2} \ , \label{eq:L2}
\end{eqnarray}
where subscripts $1,\,2$ denote the inner and outer orbits,
respectively.
The second set of angles are the arguments of periastron, $\omega_1$ and $\omega_2$ ($g_1$ and
$g_2$), which describes the position of the eccentricity vector (in the plane of the ellipse). Their conjugate momenta are the magnitude of the angular momenta vector of each orbit $G_1$ and $G_2$ (often used as  $J_1$ and $J_2$):
\begin{equation}
G_1=L_1\sqrt{1-e_1^2} \ , \quad G_2=L_2\sqrt{1-e_2^2} \ ,
\end{equation}
where $e_1$ ($e_2$) is the inner (outer) orbit eccentricity. 
  The last set of angles are 
  the longitudes of ascending
nodes, $\Omega_1$ and $\Omega_2$ ($h_1$ and $h_2$). Their conjugate momenta are 
\begin{equation}
H_1=G_1\cos{i_1} \ , \quad H_2=G_2\cos{i_2} \  ,
\end{equation}
(often denote as $J_{1,z}$ and $J_{2,z}$). Note
that $G_1$ and $G_2$ are  the magnitudes of the angular momentum
vectors (${\bf G}_1$ and ${\bf G}_2$), and $H_1$ and $H_2$ are the
$z$-components of these vectors, (recall that the  $z$-axis is chosen to be along the total angular momentum ${\bf G}_{\tot}$).
In Figure~\ref{fig:config} we show the configuration of the angular momentum
vectors of the inner and outer orbit (${\bf G}_1$ and ${\bf G}_2$, respectively) and $H_1$ and $H_2$ are the
$z$-components of these vectors, where the $z$-axis is chosen to be along the total angular momentum ${\bf G}_{\tot}$. 
This conservation of the total angular momentum $G_{\tot}$ yields a simple relation between the $z$ component of the angular momenta and the total angular momentum magnitude: 
\begin{equation}
G_{\tot}=H_1+H_2 \ .
\end{equation}

The equations of motion are given by the canonical relations  (for these equations we will use the $l,g,h$ notation):
\begin{eqnarray}
\label{eq:Canoni}
\frac{dL_j}{dt}=\frac{\partial \Ham}{\partial l_j} \ , \quad \frac{dl_j}{dt}=-\frac{\partial \Ham}{\partial L_j} \ , \\
\frac{dG_j}{dt}=\frac{\partial \Ham}{\partial g_j} \ , \quad \frac{dg_j}{dt}=-\frac{\partial \Ham}{\partial G_j} \ , \\
\frac{dH_j}{dt}=\frac{\partial \Ham}{\partial h_j} \ , \quad \frac{dh_j}{dt}=-\frac{\partial \Ham}{\partial H_j} \ ,
\label{eq:Canoni3}
\end{eqnarray}
where $j=1,2$. Note that these canonical relations have the opposite
sign relative to the usual relations \citep[e.g.,][]{Goldstein}
 because of the sign convention typically chosen for this Hamiltonian.

As apparent from the Hamiltonian Eq.~(\ref{eq:Ham2}), if the semi-major axis ratio is indeed a small parameter then for the zeroth approximation each orbit can be described as a Keplerian orbit, for which its energy is conserved. Thus, we can average over the short timescale and focus on the 
 long-term dynamics of the triple
system. This process is known as the {\it secular approximation}, where the energy (semi-major axis) is conserved and the orbits exchange angular momentum.  The short timescales terms in the Hamiltonian  depend on  $l_1$ and $l_2$, and eliminating them needs is done via a canonical transformation. The technique used is known as the Von Zeipel transformation
\citep[][]{bro59}.  In this canonical transformation, a time independent generating function is defined to  be periodic in  $l_1$ and $l_2$, which allows the elimination of   the short-period terms in the Hamiltonian and the details of this procedure are described in \citet{Naoz+11sec} Appendix A2. Eliminating these angles from the Hamiltonian yields that their conjugate momenta $L_1$ and $L_2$ are conserved [see E1.~(\ref{eq:Canoni})], thus yielding $a_1={\rm Const.~}$and $a_2={\rm Const.}$, as expected.  In the most general case  of this three body secular approximation there are only two parameters which are conserved, i.e., the energy of the system (which also means that the energy of the inner and the outer orbits are conserved separately), and the total angular momentum $G_{\tot}$. 

The time evolution for the eccentricity and inclination of the system can be easily  achieved from Equations (\ref{eq:Canoni})-(\ref{eq:Canoni3})
\begin{equation}
\frac{de_j}{dt}=\frac{\partial e_j}{\partial G_j}\frac{\partial \Ham}{\partial g_j} \ ,
\end{equation}
and
\begin{equation}
\frac{d (\cos i_j)}{dt}= \frac{\dot{H}_j}{G_j}-\frac{\dot{G}_j}{G_j} \cos i_j \ ,
\end{equation}
where $j=1,2$ for the inner and outer orbit. See full set of equations of motions in Equations (\ref{eq:g1dot8})-(\ref{eq:cosi2}).

 The lowest order of approximation, which is proportional to $(a_1/a_2)^2$  is called {\it quadrupole} level, and we find that an artifact of the averaging process results in conservation of the outer orbit angular momentum $G_2$, in other words the system is symmetric for the rotation of the outer orbit. This was coined as the ``happy coincidence" by \citet{Lidov+76}. Its significant consequence  is  that the this approximation should  be {\it only} used for  an axisymmetric outer potential such as  circular  outer orbits \citep{Naoz+11sec}.
 
The next level of approximation, the octupole, is  proportional to $(a_1/a_2)e_2/(1-e_2^2)$ (see below) and thus the TPQ approximation can be successfully applied when this parameter is small for low inclinations (see below for numerical studies). However,  close to perpendicular systems are extremely sensitive to this parameter.

 A popular  procedure which was done in earlier studies \citep[e.g.,][]{Kozai} used the ``elimination of nodes'' \citep[e.g.,][]{JM66}. This describes the a simplification of the Hamiltonian  by setting \begin{equation}
\label{eq:pi}
h_1 - h_2 = \pi \ . 
\end{equation} This relation holds in the invariable plane when the total angular momentum is conserved, such as in our case. Some studies that exploited explicitly this relation in the Hamiltonian incorrectly concluded [using Equation (\ref{eq:Canoni3})] that  the
$z$-components of the orbital angular momenta are always constant.  As showed in \citet{Naoz11,Naoz+11sec}, this  leads to qualitatively different evolution for the triple body system.  We can still use the Hamiltonian with the nodes
eliminated as
long as the  equations of motions for the inclinations are derived from
the total angular momentum conservation, instead of using the
canonical relations \citep{Naoz+11sec}.

\subsection{Physical picture}

Considering the  quadrupole level of approximation (which is valid for axisymmetric outer orbit potential) for an inner test particle  (either $m_1$ or $m_2$ goes to zero) the conserved quantities are the energy and the z-component of the angular momentum. In other words the Hamiltonian does not depend on longitude of acceding nodes ($h_1$) and thus the z-component of the inner orbit angular momentum, $H_1$, is conserved 
and the system is integrable. In this case the equal precession rate of the inner orbit's longitude of ascending nodes ($\Omega_1$) and the longitude of the periapsis ($\varpi=\Omega_1+\omega_1$) means that an eccentric inner orbit feels an accumulating effect on the orbit.  The   the resonant angle $\omega_1=\varpi_1-\Omega_1$, will librate around $0^\circ$ or $180^\circ$ which cause large amplitude eccentricity oscillations of the inner orbit. 

In that case (circular outer orbit, in the test particle approximation) the conservation of the z component of the angular momentum $j_z=\sqrt{1-e_1^2}\cos i_\tot={\rm Const.}$~yields oscillations between the eccentricity and inclination. The inner orbit will be more eccentric for smaller inclinations and less eccentric for larger inclinations.


\subsection{Circular outer body }

In this case the  gravitational potential set by the outer orbit is axisymmetric, and thus the quadrupole level of approximation describes the  behavior of the hierarchical system well. We will consider two possibility, in the first  one of the members of the inner orbit is a test particle, (i.e., either $m_1$ or $m_2$ are zero). In the second we will allow for all three masses to be non-negligible. 


\subsubsection{Axisymmetric Potential and  Inner  Test Particle  - TPQ}\label{sec:TPQ}


Following \citet{LN} we call this case the test particle approximation quadrupole (TPQ). Without loss of generality we take $m_2\to 0$, the Hamiltonian of this system is very simple  and can be written as:
\begin{equation}
\Ham=\frac{3}{8}k^2 \frac{m_1m_3}{a_2}\left(\frac{a_1}{a_2}\right)^2\frac{1}{(1-e_2^2)^{3/2}}F_{quad} \ ,
\end{equation}
where 
\begin{equation}\label{eq:TPQ_Fquad}
F_{\rm quad}=-\frac{e_1^2}{2}+\theta^2+\frac{3}{2}e_1^2\theta^2+\frac{5}{2}e_1^2(1-\theta^2)\cos (2 \omega_1) \ ,
\end{equation}
where $\theta=\cos i_\tot$ \citep[e.g.,][]{Yokoyama+03,LN}\footnote{Note that unlike the Hamiltonian that will be presented in the next section [Equation (\ref{eq:Hamquad})] this Hamiltonian only describes the test particle approximation.   }. 

\begin{figure}
\begin{center}
\includegraphics[width=0.8\linewidth]{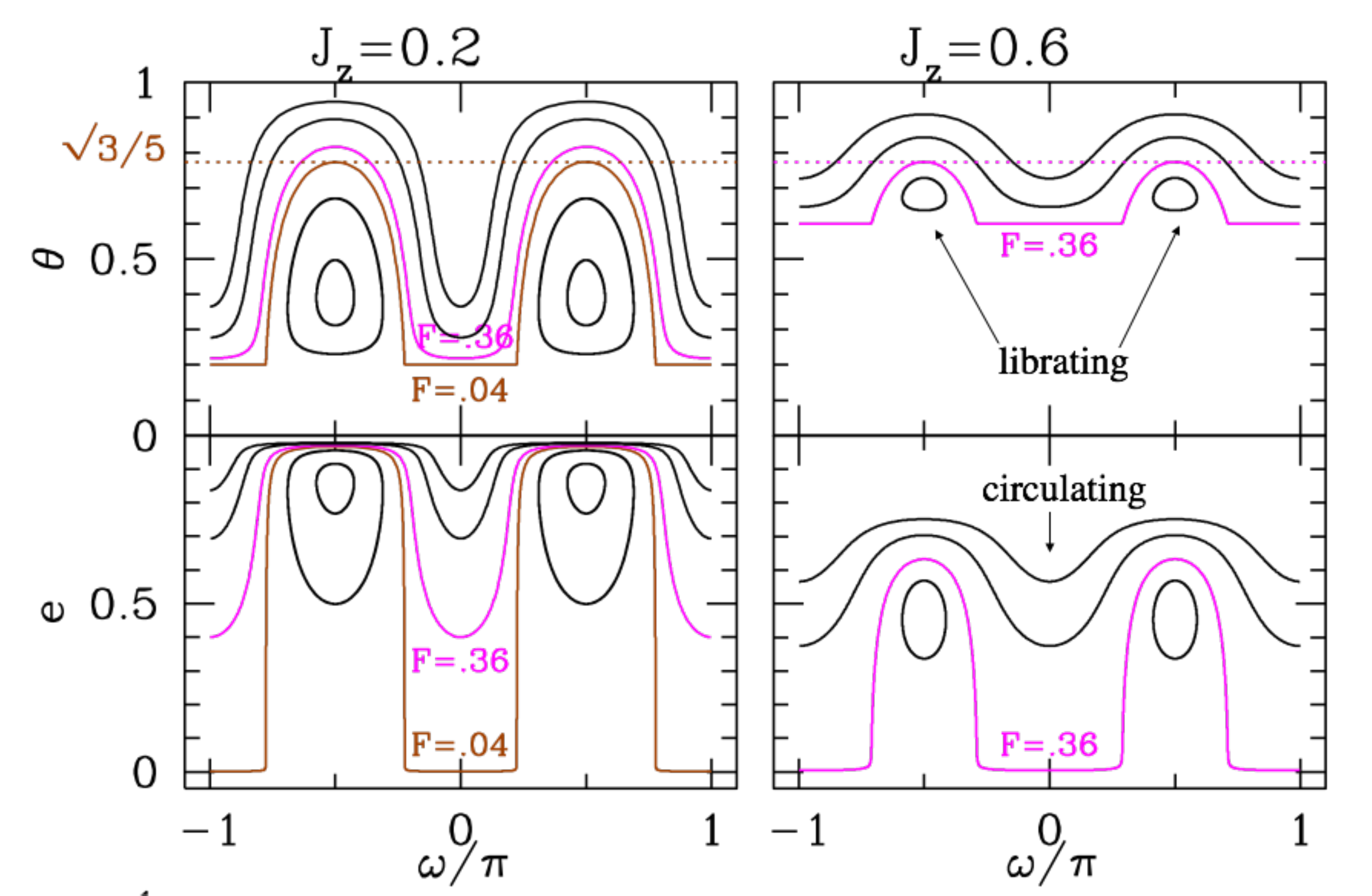}
\caption{\upshape {\bf Cross section trajectory of the TPQ in the $\theta-\omega_1$ (top panels) and $e_1-\omega_1$ (bottom panels) planes. } We define $\theta=\cos i_{\tot}$.  The  dashed horizontal lines in the top panels shows the critical inclination  for which $\theta=\sqrt{3/5}$.  The separatrix is associated with $e_{1}=0$ for $\omega_1=0$ and $\theta=\sqrt{3/5}$  for  $\omega_1=\pi/2$, as depicted in the Figure.
{\it Left panels} shows the case  for $J_z=0.2$ and $F_{\rm quad}^{\rm TP}= -1.44$ and $-.64$ (librating) and   $F_{\rm quad}^{\rm TP}= 0.04,0.36,1$ and $1.44$ (circulating). {\it Right panels} shows the case  for $J_z=0.6$ and $F_{\rm quad}^{\rm TP}= 0.25$ (librating) and   $F_{\rm quad}^{\rm TP}= 0.36,0.64$ and $1$ (circulating). Figure adopted from \citet{LN}.} \label{fig:LN_TPQ}
\end{center}
\end{figure}

At this physical setting the octupole level of approximation is zero and the inner orbit's angular momentum along the z axis is conserved ($H_1\propto j_{z,1}=\sqrt{1-e_1^2}\cos i_\tot =\rm{Const.}$), where $j_{z,1}$ is the specific z component of the angular momentum. Since both $H_1$ and $F_{\rm quad}$ are conserved, a new constant of motion can be defined.  It is convenient (for reasons that will be identified in Section \ref{sec:TPO}) to define the following constant \citep{Katz+11}:
\begin{equation}\label{eq:CKL}
C_{KL}=\frac{F_{\rm quad}}{2}-\frac{1}{2}j_{z,1}^2=e^2\left(1-\frac{5}{2}\sin i_\tot^2 \sin \omega_1^2 \right) \ ,
\end{equation}
which is a simple function of the initial conditions.
 The system is  integrable and has a well defined maximum and minimum eccentricity and inclination. To find the extreme points we set $\dot{e}_1=0$ in the time evolution equation [see Equation (\ref{eq:dote1_8}), quadrupole part] and find that the values of the argument of periapsis that satisfy this condition are $\omega_1= 0 + n\pi/2$, where $n=0,1,2,...$. Thus, the resonant  angle has two classes of trajectories, librating and circulating. On circulating trajectories, at $\omega_1=0$, the eccentricity is smallest and the inclination is largest, and visa versa for $\omega_1= \pi/2$. In  Figure \ref{fig:LN_TPQ},  librating trajectories (or libration modes) are associated with bound oscillations of $\omega_1$ and circulating trajectories (or circulation modes) are not constrained to a specific regime.   The separatrix is the trajectory which separates the two modes of behavior, as we elaborate below.
 
 The conservation of $j_{1,z}$ implies:
 \begin{equation}\label{eq:jz}
 j_{z,1}=\sqrt{1-e_{1,max/min}^2}\cos i_{1,min/max}=\sqrt{1-e_{1,0}^2}\cos i_{1,0} \ ,
 \end{equation}
 where $e_{1,0}$ and $i_{1,0}$ are the initial values. Note that in this case (TPQ) $i_1 =i_{tot}$.
 Since the energy is also conserved, plugging in $\omega_1=0$ for the circulating mode we find
  \begin{equation}\label{eq:Emin}
 E_0= 2{e_{1,min}^2}-2+ (1-{e_{1,min}^2}) \cos i_{max}^2 \ ,
 \end{equation}
 and for $\omega_1=\pm \pi/2$  in equation (\ref{eq:TPQ_Fquad}) we find:
  \begin{equation}\label{eq:Emax}
 E_0=  -3e_{1,max}^2 + (1-4{e_{1,max}^2})\cos i_{min}^2 \ ,
 \end{equation}
 where $E_0$ represents the initial conditions plugged  in  equation (\ref{eq:TPQ_Fquad}). From equations  (\ref{eq:jz}) and (\ref{eq:Emin}) one can easily find the minimum eccentricity and maximum inclination, and from  equations  (\ref{eq:jz}) and (\ref{eq:Emin}) the maximum eccentricity and the minimum inclination.  A special  and  useful case is found by setting initially $e_{1,0}=0$ and $\omega_{1,0}=0$, for this case the maximum eccentricity is 
   \begin{equation}\label{eq:e_max}
 e_{max}=\sqrt{1-\frac{5}{3}\cos^2 i_0} \ .
 \end{equation}
Solving the equations for $\cos i_{min}$ instead we can find 
  \begin{equation}\label{eq:e_max}
\cos i_{min}= \pm \sqrt{\frac{3}{5}}  \ ,
 \end{equation}
 which gives $i_{min}=39.2^\circ$ and $i_{min}=140.77^\circ$, known as Kozai angles. These angles represent the regime where large eccentricity and inclination oscillations are expected to take place. 
  The value $\cos i_{min}= \pm \sqrt{{3}/{5}}$ marks the seperatrix  depicted  in Figure \ref{fig:LN_TPQ}.
 



\subsubsection{Axisymmetric Potential Beyond the Test Particle Approximation}

In this case we still keep the outer orbit circular, thus the quadrupole level of approximation still valid, but we will relax the test particle approximation. 
The quadrupole level hamiltonian can be written as:
\begin{equation}\label{eq:Hamquad}
  \Ham_{quad} = C_2 \{ \left( 2 + 3 e_1^2 \right) \left( 3 \cos^2 i_\tot - 1 \right)  +  15 e_1^2 \sin^2 i_\tot \cos(2 \omega_1) \} \ ,
\end{equation}
where \begin{equation}\label{eq:C2}
  C_2=\frac{k^4}{16}\frac{(m_1+m_2)^7}{(m_1+m_2+m_3)^3}\frac{m_3^7}{(m_1 m_2)^3}\frac{L_1^4}{L_2^3 G_2^3} \ .
\end{equation}
Note that in this form of Hamiltonian the nodes ($\Omega_1$ and $\Omega_2$) have been eliminated, allowing for a cleaner format, however this does not mean that the z-component of the inner and outer angular momenta are constant of motion \citep[as explained in][]{Naoz11,Naoz+11sec}.

Relaxing the test particle approximation (i.e., none of the masses have insignificant mass) already allows for deviations  from the nominal TPQ behavior. This is because now $j_{z,1}$ is no longer conserved and instead the total angular momentum is. Note that the outer potential is axisymmetric  and $G_2={\rm Const}$. The system is still integrable and has well define maxima and minima for the eccentricity and inclination.  
The conservation of the total angular momentum, i.e., ${\bf G}_1+{\bf
  G}_2={\bf G}_\tot$ sets the relation between the maximum/minimum total inclination 
and inner orbit eccentricity. 
\begin{equation}
 \label{eq:ang}
L_1^2(1-e_1^2)+2L_1L_2\sqrt{1-e_1^2}\sqrt{1-e_2^2}\cos i_\tot = G_\tot^2-G_2^2 \ .
\end{equation}
Note that in the quadrupole-level approximation $G_2$, and thus  $e_2$, are
constant.  The right hand side of the above equation is set by the
initial conditions. In addition, $L_1$, and $L_2$ [see
eqs.~(\ref{eq:L1}) and (\ref{eq:L2})] are also set by the initial
conditions.  Using the conservation of energy we can write, for the
minimum eccentricity/maximum inclination  case (i.e., setting $\omega_1=0$)
 \begin{equation}
 \label{eq:Ep}
\frac{\Ham_{quad}}{2 C_2}=3\cos ^2 i_{\tot,max} (1-e_{1,min}^2)-1+6e_{1,min}^2 \ .
\end{equation}
The left hand side of this equation, and the remainder of the parameters in  Equation (\ref{eq:ang}) are
determined  by the initial conditions. 
Thus solving equation (\ref{eq:Ep}) together with (\ref{eq:ang}) gives the minimum eccentricity/maximum inclination during the system evolution as a function of the initial conditions. 
 We find a similar
equation if we set $\omega_1=\pi/2$ for the maximum eccentricity/minimum inclination :
 \begin{equation}
 \label{eq:Em}
\frac{\Ham_{quad}}{2 C_2}=3\cos ^2 i_{\tot,min} (1+4e_{1,max}^2)-1-9e_{1,max}^2 \ .
\end{equation}
Equations (\ref{eq:ang}) and (\ref{eq:Em}) give a
simple relation between the total minimum inclination and the maximum inner
eccentricity as a function of the initial conditions.

\begin{figure}
\begin{center}
\includegraphics[width=0.6\linewidth]{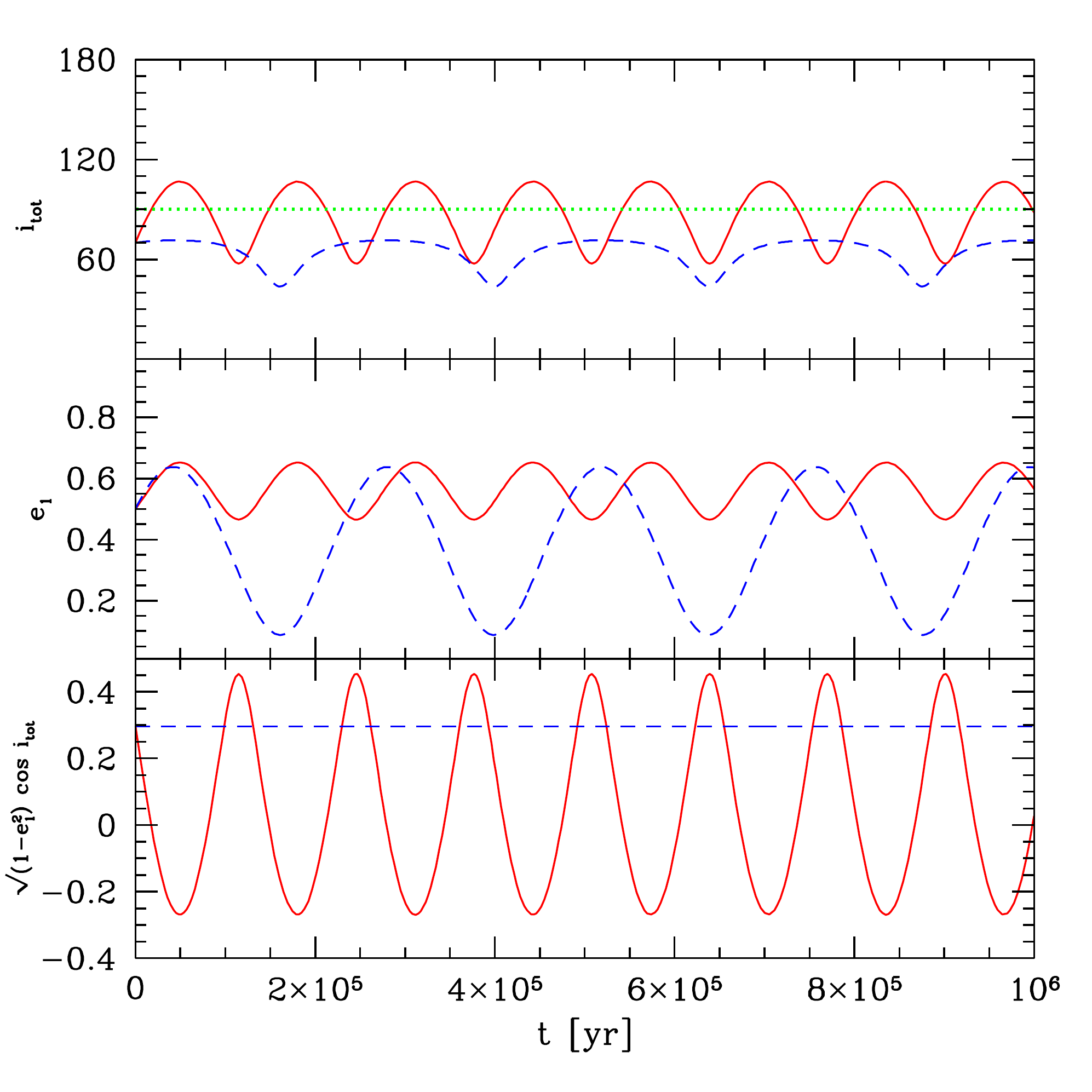}
\caption{\upshape Comparison between the  TPQ formalism (dashed blue
  lines) and the full quadrupole calculation  (solid red lines). The system has an inner binary with 
   $m_1=1.4\,M_\odot$ and  $m_2=0.3\msun$, and the
  outer body has mass $m_3=0.01\msun$. The orbit separations are  $a_1 =
  5\,$AU and $a_2 = 50\,$AU. The system was set initially with 
  $e_1 =0.5$ and $e_2 =0$, $\omega_1=120^\circ$, $\omega_2=0$ and 
  relative inclination $i_\tot=70^\circ$.  The panels show from top to bottom,
  the mutual inclination $i_\tot$, $e_1$ and
  $\sqrt{1-e_1^2} \cos i_\tot$, which in the TPQ formalism is  constant (dashed line).  Figure adopted from \citet{Naoz+11sec}.} \label{fig:quad}
\end{center}
\end{figure}

An interesting consequence of this physical picture is if the inner binary members are more massive than the third object. We adopt this example from \citet{Naoz+11sec} and consider the triple system PSR~B1620$-$26. 
 The inner binary contains a millisecond radio pulsar of
$m_1=1.4 \msun$ and a companion of $m_2=0.3\msun$
\citep[e.g.,][]{ML88}. We adopt parameters for the
outer perturber of $m_3=0.01\, \msun$ \citep{Ford00Pls} and set $e_2=0$ (see the caption for a full description of the initial conditions). Note that
\citet{Ford00Pls} found $e_2=0.45$, which  means that the quadrupole level of approximation is insufficient to represent the behavior of the system. We choose, however, to set $e_2=0$ to emphasis the point that 
even  an axisymmetric outer potential may result in a qualitative different behavior if the TPQ approximation is assumed. 
For the same reason we also adopt a higher initial value for the inner orbit eccentricity ($e_1=0.5$ compared to the measured one $e_1\sim 0.045$). The time evolution of the system is shown in Figure \ref{fig:quad}. 
In This Figure we compare the z-component of the angular momentum $H_1$  (solid red line) with
$L_1 \sqrt{1-e_1^2} \cos i_\tot$ (dashed blue line), which is the  angular momentum that would
be inferred \emph{if the outer orbit were
  instantaneously in the invariable plane}, as found in the TPQ
formalism.

\begin{figure}
\begin{center}
\includegraphics[width=0.7\linewidth]{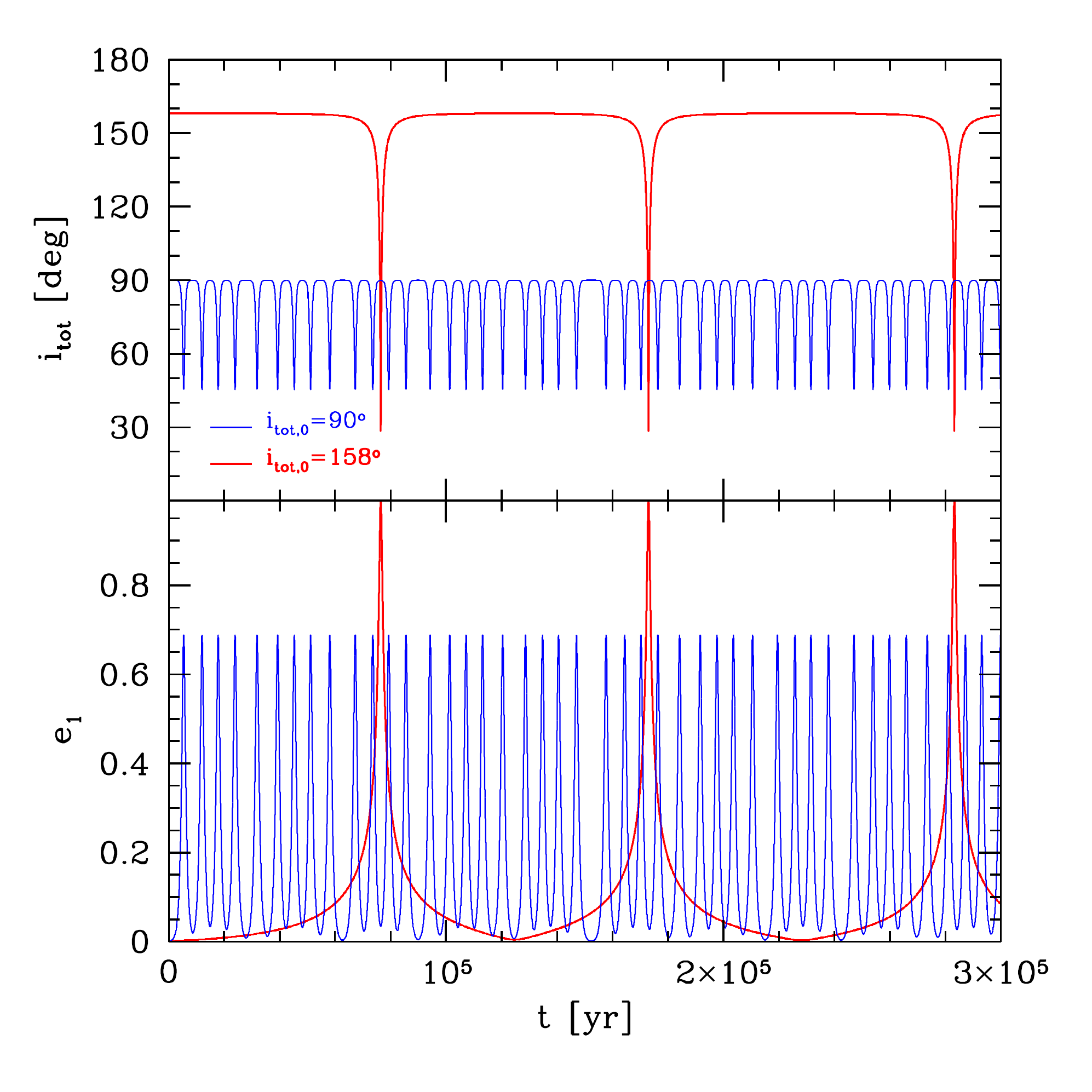}
\caption{\upshape {\bf Small mass outer perturber that induces large eccentricity excitation away from the nominal range of the Kozai angles of $39.2^\circ - 140.77^\circ$}. We consider $m_1=1$~M$_\odot$, $m_2=0.5$~M$_\odot$, $m_3=0.05$~M$_\odot$, $a_1=0.5$~AU and $a_2=5$~AU. Both outer and inner eccentricities are set initially to zero, and also set initially  $\omega_1=90^\circ$ and $\omega_2=0^\circ$. We show two examples, the first shows the eccentricity excitations for as expected initial mutual inclination of $i_{\rm tot}=90^\circ$, where in this case $i_1=25.01^\circ$ and  $i_2=64.99^\circ$. This produces eccentricity excitation with $e_{1, {\rm max}}=0.689$. We also consider an example for which the mutual inclination is set initially to be $i_{\rm tot}=158^\circ$. In this case 	$i_1=17.12^\circ$ and  $i_2=140.88^\circ$. The latter parameters are adopted from  \citet{Martin+15}, which leads to maximum inner  eccentricity of $e_{1, {\rm max}}=0.99$. Note that in both examples $i_2$ is close to the nominal Kozai angles range. } \label{fig:smallm}
\end{center}
\end{figure}

Taking the outer body to be much smaller than the inner binary (i.e., $m_3<m_1,m_2$), as done in Figure \ref{fig:quad}, yields yet another interesting consequence for relaxing the test particle approximation. In some cases large eccentricity excitations can take place for inclinations that largely deviate from the nominal range of the Kozai angles of $39.2^\circ - 140.77^\circ$. The limiting mutual inclination that can result in large eccentricity excitations can be easily found when solving Equations (\ref{eq:ang}) and (\ref{eq:Em}), since they depend on mutual  inclination, as noted by \citet{Martin+15}. This evolution is shown in Figure \ref{fig:smallm}, where large eccentricity oscillation for the inner binary is  achieved for  an initial  mutual inclination of $158^\circ$. This behavior, as expected from the Equations, is sensitive to the eccentricity of the outer orbit. 

In the circular outer orbit case, the regular oscillations of the eccentricity and inclination yields a well defined associated timescale. This can be easily achieved by considering the equation of motion of the argument of periapsis $\omega_1$ [see the part that is proportional to $C_2$ in Equation (\ref{eq:g1dot8})].
  More precisely, $t_{\rm quad} \sim G_1/C_{2}$, where $C_{2}$ is given in Eq.~(\ref{eq:C2}). Integrating between the well defined maximum and minimum eccentricity \citet{Antognini15} found a numerical factor $16/15$, and got
\begin{eqnarray}\label{eq:tquad}
t_{\rm quad}&\sim& \frac{16}{15}\frac{ a_2^3 (1-e_2^2)^{3/2}\sqrt{m_1+m_2}}{ a_1^{3/2} m_3 k} \nonumber \\ & = &\frac{16}{30\pi} \frac{m_1+m_2+m_3}{m_3}\frac{P_2^2}{P_1}(1-e_2^2)^{3/2}  \ .
\end{eqnarray}
This timescale is in a good  agreement with the numerical evolution.

\subsection{Eccentric outer orbit}

\subsubsection{ Inner Orbit's Test Particle Approximation }\label{sec:TPO}

In this approximation we will allow for an eccentric outer orbit but will restrict ourselves to take the mass of one of the inner members to zero, which yields $i_1=i_\tot$. In the test particle limit, the outer orbit is stationary and the system reduces to two degrees of freedom. The eccentric outer orbit yields the quadrupole level of approximation inadequate and thus we consider the test particle octupole (TPO) level here.    This approximation is extremely useful in gaining an overall understanding of the general hierarchical system and the EKL mechanism. 
The hamiltonian $H^{TP}$ of this system is very simple  and can be written as \citep[e.g.,][]{LN}, 
\begin{equation}
\Ham^{TP}=\frac{3}{8}k^2 \frac{m_1m_3}{a_2}\left(\frac{a_1}{a_2}\right)^2\frac{1}{(1-e_2^2)^{3/2}}\left(F_{\rm quad}+\epsilon F_{\rm oct}\right) \ ,
\end{equation}
where 
\begin{equation}
\epsilon=\frac{a_1}{a_2}\frac{e_2}{1-e_2^2} \ ,
\end{equation}
 $F_{\rm quad}$ is defined in Equation  (\ref{eq:TPQ_Fquad}), and we reiterate it here for completeness, 
 \begin{equation}
F_{\rm quad}=-\frac{e_1^2}{2}+\theta^2+\frac{3}{2}e_1^2\theta^2+\frac{5}{2}e_1^2(1-\theta^2)\cos (2 \omega_1) \ , \nonumber
\end{equation}
and
\begin{eqnarray}\label{eq:octTP}
F_{\rm oct} &= & \frac{5}{16}\left(e_1+\frac{3e_1^3}{4}\right)  [(1-11\theta-5\theta^2+15\theta^3)\cos(\omega_1-\Omega_1) + (1+11\theta-5\theta^2-15\theta^3)\cos(\omega_1+\Omega_1)]\nonumber \\
		&	& -\frac{175}{64} e_1^3[(1-\theta-\theta^2+\theta^3)\cos(3\omega_1-\Omega_1) +(1+\theta-\theta^2-\theta^3)\cos(3\omega_1+\Omega_1)] \ .
\end{eqnarray}

\begin{figure}
\begin{center}
\includegraphics[width=0.7\linewidth]{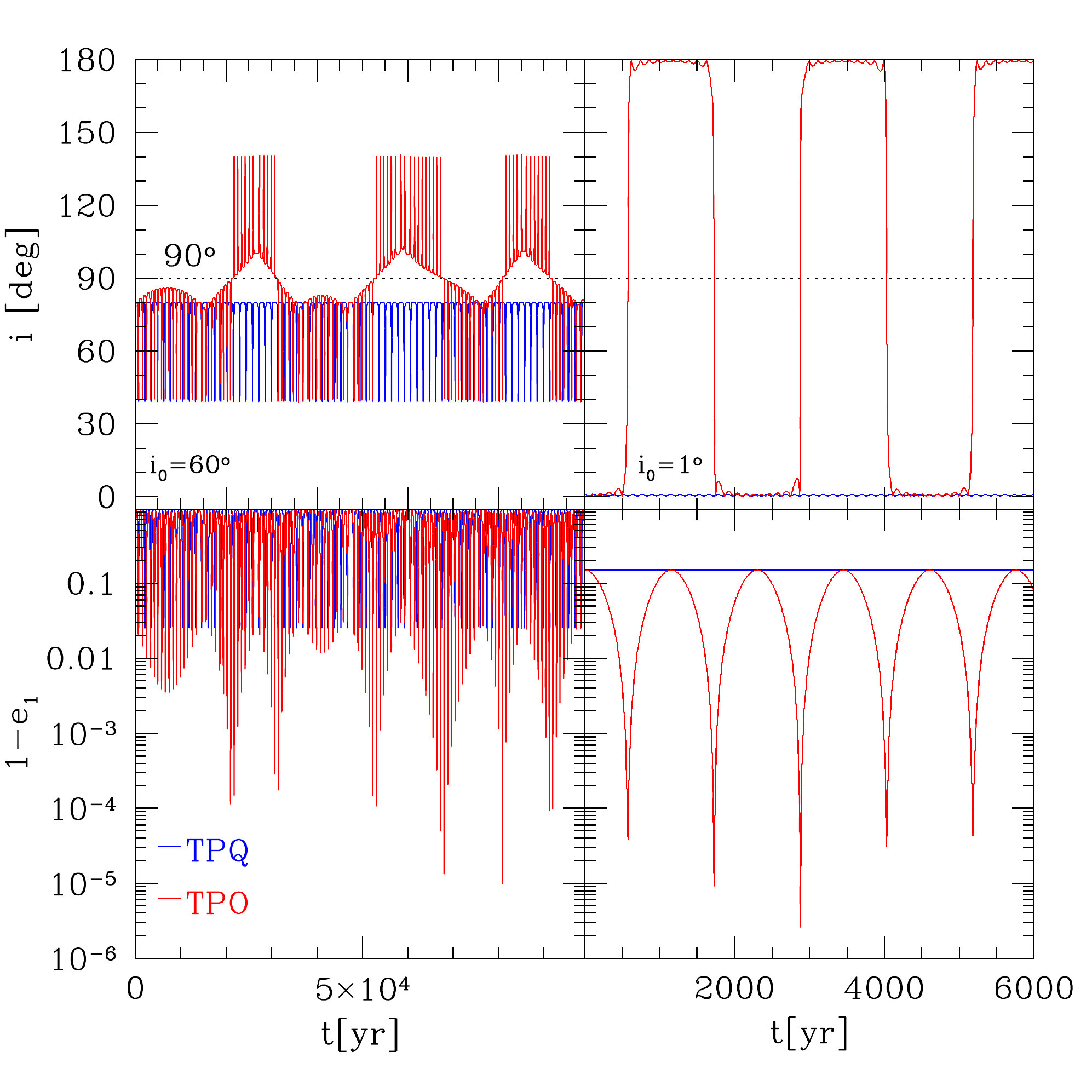}
\caption{\upshape {\bf Time evolution example of the TPO (test partial octupole) approximation (red lines) and the TPQ (test particle quadrupole) approximation (blue lines).} {\it Left panels} show high inclination flip while the {\it right panels} show the low inclination flip (see text). In this example we consider the time evolution of a test particle at $135$~AU around a  $10^4$~M$_\odot$ intermediate black hole located $0.03$~pc from the massive black hole in the center of our galaxy ($4\times10^6$~M$_\odot$).  In the left panels the system initially is set with $e_1=0.01$, $e_2=0.7$, $i=60^\circ$, $\Omega_1=60^\circ$ and $\omega_1=0^\circ$. In the right panels the system is initially set with $e_1=0.85$, $e_2=0.85$, $i=1^\circ$, $\Omega_1=180^\circ$ and $\omega_1=0^\circ$. In the top panels we show the inclination and in the bottom the inner orbit eccentricity as $1-e_1$. } \label{fig:IMBH}
\end{center}
\end{figure}

In this case the z component of the outer orbit is not conserved and the system can flip from $i_\tot <90^\circ$ to $i_\tot>90^\circ$ \citep{Naoz11,Naoz+11sec}. The flip is associated with an extremely high eccentricity  transition (see for example  Figure \ref{fig:IMBH}).  The octupole level of approximation introduces higher order resonances which overall render the system to be qualitatively different from a system at which the  quadrupole level of approximation is applicable. 
We will begin by reviewing the different effects in the systems which can be divided into two main initial inclination regimes.   

  \addcontentsline{toc}{subsubsection}{High initial inclination regime and chaos}
\subsubsection*{High initial inclination regime and chaos}

\begin{figure}
\begin{center}
\includegraphics[width=0.8\linewidth]{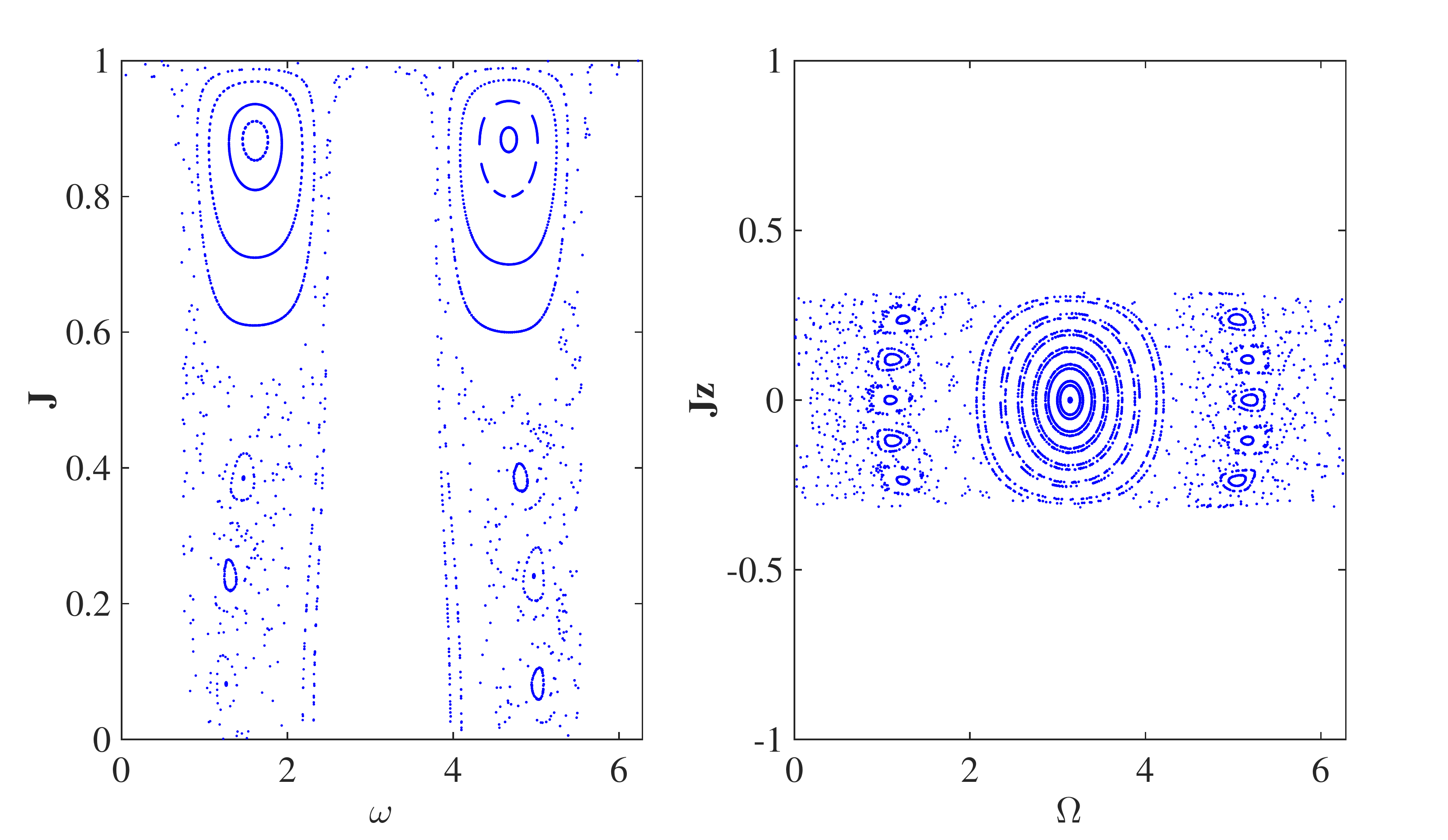}
\caption{\upshape Surface of section for $F_{quad}+\epsilon F_{oct}=-0.1$ and $\epsilon=0.1$. This initial configuration is associated with high initial inclination $i_{\tot,0}>39.2^\circ$. The quadrupole level resonances can clearly be seen  (the big islands) as well as the emergence  of high order resonances (the small islands). Figure adopted from \citet{Li+14Chaos}. } \label{fig:EKLTestP}
\end{center}
\end{figure}

When the system begins with in a high inclination regime $39.2^\circ\leq i_{\tot}\leq 140.7^\circ$ the resonance arises from the quadrupole level of approximation can cause large inclination and eccentricity amplitude modulations. Recall that this angle range is associated  with the TPQ seperatrix.   The octupole-level of approximation  is associated with high order resonances that result in extremely large eccentricity peaks, flips (see Figure \ref{fig:IMBH}) as well as chaotic  behavior (as explained below). 
%
 As can be seen from equation (\ref{eq:octTP}) these resonances arise from higher order harmonics of the octupole-level  Hamiltonian: $\omega_1\pm \Omega_1$ and $3\omega_1\pm \Omega_1$. A useful  tool to analyze this system is in the form of surface of section (see fore example Figure \ref{fig:EKLTestP}). For a two-degrees of freedom system, the surface of section projects a four-dimensional trajectory on a two-dimensional surface. The resonant regions are associated with fixed points and chaotic zones are a result of the overlap of the resonances between the quadrupole and the octupole resonances \citep{Chirikov79,Murray+97}.
 
\begin{figure}
\begin{center}
\includegraphics[width=\linewidth]{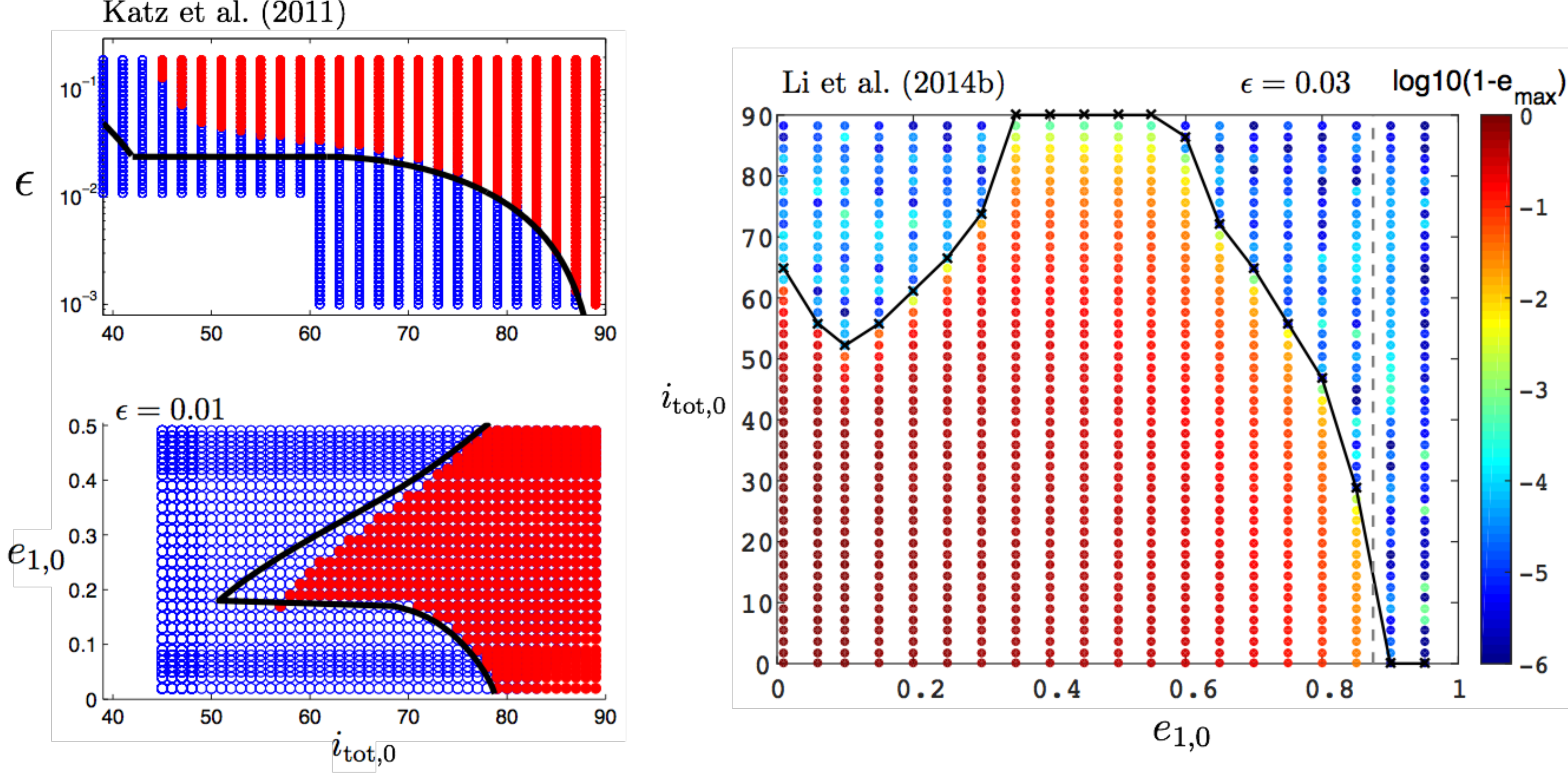}
\caption{\upshape {\bf High inclination flip parameter space} {\it Right panel} shows the results of numerical integrated systems   associated maximum eccentricity  (color coded as $1-e_1$) in the $i_{\tot,0}-e_{1,0}$ parameter space for  $\epsilon=0.03$, after $30 t_{\rm quad}$. Systems above the black line flipped.  {\it Left panel} shows the comparison with the analytical conditions derived by   \citet{Katz+11}. Open circles are the result of a numerical integration (red indicates systems that flipped and blue is for those that did not).  Here the solid lines represent the flip conditions which for $e_{1,0}\sim 0$ and $i_{\tot,0}\gsim61.7^\circ$ is reduced to equation (\ref{eq:filpHi}). The bottom left panel shows the case of $\epsilon=0.01$, and note that it shows only part of the parameter space. Left panels are adopted from  \citet{Katz+11} and right panel is adopted from  \citet{Li+13}.} \label{fig:HighFlip}
\end{center}
\end{figure}

 Figure \ref{fig:EKLTestP} shows the surface of section for   $\epsilon=0.1$ and $F_{quad}+\epsilon F_{oct}=-0.1$, which is associated  with high initial inclination $i_{\tot,0}>39.2^\circ$.  In this Figure we can identify  three distinct regions: Òresonant regions,Ó Òcirculation regions,Ó and Òchaotic regionsÓ. 
The resonant regions are associated with trajectories of which  the momenta ($J$ and $J_z$) and the angles $\omega_1$ and $\Omega_1$) undergo bound oscillations. The system is classified in a  liberation mode and the trajectories  are quasi-periodic. The libration zones in the TPQ approximation are shown in Figure  \ref{fig:LN_TPQ}, and for the TPO in Figure \ref{fig:EKLTestP}.
The circulation regions describes  trajectories for which  the coordinates are not constrained to a specific interval, and can take any value. Note that both resonant and circulatory trajectories map onto a one-dimensional manifold on the surface of section. On the contrary, chaotic trajectories map onto a two-dimensional manifold. In other words, while quasi-periodic trajectories form lines on the surface of section, chaotic trajectories are area-filling regimes. Embedded in the chaotic region, the small islands correspond to the higher, octupole order resonances, which are also quasi-periodic . 
The flip from $i_\tot <90^\circ$ to $i_\tot>90^\circ$ covers large parts of the parameter space as can be seen in Figure \ref{fig:HighFlip}  right panel. 

In some cases an analytical condition for the flip can be achieved by averaging over a quadrupole cycle  \citep{Katz+11}. 
This averaging process yields a constant of motion 
\begin{equation}\label{eq:filpHiepsilon}
\chi=f(C_{KL})+\epsilon \frac{\cos i_\tot \sin \Omega_1 \sin \omega_1-\cos \omega_1\cos\Omega_1}{\sqrt{1-\sin^2\i_\tot \sin^2\omega_1}}={\rm Const.}~ \ ,
\end{equation}
where 
  the function $f(C_{KL})$ is defined by:
\begin{eqnarray}\label{eq:filpHi}
\mathit{f}(C_{KL})&=& \frac{32\sqrt{3}}{\pi}\int^1_{x_{min}}\frac{K(x)-2E(x)}{(41x-21)\sqrt{2x+3}} dx \quad {\rm and}  
 \quad x_{min}=\frac{3-3C_{KL}}{3+2C_{KL}}\ ,
\end{eqnarray}
where $K(x)$ and $E(x)$ are the complete elliptic functions of the first and second kind, respectively. 
 For initial high inclination  a flipping critical   value for  the octupole pre-factor $\epsilon_c$ is a function of the initial inclination and the approximations takes a simple form
\begin{equation}\label{eq:filpHiepsilon}
\epsilon_c=\frac{1}{2}{\rm max}|\Delta f(y)| \ , 
\end{equation}
where $\Delta f(y)=f(y)-f(C_{KL,0})$,  $C_{KL}$ was defined in Equation (\ref{eq:CKL}) and the  subscript ``0" marks the initial conditions. We note that  $C_{KL}$ in this TPO case is no longer constant (unlike the TPQ case).  The parameter $y$ has the range $C_{KL,0}<y<C_{KL,0}+(1-e_{1,0}^2)\cos i_{\tot,0}/2$.
For cases where $e_{1,0}<<1$, i.e., $C_{KL}<<1$ and $ i_{\tot,0}\gsim61.7^\circ$ Equation (\ref{eq:filpHiepsilon}) takes a simple form:
\begin{equation}\label{eq:filpHiepsilon2}
\epsilon_c=\frac{1}{2}\mathit{f}\left(\frac{1}{2}\cos^2i_{\tot,0}\right) \ .
\end{equation}
This approximation is valid for $\epsilon\lsim0.025$. The validity of this approximation for different initial values of $e_1$ and $i_{\tot}$ are shown in the left panels in Figure \ref{fig:HighFlip}. 

\begin{figure}
\begin{center}
\includegraphics[width=0.7\linewidth]{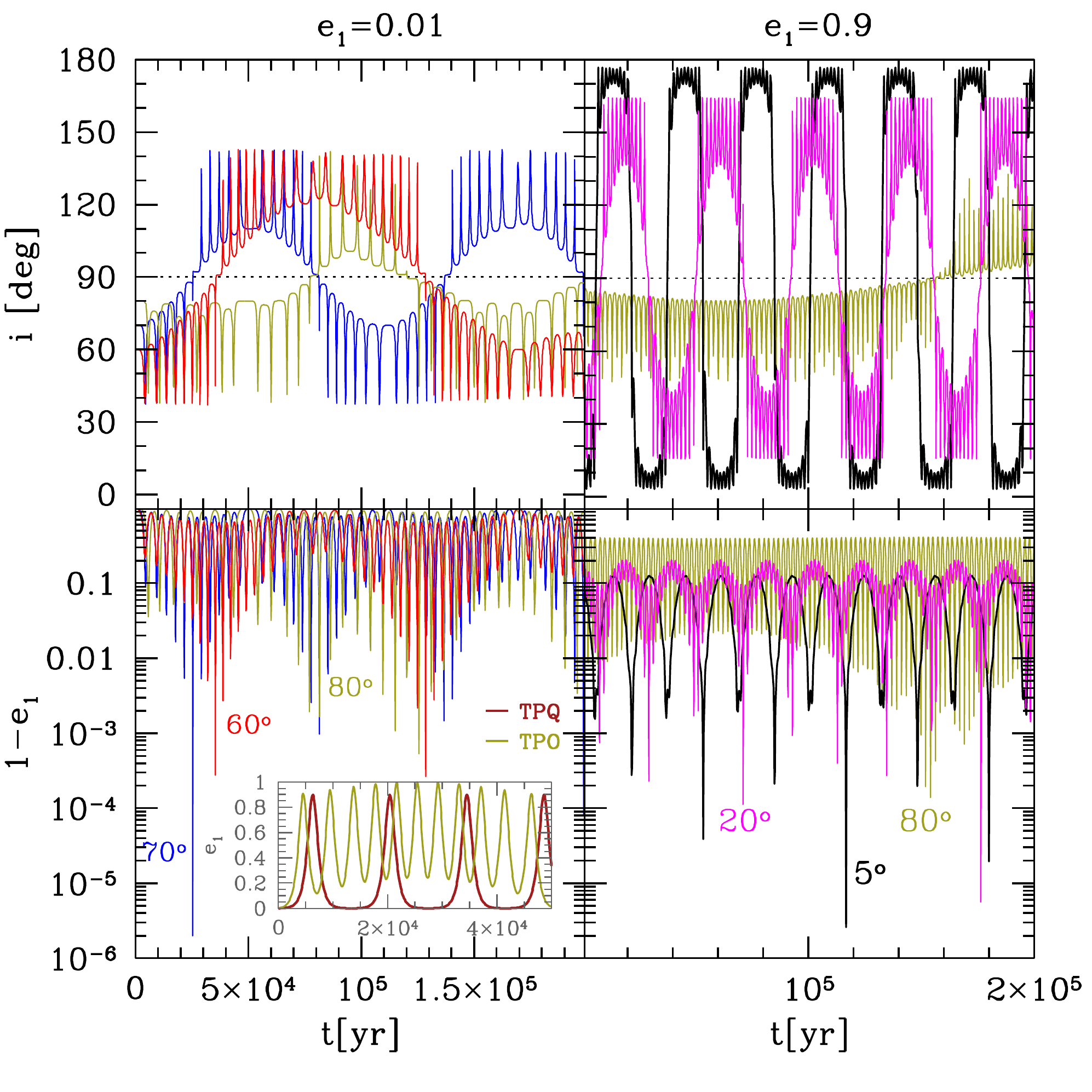}
\caption{\upshape {\bf Flip timescales}. We consider the following supermassive black hole binary system $m_1=10^7$~M$_\odot$, $m_3=10^9$~M$_\odot$ (note that in this case $m_2\to 0$).  The other parameters of this system are: $a_1=0.05$~pc  $a_2=1$~pc and $e_2=0.7$. The system is sent initially with $\omega_1=51^\circ$, $\Omega_1=165.58^\circ$ and $e_1=0.01$ for the left panels and $e_1=0.9$ for the right panels. The initial inclinations considered are colored labeled in the Figure. Note the difference in flip timescale as a function of initial inclinations. 
In the inset we show the inner orbit eccentricity $e_1$ as a function of time for the TPQ (brown line) and TPO (green) for the initial setting of $e_1=0.01$ and $i_{\tot}=80^\circ$ case, which emphasis the different  short (quadruple) timescales between the TPQ and TPO level of approximation.    
} \label{fig:InclDepdNew}
\end{center}
\end{figure}

A timescale for the high inclination oscillation or flip is difficult to quantify since the evolution is chaotic.  Furthermore, numerically it seems that the timescale for the first flip depends on the inclination (as can be seen in Figure \ref{fig:InclDepdNew}). However, an approximate analytical condition, for the regular (none chaotic) mode was achieved recently by \citet{Antognini15}, following \citet{Katz+11} formalism. This timescale has the following functional  form:
\begin{eqnarray}\label{eq:tflipHf}
t_{\rm flip}&=&\frac{256\sqrt{10}}{15\pi \epsilon}\int_{C_{KL,\rm min}}^{C_{KL,\rm max}}\frac{dC_{KL} K(x)}{\sqrt{2 (4\phi_{\rm quad}/3 +1/6+C_{KL}) } (4-11C_{KL}) \sqrt{6+4C_{KL}}}  \times\nonumber \\ 
&&\left( 1-\frac{(\chi-f(C_{KL}))^2}{\epsilon^2}  \right)^{-1/2}
 \ ,
\end{eqnarray}
where 
\begin{equation}
\phi_{\rm quad}=\frac{1}{8}\left(3 F_{\rm quad}-1\right) \ ,
\end{equation}
and note that $\phi_q$ defined in \citet{Antognini15} is simply $\phi_q=C_{KL}+j_{z,1}^2/2=4\phi_{\rm quad}/3-j_{z,1}^2/2+1/6$ in the notation used here. 
The upper limit of the integral in Equation (\ref{eq:tflipHf}) is easy to find, since for $i_\tot\to90^\circ$ the z component of the angular momentum is zero, thus
\begin{equation}
C_{KL,\rm max}=\frac{4}{3}\phi_{quad}+\frac{1}{6} \ ,
\end{equation}
and the minimum limit of the integral is found from solving $f(C_{KL,min})=\chi\pm \epsilon$. 
This timescale takes a simple form, for setting initially $e_1\to 0,\omega_1\to 0$ and $i_\tot\to 90^\circ$:
\begin{equation}\label{eq:tflipH}
t_{\rm flip} \sim \frac{128}{15\pi} \frac{a_2^3}{a_1^{3/2}}\frac{\sqrt{m_1}}{k m_3} \sqrt{\frac{10}{\epsilon} }(1-e_2)^{3/2} \quad {\rm for} \quad e_{1,0}\sim 0 \quad {\rm and} \quad i_{\tot}\sim 90^\circ \ .
\end{equation}

In the TPO level of approximation the short (quadrupole) timescales differ from the associated timescale at the TPQ level. In other words following the evolution of the same system, once by using the TPO and once using the TPQ yields different timescales, as depicted in the inset of Figure  \ref{fig:InclDepdNew}. This is because the Hamiltonian (i.e., the energy) is slightly different as the TPO includes the octupole term. Thus, the two calculations sample somewhat different values of the system energy. The difference is within a factor of a few as it represents the range of the phase space away from the seperatrix (See Figure \ref{fig:LN_TPQ} for the different oscillation's amplitudes for given initial different energy.

  \addcontentsline{toc}{subsubsection}{Low initial inclination regime}
\subsubsection*{Low initial inclination regime}

\begin{figure}
\begin{center}
\includegraphics[width=0.6\linewidth]{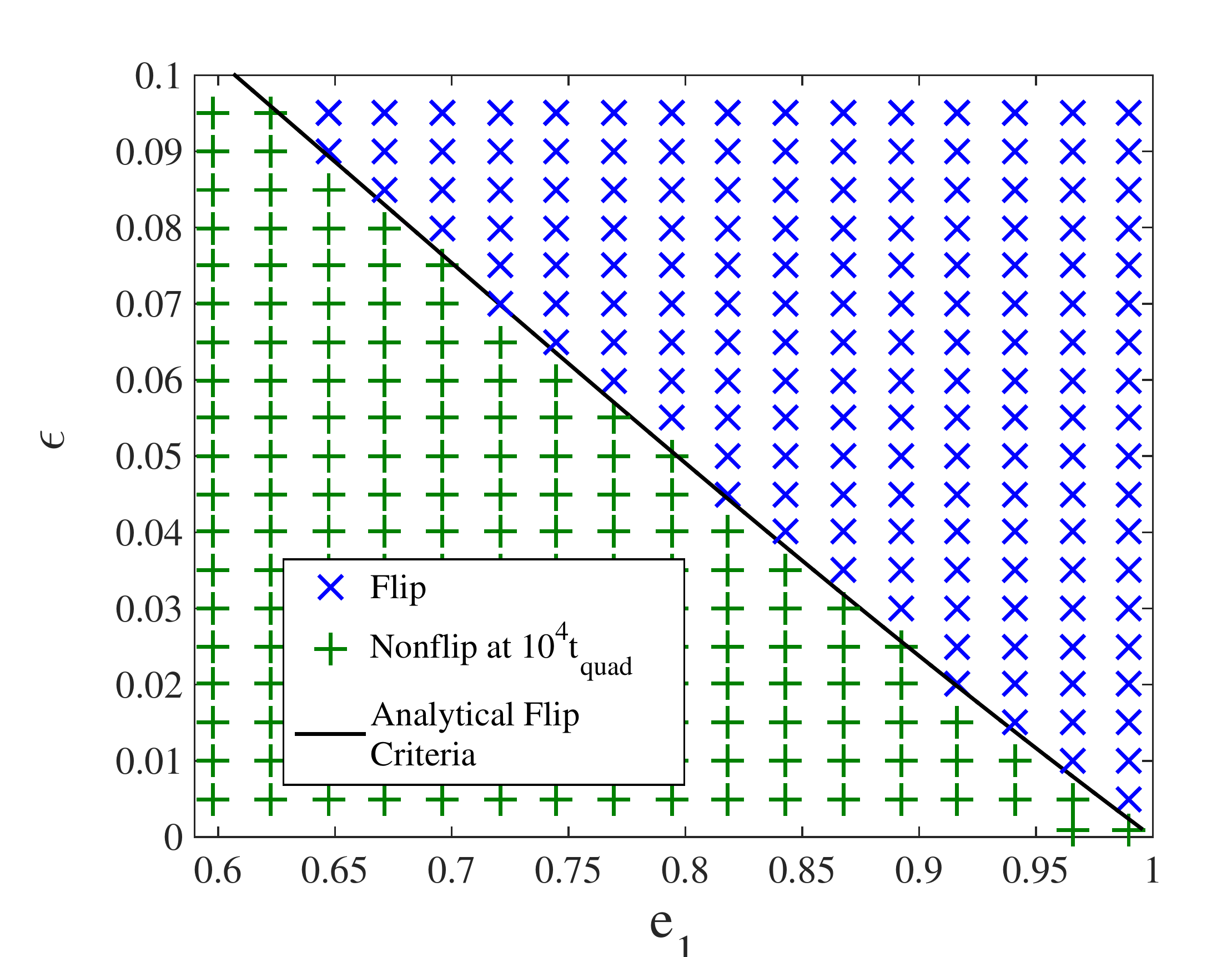}
\caption{\upshape Surface of section for $F_{quad}+\epsilon F_{oct}=-2$  and $F_{quad}+\epsilon F_{oct}=-1$ for $\epsilon=0.1$, this associated with low initial inclination $i_{\tot,0}<39.2^\circ$, Figure adopted  from \citet{Li+14Chaos}. See similar plots in \citet{Petrovich15Co}, reproducing this analysis.} \label{fig:EKLTestPCoP}
\end{center}
\end{figure}

The octupole level of approximation yields an interesting  behavior even beyond the Kozai angles. This is a result of the octupole level  harmonics, i.e.,  $\omega\pm \Omega$ and $3\omega\pm \Omega$. Since the low order resonances are missing, the co-planer flip is not associated  with chaotic behavior. Figure \ref{fig:EKLTestPCoP} shows the surface of section for two   low inclination examples, specifically   $F_{quad}+\epsilon F_{oct}=-2$  and $F_{quad}+\epsilon F_{oct}=-1$ for $\epsilon=0.1$. 

As can seen from Figure \ref{fig:IMBH} (as well as Figures \ref{fig:EKLTestP} and \ref{fig:EKLTestPCoP})  the two inclination regimes exhibit qualitative differences. The high inclination flip is driven by the quadrupole level resonance with the actual flip arrises by accumulating effects from the high order resonates. Furthermore, this flip, many times, is  associated with a chaotic behavior \citep{LN,Li+14Chaos}. On the other hand, the low inclination flip is due to a regular trajectory. In addition, this flip takes place on a much shorter timescale than the high inclination flip.

\begin{figure}
\begin{center}
\includegraphics[width=\linewidth]{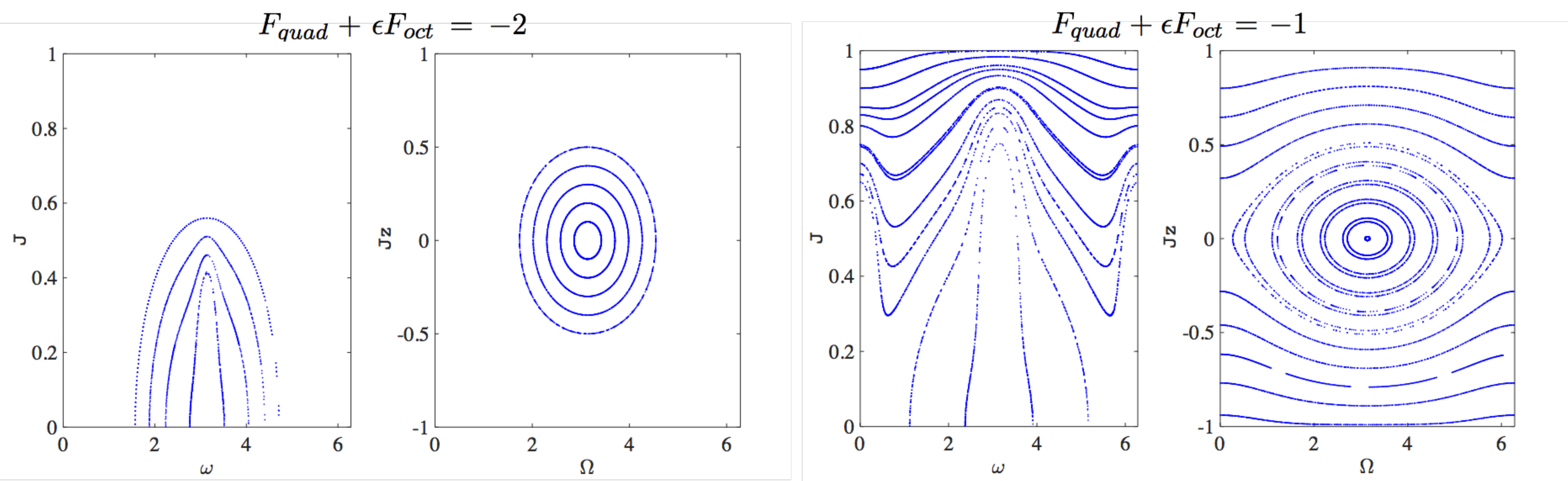}
\caption{\upshape {\bf Low inclination flip criterion:}  Comparison between the analytical expression Equation (\ref{eq:tflipL}), solid line,  and numerical integration, (green crosses mark no flip after $10^4 t_{\rm quad}$, and blue crosses systems that flipped). The system's parameters are: $m_1=1$~M$_\odot$, $m_2\to0$, $m_3=0.1$~M$_\odot$, $a_1=1$~AU, $a_2=45.7$~AU. The outer orbit eccentricity $e_2$ was changed to match the $\epsilon$ values indicated on the vertical axis.  The system was initially set with $i_\tot=5^\circ$, $\omega_1=0^\circ$, $\Omega_1=180^\circ$ and $e_1$ as indicated in the figure. Figure adopted from \citet{Li+13}.
} \label{fig:fig5_left}
\end{center}
\end{figure}

Similarly to the analytical approximation for the high inclination flip conditions, 
\citet{Li+13} achieved an analytical condition for the low inclination flip, after averaging over the flip timescale 
\begin{equation}\label{eq:filpLo}
\epsilon_c>\frac{8}{5}\frac{1-e_1^2}{7-e_1(4+3e_1^2)\cos(\omega_1+\Omega_1)} \ .
\end{equation}
Comparing this condition to the high inclination condition Equation (\ref{eq:filpHiepsilon}), also emphasis the  qualitative difference between  these two regimes.

The low inclination  regime yields a flip timescale that can be easily found by setting $i_\tot\to 0$. \citet{Li+13} found a expression for the flip timescale: 
 \begin{equation}\label{eq:tflipL}
t_{\rm flip}=\left(\int^{e_{\rm min}}_{e_{1,0}} + \int_{e_{\rm min}}^{e_{\rm max}} \right)\frac{-8}{5(4+3e_1^2)}\bigg[ \epsilon (1-e_1^2)\left(1-\frac{(F_{\rm quad}^0+\epsilon F_{\rm oct}^0-8e_1^2)^2}{25e_1^2(4+3e_1^2)^2\epsilon^2)}  \right) \bigg]^{-1/2} \ ,
\end{equation}
where $e_{1,0}$ is the initial inner orbit eccentricity and $F_{\rm quad}^0+\epsilon F_{\rm oct}^0$ is the energy that corresponds to $i_\tot=0$ and the rest of the initial conditions. The reason for the two integrals is because if initially $\sin (\omega_1+\Omega_1) >1$, the inner orbit eccentricity, $e_1$,  decreases before it increases, otherwise if  $\sin (\omega_1+\Omega_1) <1$ $e_{\rm min}=e_{1,0}$.

\begin{figure}
\begin{center}
\includegraphics[width=0.85\linewidth]{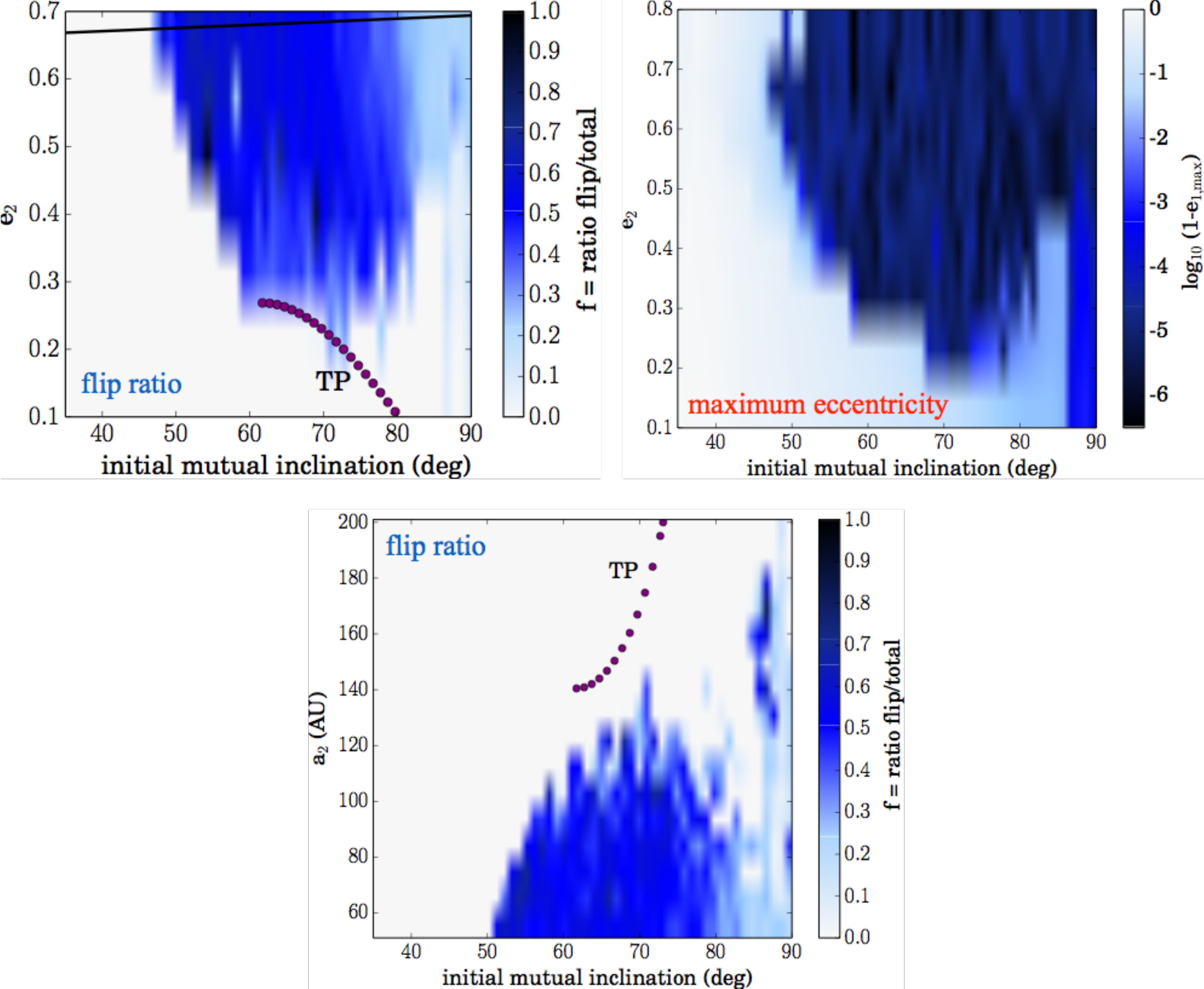}
\caption{\upshape {\bf  Flip and maximum eccentricity parameter space in two hierarchical planets configuration}. The color describes the maximum eccentricity reached over integration time of $\sim 5000 t_{\rm quad}$ (right top panel) and the flip ratio, defined as  the time the total inclination spends over $90^\circ$ from the entire integration time (the other two panels). The top two panels show the phase space corresponding to $e_{\rm max}$ (right) and the flip ratio (left) as a function of the initial outer orbit eccentricity ($e_2$) and the initial mutual inclination. Note that both follow exhibit interesting behavior at similar parts in the parameter space. However, for initial large inclination $80^\circ-90^\circ$, the flip is suppressed. The system considered here has the following parameters: $m_1 = 1$~M$_\odot$, $m_2=1$~M$_J$, $m_3 = 6$~M$_J$, $a_1 = 5$~AU, and $a_2 = 61$~AU. The bottom panels shows the flip ratio in the initial $a_2$--$i_{\rm tot}$ phase space. The system considered in this panel has the same parameters as the top two panels, but with $e_2=0.5$ and varying $a_2$. The flip condition for the TPQ, following the condition in Equation (\ref{eq:filpHiepsilon2}) is shown in purple dots. The TPQ analysis for the top left (bottom) panel suggests that all systems above (below) the ``TP" dotted line are expected to flip,. The solid line represents the stability condition, see Equation (\ref{eq:Mar}).   Figure adopted from \citet{Tey+13}.
} \label{fig:2plTey}
\end{center}
\end{figure}


\subsubsection{Beyond the Test Particle Approximation}

Relaxing the test particle approximation  leads to some qualitative differences. The first is that now one of the inner bodies can torque the outer body, and thus suppress the flip. This also causes a shift in the parameter space of the flip condition and the extreme eccentricity achieved compared to the TPQ case (see Figure \ref{fig:2plTey}).  While the value of the maximum of $e_1$  is similar to that in the TPQ case, large eccentricity excitations may take place in different parts of the parameter space  (compare Figure  \ref{fig:2plTey} to  Figure \ref{fig:HighFlip}). In particular, in the high inclination regime, the flips and the large eccentricity excitations of the TPQ case are concentrated around  $i_{\rm tot}=90^\circ$ but in the full case they can shift to lower mutual inclinations and tap to larger range of inclinations (Figure \ref{fig:2plTey}). This is mainly because the outer orbit is being torqued by the inner orbit. \citet{Tey+13}  studied the effect of a similar mass companion and showed that if the outer body mass is reduced to below twice the smallest mass of the inner orbit, the flip and large eccentricity excitations are suppressed for large parts of the parameter space. 

\begin{figure}[h]
\begin{center}
\includegraphics[width=8.5cm,clip=true]{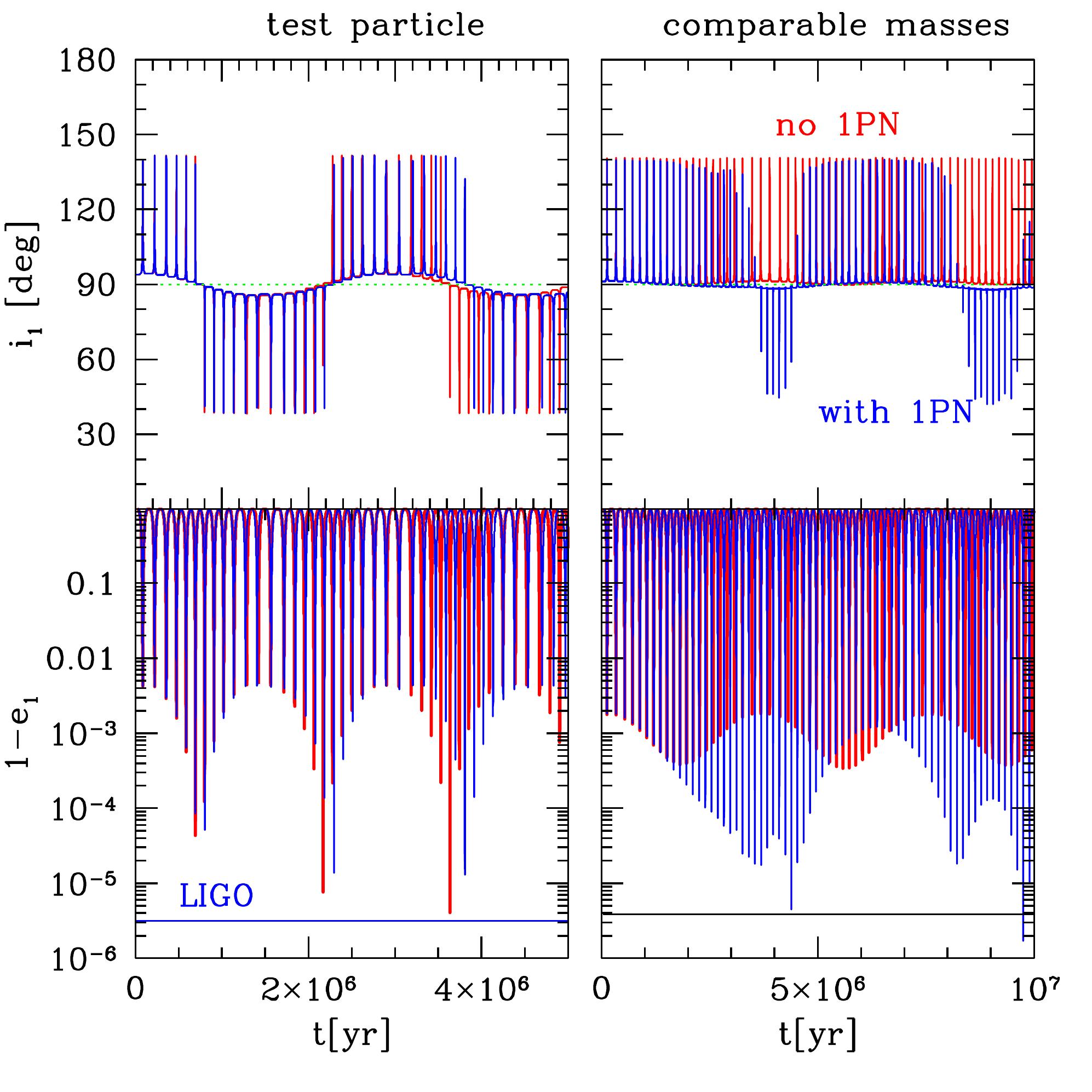}
\caption{ \upshape{\bf Comparison between test particle approximation and comparable mass system in the presence of general relativity}. The systems in the right and left panel have same parameters and initial conditions apart from $m_2$ which is set to zero in the left panels and $m_2=8$~M$_\odot$ in the right panels. 
The other parameters are: $m_1=10\,M_\odot$, $m_3 = 30\,M_\odot$, $a_1=10$~AU, $a_2=502$~AU, $e_1 =0.001$, $e_2 =0.7$, $\omega_1=\omega_2=240^\circ$ and $i_\tot=94^\circ$. Red lines corresponds to pure Newtonian evolution, and blue lines include general relativity effects (1st post newtonian expansion, to the inner and outer orbits). The horizontal lines   are the minimum eccentricity corresponding to the detectable LIGO frequency range (horizontal lines in the bottom panels).
General relativity corrections help to further increase the eccentricity and lead to orbital flips for the inner binary for comparable masses.  Figure adopted from \citet{Naoz+12GR}
} \label{fig:e1_excite}
\end{center}
\end{figure}

The system's hamiltonian is (here again the nodes were eliminated for simplicity, but the z-component of the angular momenta are not conserved):
\begin{equation}
\label{eq:totHam}
\Ham=\Ham_{quad}+\epsilon_M\Ham_{oct} \ ,
\end{equation}
where $\Ham_{quad}$ is define in equation (\ref{eq:Hamquad}) and we copy it here for completeness 
\begin{equation}\label{eq:Hamquad2}
  \Ham_{quad} = C_2 \{ \left( 2 + 3 e_1^2 \right) \left( 3 \cos^2 i_\tot - 1 \right)  +  15 e_1^2 \sin^2 i_\tot \cos(2 \omega_1) \}  \ , \nonumber
\end{equation}
the octupole level approximation is:
\begin{eqnarray}
  \Ham_{oct}& =& C_2 \{ \left( 2 + 3 e_1^2 \right) \left( 3 \cos^2 i_\tot - 1 \right)  + 15 e_1^2 \sin^2 i_\tot \cos(2 \omega_1) \}  \nonumber \\ \nonumber
  &+& C_3 e_1 e_2 \{ A \cos \phi +  10 \cos i_\tot \sin^2 i_\tot  (1-e_1^2) \sin \omega_1 \sin \omega_2 \} \nonumber \ ,
\end{eqnarray} 
where
\begin{eqnarray}
\label{eq:C3}
C_3&=&-\frac{15}{16}\frac{k^4}{4}\frac{(m_1+m_2)^9}{(m_1+m_2+m_3)^4}\frac{m_3^9(m_1-m_2)}{(m_1m_2)^5}\frac{L_1^6}{L_2^3G_2^5} \nonumber \\
&=& -C_2\frac{15}{4}\frac{\epsilon_M}{e_2} \ 
\end{eqnarray}
and
\begin{equation}\label{eq:epslionM}
\epsilon_M=\frac{m_1-m_2}{m_1+m_2}\frac{a_1}{a_2}\frac{e_2}{1-e_2^2} \ .
\end{equation}
and  
\begin{equation}
\label{eq:A8}
A= 4+3e_1^2-\frac{5}{2}B\sin i_\tot^2 \ ,
\end{equation}
where
\begin{equation}
B=2+5e^2_1-7e_1^2\cos(2\omega_1) \ ,
\end{equation}
and
\begin{equation}
\cos \phi=-\cos \omega_1\cos \omega_2 -\cos i_\tot \sin \omega_1 \sin \omega_2 \ .
\end{equation}
The latter equation emphasis one of the main difference that arise from relaxing the test particle approximation. In cases for which $m_1\sim m_2$ the contribution from the octupole lever of   approximation can be negligible. This can be seen in the example in Figure \ref{fig:e1_excite} for a system where the only difference between the left and right panels is setting $m_2=0$ in the left panels and $m_2=8$~M$_\odot$ in the right panels ($m_1=10\,M_\odot$). In the pure Newtonian regime, (red lines) the EKL behavior is suppressed (no flips or eccentricity peaks).  The complete set of equation of motions can be found in Section \ref{sec:eqs}.

%
%


\section{The validity of the approximation and the stability of the system}\label{sec:valid}

\begin{figure}[tb]
\begin{center}
\includegraphics[width=\linewidth]{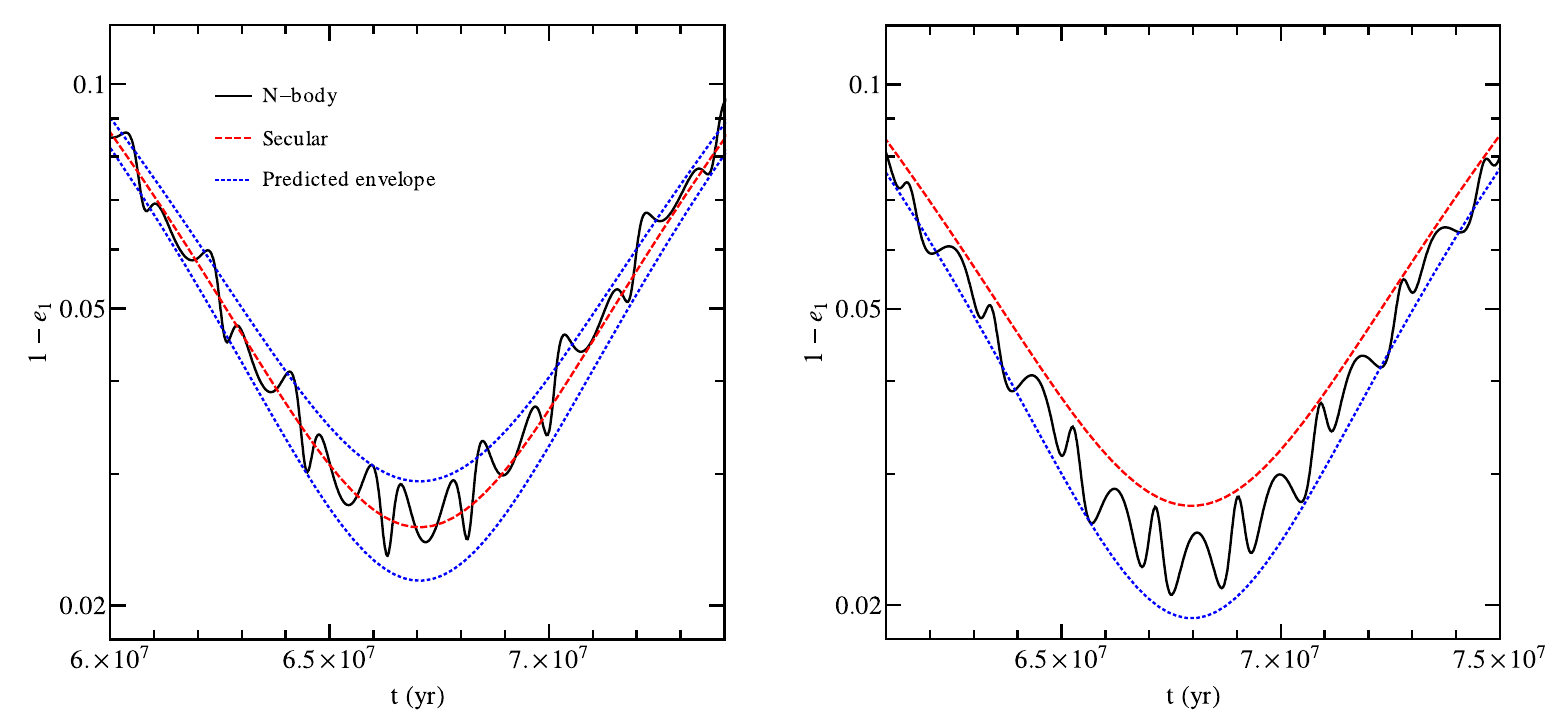}
\caption{ \upshape{\bf Comparison of the eccentricity excitations}. The Figure considers the results from the secular approximation (red lines), and N-body (black lines) and the predicted change from Equation (\ref{eq:REO}). The system considered has the following parameters: $m_1=10^7$~M$_\odot$, $m_2=10^5$~M$_\odot$, $m_3=10^7$~M$_\odot$, $a_1=1$~pc, $a_2=20$~pc, $e_1=0.1$,  $e_2=0.2$, $i_{\rm tot}=80^\circ$. Left panel was initialized with $\omega_1=\omega_2=0^\circ$ and the right panel was initialized with $\omega_1-\omega_2=90^\circ$. Figure adopted from \citet{Antognini+13}.
} \label{fig:DAeffects}
\end{center}
\end{figure}

The secular approximation described here utilize averaging over the short orbital timescales, and thus any modulations over these times are washed out. \citet{Ivanov+05}, \citet{Katz+12},  \citet{Antognini+13},  \citet{Antonini+14} and \citet{Bode+14}  showed that the inner orbit undergoes   rapid  eccentricity oscillations near the secular value (see for example Figure \ref{fig:DAeffects}). \citet{Ivanov+05} found the change in angular momentum during an oscillation as 
\begin{equation}\label{eq:REO}
\frac{\Delta G_1}{\mu_1}=\frac{15}{4} \frac{m_3}{m_1+m_2}\cos i_{\rm min} \left( \frac{a_1}{a_2}\right)^2 k \sqrt{m_3 a_2} \ ,
\end{equation}
where $\mu_1$  is the reduced mass of the inner binary, and $ i_{\rm min}$ is the minimum inclination reached during the oscillation.
These rapid eccentricity oscillations  happen because the value of the inner orbit angular momentum goes to zero (i.e., extreme inner orbit eccentricity) on shorter timescale than the inner  orbital period. In that case the averaging is not suffice and the secular approximation underestimates the maximum eccentricity that the system can reach. Assuming a
fixed outer perturber and adopting an instantaneous quadrupole torque,  \citet{Antonini+14} took the limit of $e_1\to 1$ and found a simple form to the condition for which the averaging is valid
\begin{equation}\label{eq:ava}
\sqrt{1-e_1}\gsim 5\pi \frac{m_3}{m_1+m_2}\left( \frac{a_1}{a_2(1-e_2)} \right)^3  \ ,
\end{equation}
(using slightly different settings, \citet{Bode+14} found a similar condition).
Thus, if during the evolution the specific angular momentum becomes smaller than the right hand side of this equation, the angular momentum goes to zero on shorter timescale than the inner orbital timescale. The immediate  consequences of this is that the inner binary maximum eccentricity will be  larger than the value the secular approximation predicts. 

 Recently, \citet{Luo+16}  showed that these rapid short-timescale oscillations  can accumulate over long timescales and lead also to deviations from the flip conditions discussed in \S \ref{sec:TPO}, as described in Equation (\ref{eq:filpHiepsilon}). They found that the double averaging procedure fails when the  mass of the tertiary $m_3$ is large compared to the mass of the inner binary, similarly to the condition in Equation (\ref{eq:ava}).

Another consequence of large eccentricities is the stability of the system. A long term stability condition that is often used in the literature is the one give by \citet{Mardling+01}, which has the following form
\begin{equation}\label{eq:Mar} \frac{a_2}{a_1} > 2.8\left(1+\frac{m_3}{m_1+m_2}\right)^{2/5}\frac{ (1+e_2)^{2/5}}{ (1-e_2)^{6/5}} \left(1-\frac{0.3 i_{\tot}}{  180^\circ}\right)     \ .  \end{equation}
Although this criterion was generated for similar mass binaries, and the inclination was added ad hock, it is often used for large range of masses. 
A criterion which takes into account both having the outer orbit be  wider than the inner one, and the validity  of secular approximation 
\begin{equation}\label{eq:epsilon}
\epsilon=\frac{a_1}{a_2 }\frac{e_2 }{1-e_2^2}<0.1 \ .
\end{equation}
 This is numerically similar to the \citet{Mardling+01} stability criterion [Equation~(\ref{eq:Mar})], for large range of mass system,  \citep[as shown in][]{Naoz+12GR}.  
 
The stability of a two planet system with low mutual inclination was studied in  \citet{Petrovich152p}, using N body integration. Assuming that $m_1$ is a stellar mass object and $m_2$ and $m_3$ are planetary mass  objects he found a stability criterion of the form: 
\begin{equation}\label{eq:2p}
\frac{a_2(1-e_2)}{a_1 (1+e_1)} > 2.4 \bigg [{\rm max}\left(\frac{m_2}{m_1},\frac{m_3}{m_1}\right) \bigg ]^{1/3} \sqrt{\frac{a_2}{a_1}} +1.15  \ .
\end{equation}
Systems that do not satisfy this condition (by a margin factor of $\sim 0.5$) may become unstable. Specifically,  \citet{Petrovich152p} found that systems  for which $m_2/m_1 > m_3/m_1$ will most likely result in planetary ejections while systems for which $m_2/m_1<m_3/m_1$ may slightly favor collisions with the host star.

The eccentricity excitations, both in the secular approximation and in its deviations, are extremely large (see  Figures \ref{fig:HighFlip} and \ref{fig:2plTey}). This implies that in some cases the inner orbit can reach such a small pericenter distance $R_{Lobe}$ so one of the objects may cross its Roche-limit (in the case where $m_2<m_1$):
\begin{equation}\label{eq:RLoeb}
R_{Lobe}=\eta R_{2}\left(\frac{m_{2}}{m_1+m_2}\right)^{-1/3} \ ,
\end{equation}
where $\eta$ is a numerical factor of order unity.

Considering the definition of the Roche limit, we can also ask when the eccentricity of the inner orbit becomes so large such that the tertiary captures a test particles that is orbiting around the primary ($m_1,m_3 >> m_2)$, which can be written as:
\begin{equation}\label{eq:Roche2}
a_{1}(1+e_{1}) = {\tilde \eta} a_{2} (1-e_{2}) \left(\frac{m_1}{m_3}\right)^{1/3} \ ,
\end{equation}
where ${\tilde \eta}$ is of order of unity and is of different value from $\eta$ in Equation (\ref{eq:RLoeb}). 
A test particle initially around  $m_1$ with larger separations will feel a larger gravitational force from $m_3$.  Using the definition of $\epsilon$, \citet{NaozSilk} found  the mass ratio that will result in a stable configuration as a function of the binary mass ratio, i.e., 
\begin{equation}\label{eq:Stab}
\frac{m_3}{m_1}=\left( {\tilde \eta}\frac{e_{2}}{\epsilon (1+e_{1})(1+e_{2})}\right)^3 \ .
\end{equation}
Thus for mass ratios for that are larger than the right hand side the approximation breaks down and the test particle may be captured by $m_3$ \citep[some consequences are discussed at][]{Li+15}.



\section{Short range forces and other astrophysical effects}\label{sec:short}

The Newtonian evolution of the secular hierarchical three body system has proven to be very useful in modeling and analyzing many astrophysical systems. In realistic systems there are several short range forces and astrophysical affects that can significantly alter the evolution of the system. For example, some short range forces, such as tides and general relativity induce precession of the periapse which strongly depends on the orbital eccentricity. If the orbit  precesses due to the short range force to the opposite direction than the one induced by the Kozai-Lidov mechanism, further excitations of the eccentricity can be suppressed.  In the limiting case, the precession is so fast compared to quadrupole-level precession that  the  inner orbit initial eccentricity remains constant. In fact, as will be discussed below, in some cases, the eccentricity excitation in the presence of  short range force can be estimated analytically.  Since in the Kozai-Lidov mechanism eccentricity is being traded for inclination, once the eccentricity can not be excited, the oscillations in the inclination are limited in a similar way. 

\subsection{General Relativity} 

\begin{figure}
\begin{center}
\includegraphics[width=0.75\linewidth]{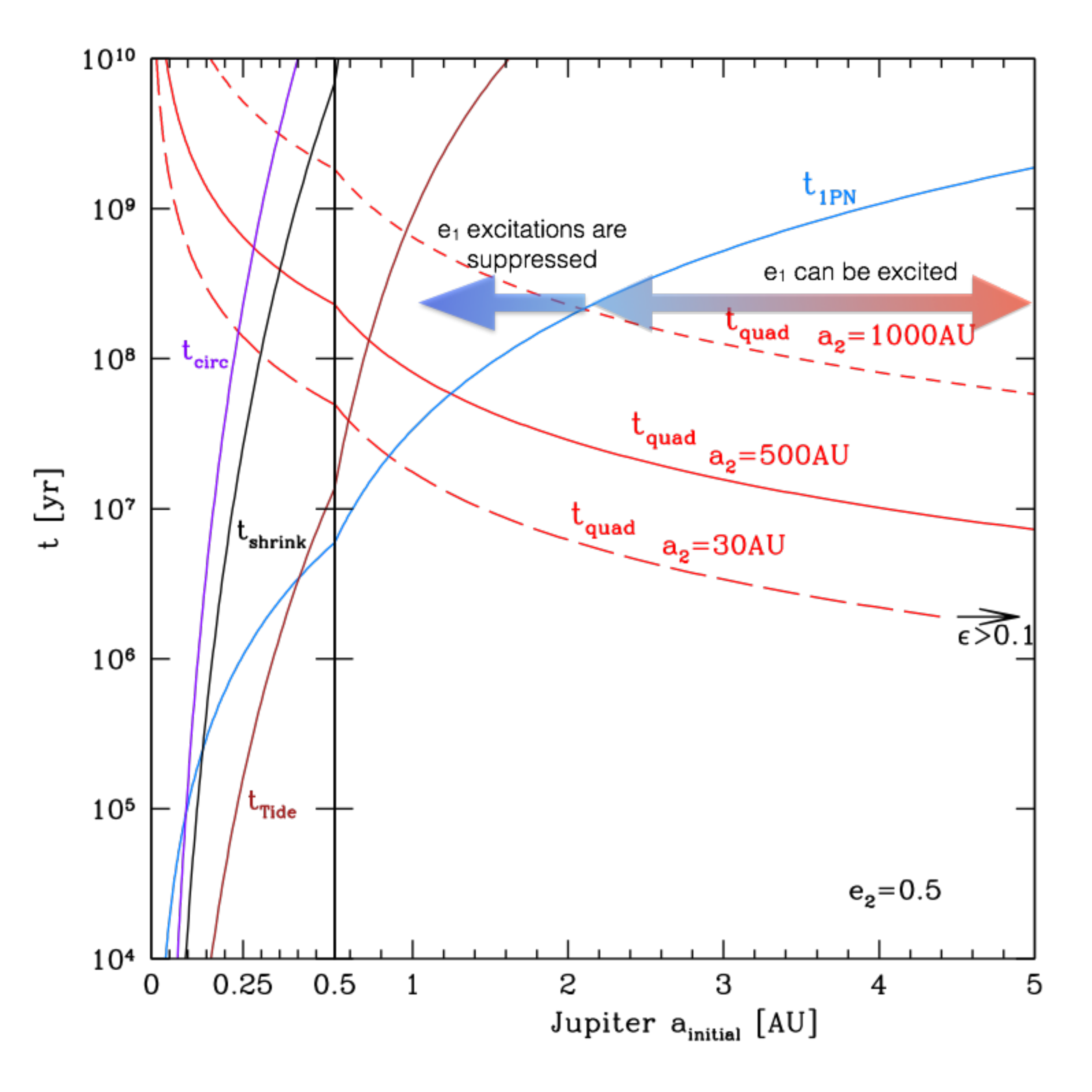}
\caption{ \upshape{\bf Relevant timescales for a Jupiter system}.  The system considered here is a Jupiter mass planet at different initial separations $a_{\rm iniital}$ from a $1$~M$_\odot$ star. We consider the quadrupole  timescale [Equation (\ref{eq:tquad})] for a stellar perturber ($m_3=1$~M$_\odot$) at $a_2=1000$~AU and $a_2=500$~AU (short-dash and solid red lines, respectively), as well as the case of a Jupiter perturber at  $30$~AU (long-dashed red line).  $e_2=0.5$ in all these cases. We also consider the   precession of the inner orbit due to  general relativity, according to Equation (\ref{eq:tGR}), blue line. The crossing point between the blue and red lines roughly separates between the different behaviors, as depicted by the arrows. We also consider the precession  due to oblate objects form static tides (Equation (\ref{eq:tTF}), brown line), and the typical timescales to circularize and shrink the orbit (purple and black lines, respectively) according to the equations in Section \ref{sec:tides} while adopting $T_{V,1}=50$~yr and $T_{V,2}=1.5$~yr. } \label{fig:Timescales}
\end{center}
\end{figure}

 The fast precession of the perihelion of the inner orbit due to GR effects takes place on the opposite direction of the quadrupole precession. Therefore, as mentioned before, the  inner orbit extremely high eccentricity excitations are suppressed, and thus are the inclination flips as well.. For example, in the current location of most hot Jupiters, further eccentricity excitations are suppressed due to fast general relativity precession (and tides) compared to the quadrupole precession. Thus, Hot Jupiters have decoupled from their potential  pertrubers, and do not flip anymore. On the other hand, the $t_{\rm quad}$ timescale is much shorter compared to the general relativity precession   in asteroid and Kuiper belt binaries.

The precession of the inner orbit due to  general relativity has a simple form 
\begin{equation}\label{eq:GR1}
\frac{d\omega_1}{dt} \bigg |_{\rm 1PN,inner} = \frac{3 k^{3} (m_1 + m_2)^{3/2} }{a_1^{5/2} c^2 (1 - e_1^2)} \ ,
\end{equation}
where the subscript $1PN,inner$ indicates that precession is due to first Post Newtonian (PN) expansion for the inner orbit \citep[see][for a general derivation]{Misner+73}. A similar expression can be written to outer orbit general relativity precession, although this, typically, has little effect.  Expanding the 1st PN three body Hamiltonian in semi-major axes ratio up to the octupole level of approximation reveals another term which describes the general relativity interaction between the inner and outer orbits \citep{Naoz+12GR}. In many cases, where the leading Newtonian terms are important, this interaction term can be neglected. The inner orbit GR precession  timescale can be estimated simply as \citep{Naoz+12GR}:
\begin{equation}\label{eq:tGR}
t_{\rm 1PN,inner} \sim 2\pi \frac{a_1^{5/2} c^2 (1-e_1^2) }{ 3 k^3 (m_1+m_2)^{3/2}} \ .
\end{equation}
If this timescale is shorter than the quadrupole timescale Equation (\ref{eq:tquad}) eccentricity excitations are suppressed \citep[this was noted in many studies before, e.g.,][]{Ford00,Dan,Naoz+12GR}.  For example, Figure \ref{fig:Timescales}, depicts the relevant timescales for a Jupiter around a $1$~M$_\odot$ star. Different  pertrubers induce quadrupole precessions which are compared to the   general relativity precession, Equation (\ref{eq:tGR}). For example a planetary companion at $30$~AU cannot excite the eccentricity of a Jupiter that  formed at $0.5$~AU (a closer companion can), however,  a  companion can excite the eccentricity  of a $1$~AU Jupiter which may result in the formation of a Hot Jupiter (see below).

\begin{figure}[tb]
\begin{center}
\includegraphics[width=\linewidth]{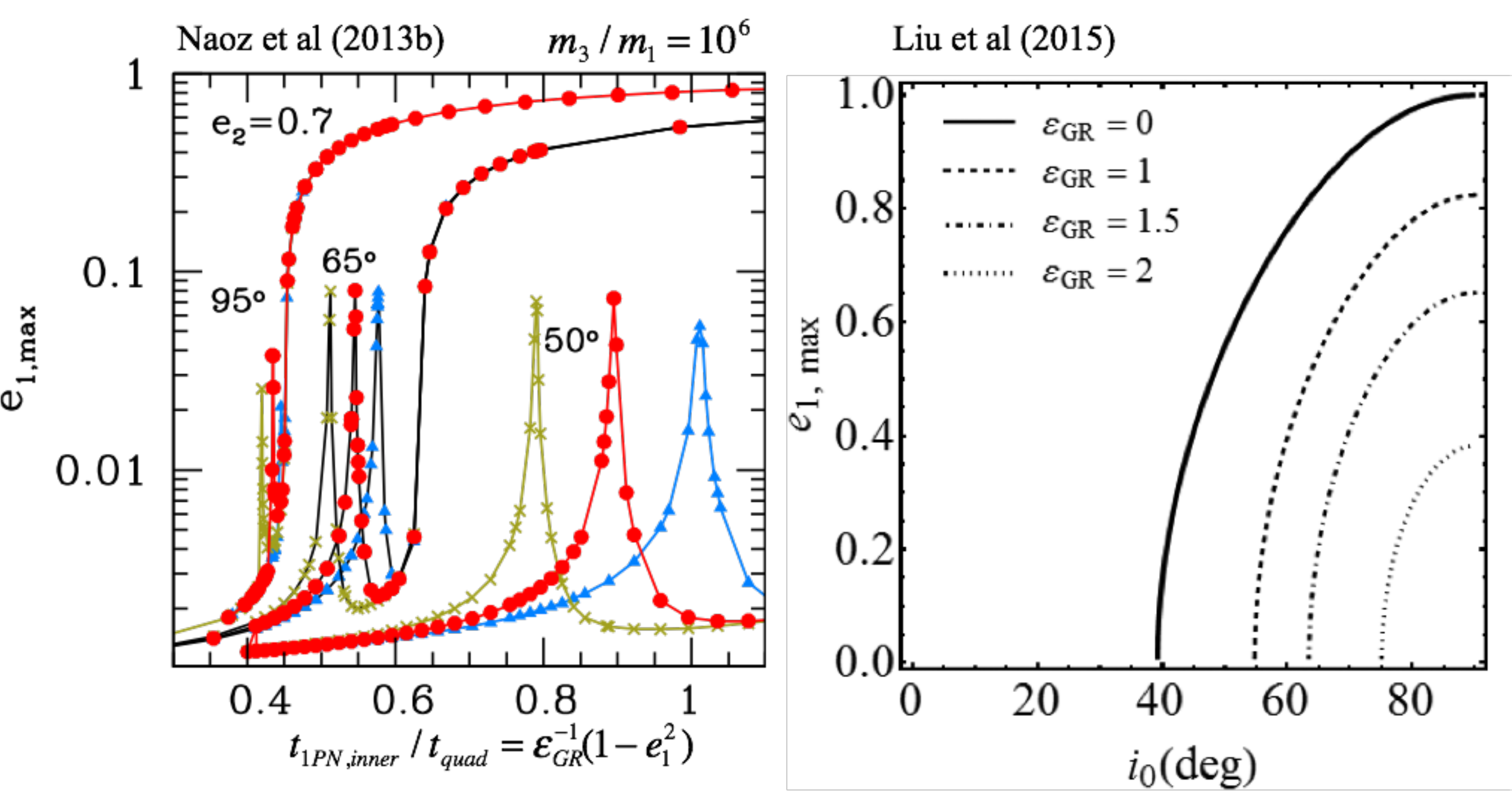}
\caption{ \upshape{\bf General relativity (1PN) effects on the hierarchical three body system}. {\it Left panel:} shows the  emergence of resonant-like eccentricity excitations in the $e_{1,max} - \varepsilon_{GR}^{-1}$ plane for different initial inclinations.  We consider the secular Newtonian evolution and the  PN evolution including terms only up to $\mathcal{O}(a_1^{-2})$ (inner orbit precession, blue triangles), $\mathcal{O}(a_2^{-2})$ (outer orbit precession, green crosses)  and the interaction term (red squares). The location of the resonance shift when including additional 3-body 1st PN terms.  The system is initialized with $e_1  =0.001$,  $\omega_2=0^\circ$ and $\omega_1=240^\circ$ and with mutual inclination corresponding to (from left to right) $95^\circ$, $65^\circ$ and $50^\circ$. The other parameters are $m_3/m_1=10^6$, $m_2\to 0$ and $e_2=0.7$.  Figure adopted from \citet{Naoz+12GR}. {\it Right panel:} shows the analytical solution for the maximum eccentricity in the $e_{1,max} - i_0$ plane for different values of $ \varepsilon_{GR}$ (note that  $\varepsilon_{GR}\to 0$ means no PN contribution). This calculation considers only the inner orbit precession for small $\varepsilon_{GR}$ and high inclination test particle orbit. Figure adopted from \citet{Liu+15}.
} \label{fig:GReffects}
\end{center}
\end{figure}

The relation between the timescales can be estimated by \citep[e.g.,][]{Naoz+12GR}
\begin{equation}
\frac{t_{\rm 1PN,inner}}{t_{\rm quad}}=\frac{ a_1^4} {3a_2^3 }\frac{(1-e_1^2)m_3  c^2 } {(1-e_2^2)^{3/2}  (m_1+m_2)^2 k^2} =\varepsilon_{GR}^{-1} (1-e_1^2)\ ,
\end{equation}
where we also introduced the parameter $\varepsilon_{GR}^{-1}$ defined in \citet{Liu+15}. 
When the two timescales are similar to one another a resonant like behavior emerges  \citep{Ford00,Naoz+12GR}.   An example for this  behavior is shown in the left panel of Figure \ref{fig:GReffects} for different initial mutual inclination and setting initially $e_1\to 0$. The value of this eccentricity can be estimated analytically, and have a simplified equation for large eccentricity excitations \citep{Liu+15}
\begin{equation}\label{eq:Liu1}
\left(\frac{\varepsilon_{GR}}{\sqrt{1-e_1^2}}  \right)_{e_1=e_{1,max}}\approx \frac{9}{8}e_{1,max}^2\frac{ j_{1,min}^2-5 \cos^2 i_0/3}{j_{1,min}^2} \ ,
\end{equation}
where we remind the reader that $j_{1,min}=\sqrt{1-e_{1,max}^2}<<1$. This behavior is shown in the right panel of Figure \ref{fig:GReffects}  \citep[see also][]{Dan}.   The general expression of Equation (\ref{eq:Liu1}), which is  valid for all values of $e_{1,max}$, can be found in Eq. 50 in  \citet{Liu+15}\footnote{Note that it has a typo and  the  $3/5$ in that equation should be $5/3$, Liu et al private communication.}. As shown in this latter study, given an extra short range force, such as $\varepsilon_{GR}$, the maximum eccentricity can be predicted for the octupole level of approximation, by considering the perpendicular case of the quadrupole level of approximation.   

Interestingly, even if the GR precession timescale is longer than the quadrupole timescale $t_{\rm 1PN,inner}>t_{\rm quad}$ general relativity can have significant  implications on the dynamical evolution. Specifically, if $t_{\rm quad}<t_{\rm 1PN,inner}\lsim t_{\rm oct}$ general relativity precession can re-trigger the EKL behavior for similar mass inner binaries. This can be seen in the right hand side example of Figure  
\ref{fig:e1_excite}, where we compare between the pure Newtonian case (red lines) and the case which includes general relativity precession for the inner orbit (blue lines). 
As depicted, including general relativity effects  re-trigger the EKL behavior.   

In the secular approximation general relativity  effects are typically being taking into account  by  only including the {\it inner} body  precession [Equation (\ref{eq:GR1})]. Sometimes the outer orbit precession is also being taken into account (simply replace $1$ with $2$ in  Equation (\ref{eq:GR1})), this mainly affects the position of the $t_{\rm quad}\sim t_{\rm 1PN,inner}$ resonance \citep[e.g.,][and see left panel of Figure \ref{fig:GReffects}]{Naoz+12GR}. 

In some astrophysical settings higher PN orders of the inner orbit are important \citep[e.g.,][]{MH02,Bla+02,Wen,Seto13,Antognini+13}. In some cases the general relativity (1PN) term that describes the interactions between the inner and outer orbits may have some effects \citep{Naoz+12GR}. However, as shown by \citet{Will14a,Will14b} when GR effects between the two orbits become more important, the gravitational weak field approximation is no longer valid, which results in deviations of the  dynamics compared to the double averaging process.

\subsection{Tides and rotation}\label{sec:tides}

Similarly to the  suppression of eccentricity excitations  due to  general relativity precession, precession of the nodes due to oblate objects form static tides, or rotating objects, can cause similar affect. \citet{Mazeh+79} first included tidal effects to the hierarchical triple dynamical evolution (in the TPQ case and assuming small mutual inclinations).  This was then generalized in a series of papers by \citet{1998KEM}, \citet{1998EKH} and \citet{Egg+01}, based on \citet{Hut} equilibrium and static tides formalism. The strength of the equilibrium tide recipe presented here  is that  it is self consistent  with the secular approach. Furthermore, assuming polytropic stars this recipe has only one dissipation parameter for each member of the binary. In other words, tides can be considered for  both members of the inner orbit. Using this description one is able to follow the precession of the spin of the star and the planet  due to oblateness and tidal torques. We provide the set of equations in Section \ref{sec:tides}. Different choices of the tidal model can result in quantitatively different results, such as the relevant separations at which eccentricity excitations are suppressed, and the time evolution of the circularization and orbital shrinking process.  


\begin{figure}[h]
\begin{center}
\includegraphics[width=\linewidth]{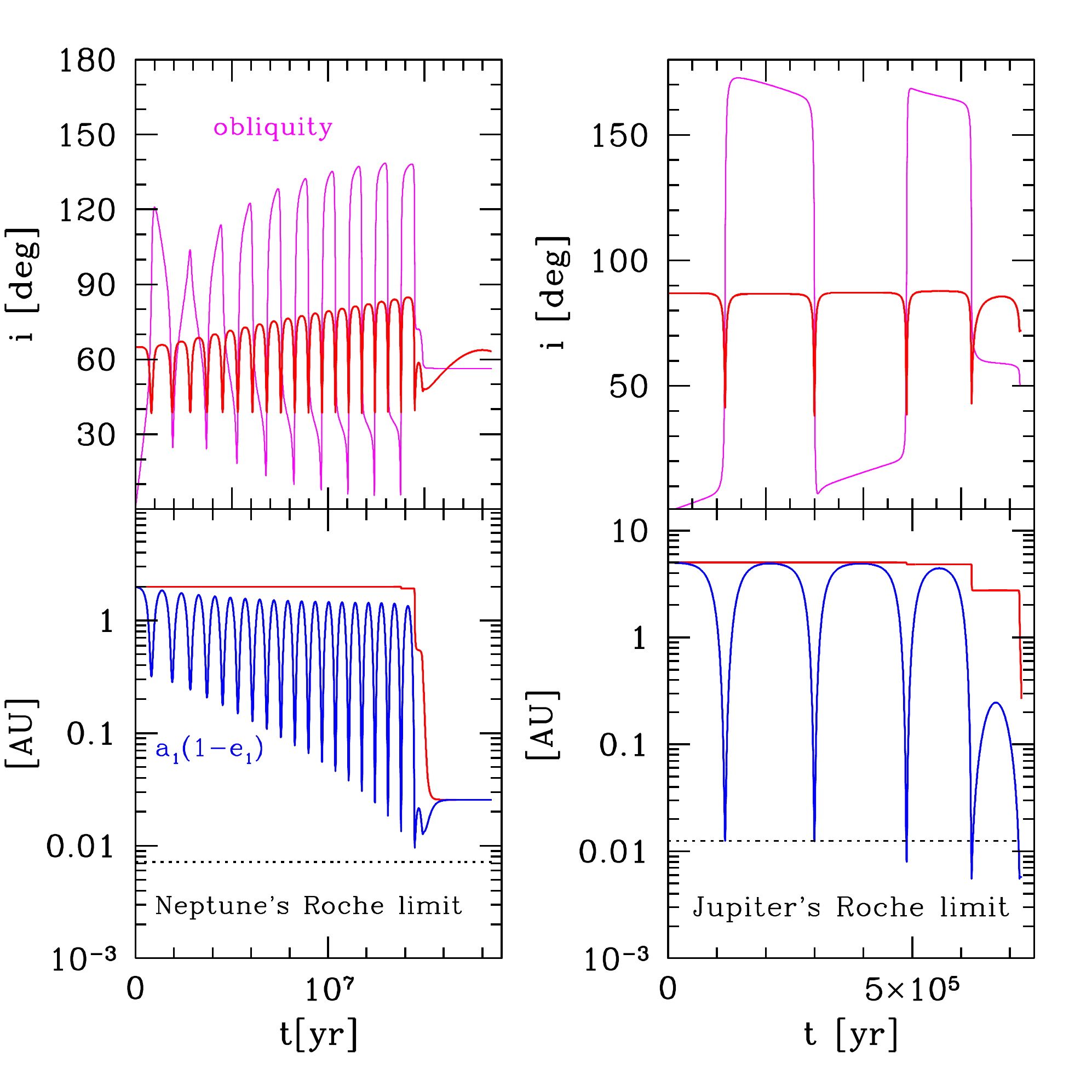}
\caption{\upshape {\bf Tidal disruption (right panel) and circularization and shrinking the orbits due to tides (left panel)}. Top panels show the systems' mutual inclination (red line), and obliquity (magenta lines). Bottom panels shows the semi-major axes (red lines) and pericenter distances (blue lines) in AU. Also shown in dashed lines are the pericenter at which tidal disruption takes place according to Equation (\ref{eq:RLoeb}), adopting $\eta=2.7$ \citep[e.g.,][]{Guillochon+11,Liu+13}.   Left panels considers a Neptune  around a  $0.32$~M$_\odot$ M dwarf star, initially set at $a_1=2$~AU, and $e_1=0.01$. The third object is a brown dwarf with $m_3=10$~M$_j$ at $50$~AU, with $e_2=0.52$. The orbits have initially $\omega_1=\omega_2=0^\circ$ and mutual inclination of $65^\circ$. The spin periods of the star and plant were assumed to be $4.6$~days and $1$~day, respectively.   Right panels  consider a Jupiter mass planet at a $5$~AU separation from a  $1$~M$_\odot$ star with a $1$~M$_\odot$ stellar companion at $200$~AU. The system initially sets with $e_1=0.001$, $e_2=0.75$, $\omega_1=\omega_2=0^\circ$ and $i=87^\circ$.  The spin periods of the star and plant were assumed to be $24$~days and $10$~day, respectively.   Both systems start initially  aligned (i.e., zero obliquity for both the planet and the star) and  $T_{V,1}=50$~yr and $T_{V,2}=1.5$~yr. 
} \label{fig:MdwarfH}
\end{center}
\end{figure}


During the system evolution,  the EKL mechanism can cause large  eccentricity excitations for the inner orbit (for example, see Figures \ref{fig:HighFlip} and \ref{fig:2plTey}). Thus, on one hand,   the nearly radial motion of the binary   drives the two inner binary members   to merge, while on the other hand,  the tidal forces tend to shrink  and circularize  the orbit, see  Figure \ref{fig:MdwarfH} right and left panels, respectively.  If during the evolution the tidal precession timescale (or the GR timescale) is similar to that of  the quadrupole timescale (which is the shortest secular timescale, Equation (\ref{eq:tquad})), further eccentricity excitations are suppressed. 
 In this case  tides can shrink the binary semi-major axis and form a tight  binary decoupled from the tertiary companion. In other words, the precession timescale associated with the gravitational perturbations from the tertiary is slower than the short range precession timescales. The final separation may remain on a stable orbit\footnote{Note that tides always tend to shrink the binary separation, but this happens on much longer timescale.}. An example of this behavior is shown in the left panels of Figure  \ref{fig:MdwarfH}.
 However,  if
the eccentricity is excited on a much shorter timescale than the typical tidal (or GR) precession timescale (but, of course  still long enough so the secular approximation is valid), the orbit becomes almost radial and tidal precession does not have enough time to affect the evolution. In this case the peri-center distance may be shooter than the Roche limit of at least one of the binary members (see Eq.~(\ref{eq:RLoeb})).  An example of this behavior is shown in the right panels of Figure  \ref{fig:MdwarfH}. 

\begin{figure}[h]
\begin{center}
\includegraphics[width=\linewidth]{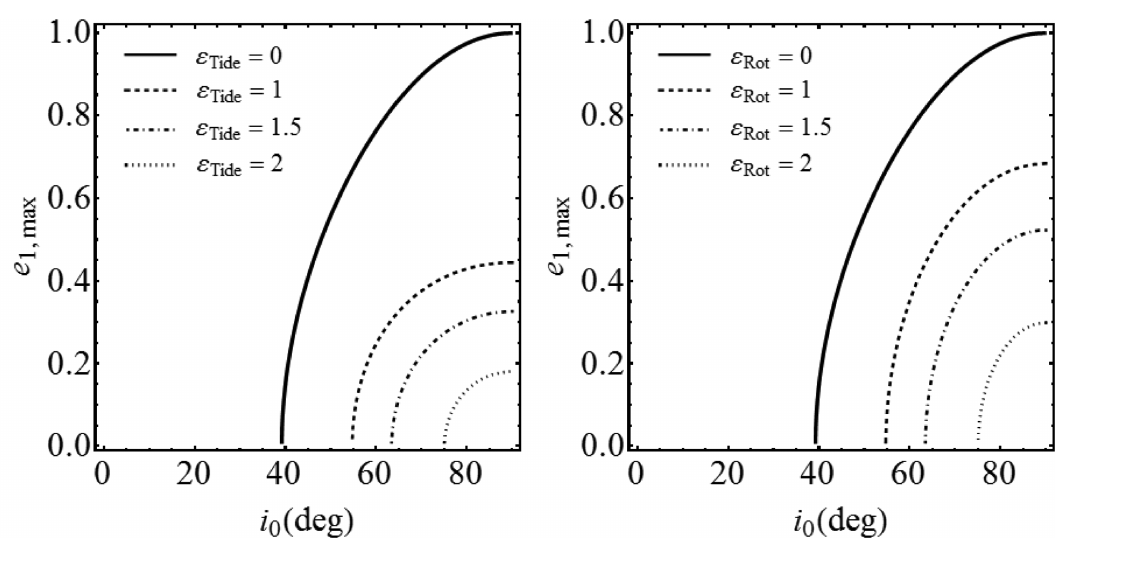}
\caption{\upshape {\bf Maximum eccentricity in the presence of rotation (right panel) and tides (left panel)}. Figure adopted from \citet{Liu+15}.
} \label{fig:LiuRotTide}
\end{center}
\end{figure}

The typical timescales associated with these precessions are (see equations (\ref{eq:TF1})-(\ref{eq:XYZ}) for the source of these timescales)
\begin{equation}\label{eq:tTF}
t_{\rm Tide} \sim  \frac{a_1^{13/2} m_2 (1-e_1^2)^5}{\sqrt{k} k_{L,2} f_T(e) m_1 (m_1+m_2) R_2^5}\ 
\end{equation}
and 
\begin{equation}
t_{\rm Rot} \sim  \frac{ \sqrt{k}  a_1^{7/2} m_2 (1-e_1^2)^2}{k_{q,2} \Omega^2_{s,2} \sqrt{_1+m_2} R_2^5}\ 
\end{equation}
for tidal and rotational  precessions respectively.  We define 
\begin{equation}
f_T(e_1)=1+\frac{3}{2}e_1^2+\frac{1}{8}e_1^4
\end{equation}
and $R_1$ and $\Omega_{s,2}$ are the radius and spin rate of $m_2$. Furthermore, $k_{L,2}$ is its Love parameter and $k_{q,2}$ the apsidal motion constant.  
Similarly to the GR case,  \citet{Liu+15}  defined $\varepsilon_{\rm Rot}=t_{\rm quad}/t_{\rm Rot} (1-e_1^2)^2$ and  $\varepsilon_{\rm Tide}=t_{\rm quad}(1-e_1^2)^5 /(t_{\rm Tide} f_T(e_1))$. With these definitions, Equation (\ref{eq:Liu1}) can be generalized \citep{Liu+15}
\begin{equation}\label{eq:Liu2}
\left(\frac{\varepsilon_{GR}}{\sqrt{1-e_1^2}} + \frac{1}{15} \frac{\varepsilon_{\rm Tide}}{(1-e_1^2)^{9/2}} \tilde{f}(e_1)+ \frac{1}{3} \frac{\varepsilon_{\rm Rot}}{(1-e_1^2)^{3/2}}\right)_{e_1=e_{1,max}}\approx \frac{9}{8}e_{1,max}^2\frac{ j_{1,min}^2-5 \cos^2 i_0/3}{j_{1,min}^2} \ ,
\end{equation}
where 
\begin{equation}
\tilde{f}(e_1)=1+3e_1^2+\frac{3}{8}e_1^4 \ .
\end{equation}
Note that here we used the $\varepsilon$ notation introduced in \citet{Liu+15}, to avoid confusion with their definition of $\dot{\omega}$ which is different than the one used in this review.  In Figure \ref{fig:Timescales} we show the tidal precession timescale compared to the other relevant timescales for a Jupiter around a Sun like star.  The maximum eccentricity that can be achieved as a function of inclination for a test particle approximation, and $e_{1,0}\sim 0$ is shown in Figure \ref{fig:LiuRotTide}.

\section{Applications}\label{sec:app}

 There are a few main general commonalties between all applications discussed below. The first is the possible outcome due to eccentricity excitation of the inner orbit. As shown in Figure \ref{fig:MdwarfH}, these high eccentricities can result in tidal evolution which will lead to tight inner binary, or it will result in Roche limit crossing. For a different astrophysical setting this can result in mergers, collisions, tidal disruption events, supernova etc. Another general outcome is that an initial   isotropic distribution of inclination of triple systems is not conserved. In the following we review a few examples of these applications to different astrophysical systems.  

\subsection{Solar system}

\begin{figure}
\begin{center}
\includegraphics[width=0.7\linewidth]{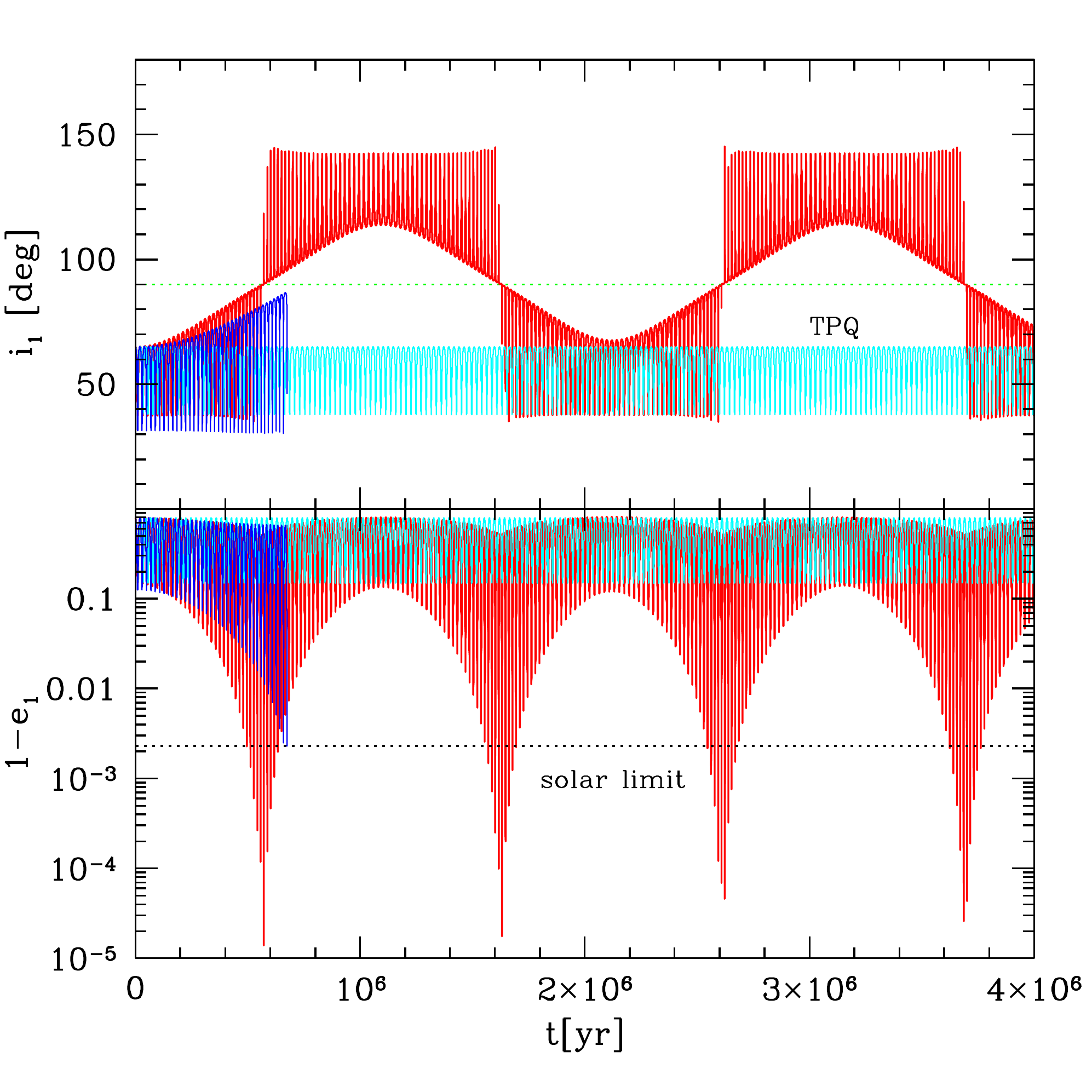}
\caption{\upshape Kozai's (1962) study of secular evolution of an asteroid due to Jupiter's gravitational  perturbations. The system is set with $m_1
  =1\,\msun$, $m_2\to0$ and $m_3 =1$~M$_J$, with $a_1 =2$~AU and $a_2
  =5$~AU. We initialize the system  with $e_1 =0.2$, $e_2
  =0.05$, $\omega_1 =\omega_2 =0^\circ$ and $i_\tot =65^\circ$. We consider the TPQ
  evolution (cyan lines) and the EKL evolution (red lines).  The thin
  horizontal dotted line in the top panel marks the $90^\circ$
  boundary. The result of an
  $N$-body simulation (blue lines) is also shown. The thin horizontal dotted line in
  the bottom panel marks the eccentricity corresponding to a collision
  with the solar surface, $1-e_1 = R_\odot / a_1$. At this instance we have stopped the numerical integration. Figure adopted from \citet{Naoz+11sec}.} \label{fig:Kozai}
  \end{center}
\end{figure}

 \citet{Kozai}  
 studied the secular dynamical evolution of 
an  asteroid, at $2$~AU, due to
Jupiter's gravitational  perturbations in the framework of the TPQ approximation. He showed that the asteroid undergoes large eccentricity and inclination oscillations. 
Considering the  hierarchical nature of the approximation, we note that the system is in fact not valid to be addressed by secular approximation. The semi-major axes ratio between the asteroid ($2$~AU) and Jupiter ($5$~AU) yields rather large value ($a_1/a_2=0.6$), which suggests that the hierarchical approximation is not valid. Furthermore, \citet{Kozai} assumed the  Jupiter's
eccentricity is strictly  zero. Taking into account Jupiter's eccentricity $\sim 0.05$ leads to non-negligible contribution from the octupole level of approximation $ \epsilon = 0.03$, this suggest that the EKL mechanism may significantly alters the
evolution of the asteroid. This is shown in Figure \ref{fig:Kozai} which considers  the TPQ approximation (cyan lines) but also consider the EKL evolution (red lines). The latter show that the TPQ approximation  is rather inadequate to address this problem. Furthermore, as mentioned, Jupiter is not far away enough to unitize the hierarchical  approximation for this problem, which can be seen from the  $N$-body
simulation result, using the {\tt Mercury} software package
\citep{Mercury}. We used both Bulirsch-Stoer and symplectic
integrators \citep{WH91}. This calculation shows that indeed the asteroid may impact the sun, and that the actual evolution of the system is closer in behavior  to the EKL (TPO in this case) than the TPQ approximation.

 As mentioned above, the TPQ approximation can successfully describe  the evolution for a verity of test particle systems in the solar system. For example, it was used to explain the  inclinations of gas giant satellites and Jovian irregular
satellites \citep[e.g.,][]{Kin+91,vas99,Carr+02,Nes+03,cuk+04,KN07}. Furthermore, the importance of secular interactions for the dynamics of comets and other test particles in the solar system was noted in several studies \citep[e.g.,][]{Kozia79,Quinn+90,Bailey+92,Thomas+96,Duncan+97,Gronchi+99,Gomes+05,Tamayo+13}.   
Another interesting example of the application of three body dynamics relates to binary minor planets. Observations suggests that Near Earth asteroid binaries are common (about $15\%$ $r_{NEA}>300$m \citep{Pravec+06,Margot+15} and perhaps as high as $63\%$, for a larger range of sizes \citep{Polishook+06}). Furthermore, about $15\%$ of asteroids and high multiples reside in binaries \citep{Pravec+06} and \citet{Nesvorny+11} suggested that  the binary fraction in the Kuiper belt can be as high as $40\%$. In all of these cases a natural third body is simply the Sun, which gravitationally perturb the binary orbit. 
 \citet{PN09} and \citet{Naoz10obs}
have studied the evolution of binary minor planets, in the frame work of TPQ, and showed that the dynamical evolution largely affects the observed orbital  distribution of these objects. Specifically they showed that in the inclination--separation phase space there is a regime associated with high mutual inclination which is   devoid of eccentric wide binaries. Eccentricity excitations in this regime, due to the Sun's gravitational  perturbation, can lead to shrinking, and circularizing of the binary's orbit, or even lead to binary coalescence's.   Furthermore,   \citet{Kin+91},
\citet{vas99}, \cite{Carr+02}, \citet{Nes+03}, \citet{cuk+04} and
\citet{KN07} suggested that secular interactions and Kozai oscillations  may explain the
significant inclinations of gas giant satellites and Jovian irregular
satellites.
Binaries that are closer to the sun, such as binary asteroids and near Earth binaries will be sensitive to a wider range of physical effects, and specifically the induced precession of the binary due to an oblate object may suppress eccentricity excitations  \citep{Fang+12}. Another, potentially important mechanism, is the YORP effect which can significantly alter the spin of asteroids and near earth objects \citep[e.g.,][]{Polishook+09}. This in turn can result in even larger effects on the precession due to rotation.

\subsection{Planetary systems}

 Recent ground and space based observations have transformed our understanding of the properties of exoplanetary systems. The detection of several thousand planets and planet candidates have revealed many puzzles that  challenge traditional planet formation theories and generated many new ideas. One of the greatest mysteries in the last two decades lays in a class of giant planets called ``Hot Jupiters." These are a Jupiter size planets that are found in extremely short period orbits around their host stars (i.e. periods of a few days or less).
Most theories posit that these planets still form on larger ($> AU$) scales, like in the solar system, but move inwards to short orbital periods. Thus, a  migration mechanism is needed to reduce the angular momentum of these planets by two orders of magnitude (from few AUs to about few percent of an AU).
  Broadly speaking, there are two main channels considered in the literature to achieve this. In the first channel, planets form in the disk, and in some cases, angular momentum exchange between the planets and the protoplanetary disk  can produce inward migration \citep[e.g.,][]{Lin+86,Mass+03}. In the second channel, planets also formed in the disk, but dynamical interactions between multiple planets or a stellar companion greatly affect the final orbital configuration of the system, through a variety of mechanisms such as planet-planet scattering \citep[e.g.,][]{RF96}, EKL (see below), or secular chaos \citep{Lithwick+12,Hansen+15}. The role of planet or stellar dynamical interactions is motivated by the presence of substantial eccentricities amongst the more distant Jovian population, and the discovery  of
high obliquities \citep[misalignments between planetary orbital and host star spin directions, e.g.,][]{Albrecht+12}. Both of these features would tend to be damped by the dissipative interactions with a 
protoplanetary disk and have spawned an interest in processes that can lead to migration through predominantly dynamical interactions.

  The first application of three body secular interaction to a planetary system began with the detection of 16~Cyg B \citep{Cochran+96}, where 
\citet{Hol+97} and \citet{Mazeh+97} attributed  its high eccentricity ($e\sim 0.63$) to the Kozai-Lidov mechanism (in the framework of the TPQ approximation).
They also showed that the planet spends about $\sim 35\%$ of its lifetime in a high eccentric orbit $e>0.6$. In subsequent nominal studies by \citet{Wu+03}, \citet{Wu+07} and \citet{Dan} the consequences of the TPQ approximation in forming Hot Jupiters in stellar binaries was investigated in greater detail including GR and tides. 
As the orbit evolves dynamically due to gravitational perturbation from the outer orbit the planet's orbit becomes eccentric and the planet spends long times around the host star. At these intervals tides on the planet and on the star affect  the orbit, which tends to circularize and shrink it.  This scenario was suggested as a possible formation channel for Hot Jupiter without the need for disk migration \citep{Lin+86}.

As an aftermath of using the TPQ approximation these studies found that in order to form Hot Jupiters the initial mutual inclination needs to be rather close to perpendicular \citep[$90^\circ\pm 3^\circ$ e.g.,][]{Dan}.  An important outcome from these calculations  was the prediction of retrograde Hot Jupiters (i.e., obliquities larger than $90^\circ$)  \citep{Dan,Wu+07}.  The recent detections of retrograde Hot Jupiters \citep[e.g.,][]{Tri+10,Winn+10b,Albrecht+12} resulted in a new interest in the possibility that secular three body interactions presented in this field.


\begin{figure}
\begin{center}
\includegraphics[width=0.9\linewidth]{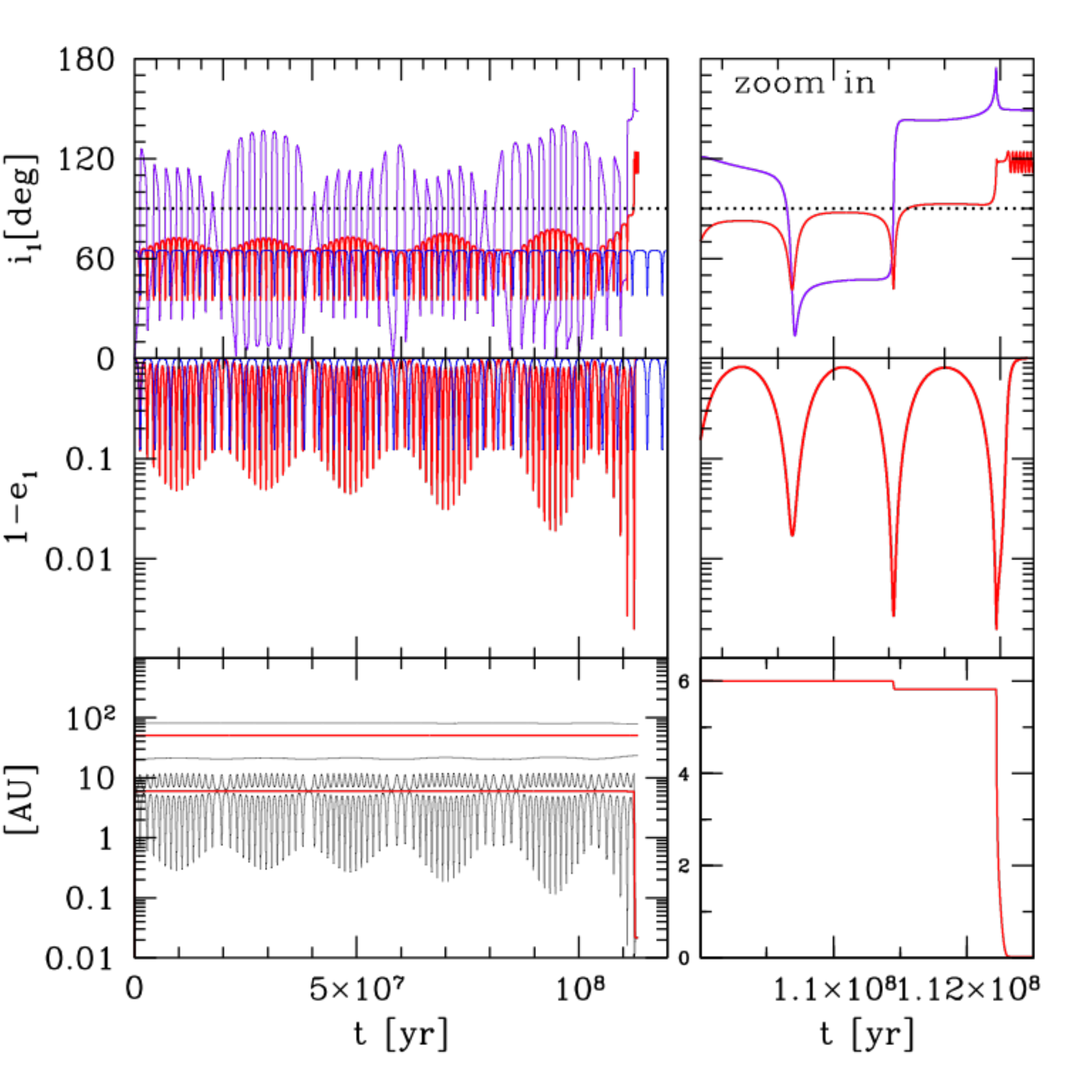}
\caption{ \upshape{\bf Hot Jupiter formation in a two planet system}  Left panels show the full evolution and the right panel show a zoom in on the final three quadrupole cycles. We consider the full, octupole level evolution which includes GR and tides evolution (red lines), the quadrupole level, including  GR and tides (blue lines). 
Top panels show the inclination of the system of the  full, up to the octupole level evolution which includes GR and tides  (red line), and the inclination for the   quadrupole level, including  GR and tides (blue lines). In purple we show the obliquity. Middle panels show the eccentricity as $1-e_1$ (again red lines are for the octupole and the blue lines are for the quadrupole). Bottom panels show the semi-major axes for the outer (top) and inner (bottom) binaries (red lines)  and their apo- and peri-centers (grey lines). Note that the left bottom panel is log scale while the right bottom panel is linear scaled. The system parameters are: $m_1=1$~M$_\odot$, $m_2=1$~M$_j$, $m_3=3$~M$_j$, $a_1=6$~AU, $a_2=61$~AU, $e_1=0.001$, $e_2=0.6$, $\omega_1=45^\circ$, $\omega_2=0^\circ$ and $i_{\rm tot}=71.5^\circ$. The system started with zero obliquity and the spin periods of the star and the planet are $25$~days and $10$~days, respectively.  The viscous  times here are $t_{V,1}=5$~yr and $t_{V,2}=1.5$~yr, the spin period of the star was assumed to be $25$~d. Figure adopted from \citet{Naoz11}, but shows the evolution of the obliquity.
     } \label{fig:HJnature}
  \end{center}
\end{figure}

\begin{figure}
\begin{center}
\includegraphics[width=\linewidth]{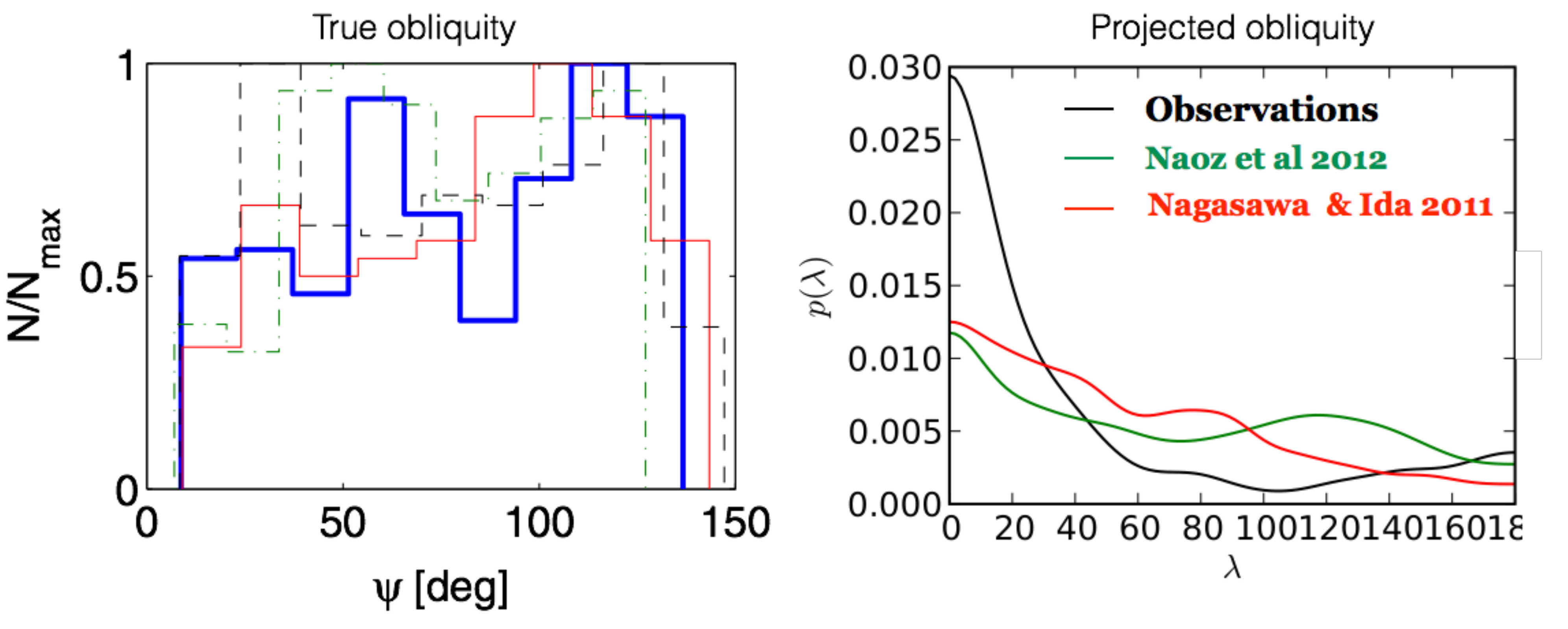}
\caption{ \upshape{\bf Hot Jupiter obliquity distribution in stellar binaries}. Left panel shows the true obliquity ($\psi$) distribution, as a result from  fiducial Monte Carlo simulations by \citet{Naoz+12bin} (blue line), for wide range of companion initial separations and setting planetary viscous tides to be $t_{V,2}=1.5$~yr. This distribution has a characteristic cut-off near $140^\circ$. This limit arises from the Kozai angles (the seperatrix $\sim 140^\circ$) for which the large oscillations take place. Also shown are the results from the Monte Carlo simulations with different settings. In particular, the dashed black line represents  a companion at $a_2=1000$~AU, and   thin solid red line represents a companion with $a_2=500$~AU.  In both cases the planetary viscous tides is set to be: $t_{V,2}=1.5$~yr. Also over-plotted is a Monte-Carlo simulation for a companion separation of  $a_2=500$~AU with $t_{V,2}=0.015$~yr,  dot-dashed green line.  Right panel shows the projected obliquity from \citep{Naoz+12bin} Monte-Carlo simulations, as well as the observations (as for 2012) {\it exoplanets.org}, and the projected obliquity of \citet{Nag+11}.  The stellar spin-period assumed for these figures was $25$~d. Different Roche-limit estimates do not change this result  \citep[e.g.,][]{Petrovich15He}, however, different stellar spin-periods or evolution of the spin period may result in deviation form this distribution  \citep{Storch+14}. Figure adopted from \citet{Naoz+12bin}. 
     } \label{fig:HJobliq}
  \end{center}
\end{figure}

The formation of Hot Jupiters via the EKL mechanism, including GR and tides for two planet systems was studied in \citet{Naoz11}, see  Figure \ref{fig:HJnature}. A simplified Monte-Carlo for initially an aligned Jupiter  in a two planet system resulted in a nearly uniform obliquity distribution, as well as nearly uniform mutual inclination distribution. Similar results for the inclination and obliquity distributions were achieved for the formation of Hot Jupiters in  stellar binary systems (effectively repeating the analysis by \citet{Dan}  but for the EKL mechanism and exploring larger range of orbital parameters). The obliquity distribution is shown in Figure \ref{fig:HJobliq} left panel. Projecting the resulted obliquity angles on the sky (see right panel of Figure \ref{fig:HJobliq}) allows for direct comparison with observations \citep[e.g.,][]{MJ}.  \citet{Naoz+12bin} performed a bayesian analysis that treats the complete obliquity distribution as a sum of contributions from an aligned component, an EKL component, and planet--planet scattering component \citep[adopting][]{Nag+11}. They found  that the EKL most likely accounts for $\sim 30\%$ of the observed systems and planet--planet scattering contributes about $\sim 10\%-20\%$, independently of the formation rate.  That analysis also showed that EKL produces between $60\%$ to $80\%$ of large obliquity angles. 
These values are consistent with complementary  analyses that showed that Hot Jupiters are likely to have a far away companion \citep[e.g.,][]{Knutson+14,Ngo+15,Wang+15}. 


It was later shown, in the frame work of hierarchical triple system, that the behavior of the obliquity angle is chaotic in nature \citep{Storch+14,Storch+15}. 
The planetary orbital angular momentum vector precesses around the total angular momentum at a rate which is inversely proportional to the quadrupole timescale $\sim t_{\rm quad}^{-1}$. Due to the rotation-induced stellar quadrupole, the planet induces precession in the stellar spin axis which is proportional to the stellar spin's angular momentum. As shown by \citet{Storch+14}, when the latter precession spin is larger than the orbital precession rate, the stellar spin axis  follows ${\bf G}_1$ adiabatically, while maintaining an approximately constant obliquity. In the other extreme case, when the maximal spin precession rate  is always smaller than the orbital precession rate the spin axis effectively precesses around the total angular momentum (about which ${\bf G}_1$ is precessing). In the intermediate regime, \citet{Storch+14} showed that a secular resonance occurs, which leads to complex and chaotic spin evolution. Short range forces can further complicate the obliquity evolution, and  affect the formation of Hot Jupiters \citep{Storch+14,Storch+15}.

\begin{figure}
\begin{center}
\includegraphics[width=0.6\linewidth]{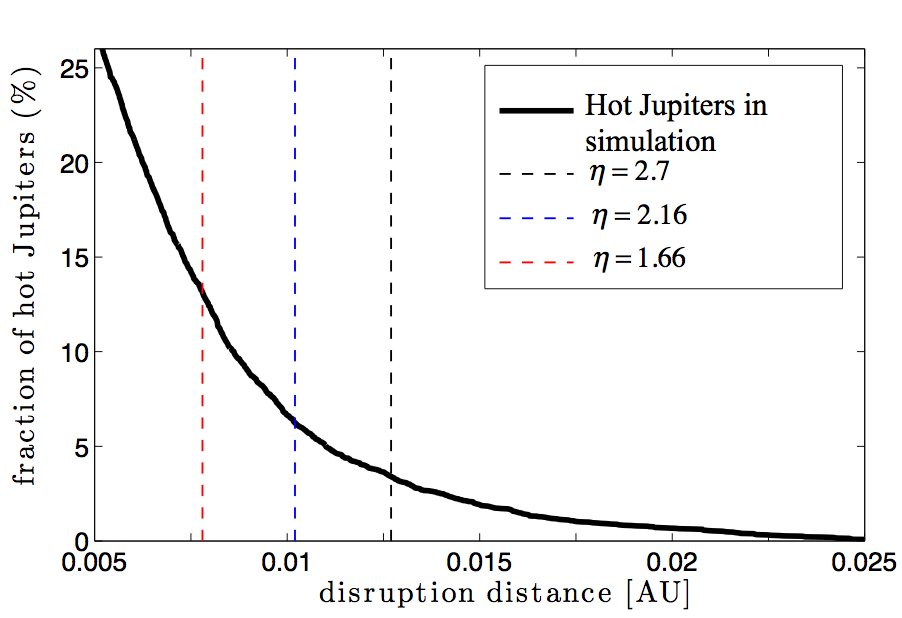}
\caption{ \upshape{\bf Fraction of Hot Jupiters in stellar binaries} The fraction of hot Jupiters formed in the fiducial Monte Carlo simulation by \citet{Petrovich15He} as a function of the disruption distance. The vertical lines indicate different disruptions distances parametrized by $\eta$ in Equation  (\ref{eq:RLoeb}). The different  values ($\eta=1.66, 2.16$ and $2.7$)   correspond to the values adopted in three different studies:   \citet{Naoz+12bin}, \citet{Wu+07} and \citet{Petrovich15He},  respectively. 
Figure adopted from \citet{Petrovich15He}. } \label{fig:HJfrac}
  \end{center}
\end{figure}

The large eccentricity excitations induced via the EKL mechanism can result in a nearly radial motion and drive the planet into the star (as illustrated in Figure \ref{fig:MdwarfH}, right panel). Thus, the formation fraction of Hot Jupiters is highly sensitive to the disruption distance (as shown in  Figure \ref{fig:HJfrac}, vertical lines are based on Eq.~(\ref{eq:RLoeb})). For lower mass planets, such as rocky planets, tides (or GR, or quadrupole moments from fast rotating stars) are largely ineffective to stop the EKL's nearly radial motion, resulting in high probability of tidal disruption \citep[e.g.][]{Lanza+14,Rice15}.    
Apart from tidally disrupting the planet, binary companions can also lead to large instabilities, which may result in swapping planets between the stars  \citep[e.g.,][]{Kratter+12,Moeckel+12}. In addition, as the star evolve beyond the main sequence, the existence of a companion (either a star, brown dwarf or a planet) can lead to ejection of planets \citep[e.g.,][]{Veras+12,Veras+13,Veras+14} or engulfment of the inner most planet \citep[e.g.,][]{Li+14Kepler56,Frewen+15}. 

The observational studies that showed that Hot Jupiters are likely to have far away companion  \citep[e.g.,][]{Knutson+14,Ngo+15,Wang+15} promoted further investigations of two planet systems. As shown in Figure \ref{fig:2plTey}, a similar mass perturber yields large eccentricity excitations with suppression of large eccentricities for large mutual inclinations  \citep{Tey+13}. Therefore an inclined planetary perturber can lead to short period oblique planets \citep{Naoz11,Li+14Kepler56}. If large eccentricities are generated, according to the stability criterion in Equation (\ref{eq:2p}) the inner planet can either be ejected from the system or collide with the host star. In some cases, when the forced eccentricity from the perturber causes the orbit to shrink, the orbit reaches a semi-major axis for which tidal precession is comparable to the quadrupole timescale (as noted in Section \ref{sec:tides}). This suppresses further circularization and shrinking of the orbit, which may lead to the formation of eccentric warm Jupiters \citep{Dawson+14}.

 Recently, the Kepler mission detected several circumbinary planetary systems \citep{Doyle+11,Orosz+12,Orosz+12N,Welsh+12,Welsh+15,Kostov+13,Kostov+14,Schwamb+13}. These systems are composed of a stellar binary on an orbit with a typical period of $7.5$ to $41$~days orbited by a planetary companion on a much longer period ($\sim 50 - 250$~days). Interestingly, 
no transiting planets have been found around more compact stellar binaries ($\lsim 7$~days period), although these binaries are abundant  in nature  and in Kepler eclipsing binary data \citep{Raghavan+06,Raghavan+10,Tokovinin14a}. Two main questions about  circumbinary planets were addressed recently in the literature. One considered the apparent absence of circumbinary planets around compact stellar binary, and the other was about the configuration of the planetary orbit. Starting with the former,  the formation of compact stellar binaries is often associated with dynamical evolution, which involves a tertiary \citep[e.g.,][and see below]{NF}. In the context of this channel, it was suggested that the outer perturber that drives the two stars into a tight  orbit may also impact the planetary companion around the inner two stars and may result in a   large   eccentricity planetary orbit leading to ejection or colliding with the inner stars. However, circumbinary planets around compact binaries may still exists but they probably will end up to be misaligned with the inner stellar orbit \citep[e.g.][]{Hamers+15c,Martin+15cp,Munoz+15}. In fact the misalignment can be generated simply due to eccentricity  and  inclination  oscillations on the inner orbit, from a test particle as shown in Figure \ref{fig:smallm}, \citep[e.g.,][]{Martin+15}. Therefore, since many of the Kepler binary  detections are eclipsing binaries, if a misaligned systems around the stellar systems were to exist, they are presently hidden from the current Kepler detection methods. This may imply that circumbinary planets are rather abundant, perhaps even more than planets around single stars (e.g., \citet{Armstrong+14,Martin+15cp,Martin+15cpf,Martin+15} but see, \citet{Deacon+15}).

\subsection{Stellar systems}

 Most massive stars reside in a binary configuration \citep[$\gsim 70\%$  for massive stars; see][]{Raghavan+10}. It seems that stellar binaries are responsible for diverse astrophysical phenomena, from Type Ia supernovae to X-ray binaries.  However, observational campaigns have suggested that 
  probably many of these binaries are in fact triples  \citep[e.g.,][]{T97,Tok08,Eggleton+07}. 
 \citet{T97} showed that
$40\%$ of binary stars with period $<10$~d in which the primary is a
dwarf ($0.5 -1.5\,M_{\odot}$) have at least one additional
companion. He found that the fraction of triples and higher multiples
among binaries with period ($10-100\,$d) is $\sim10\%$. Moreover,
\citet{Pri+06}  surveyed a sample of contact binaries, and noted
that among 151 contact binaries brighter than 10 mag., 42$\pm5\%$ are
at least triple. Furthermore, a recent analysis of eclipse time variation curves of {\it Kepler} binaries showed that indeed a substantial fraction of these binaries have a third body \citep{Borkovits+16}. Thus, it seems that triple stars are abundant in our galaxy.
 From dynamical stability arguments
these must be hierarchical triples, in which the (inner) binary is
orbited by a third body on a much wider orbit.

\begin{figure}
\begin{center}
\includegraphics[width=\linewidth]{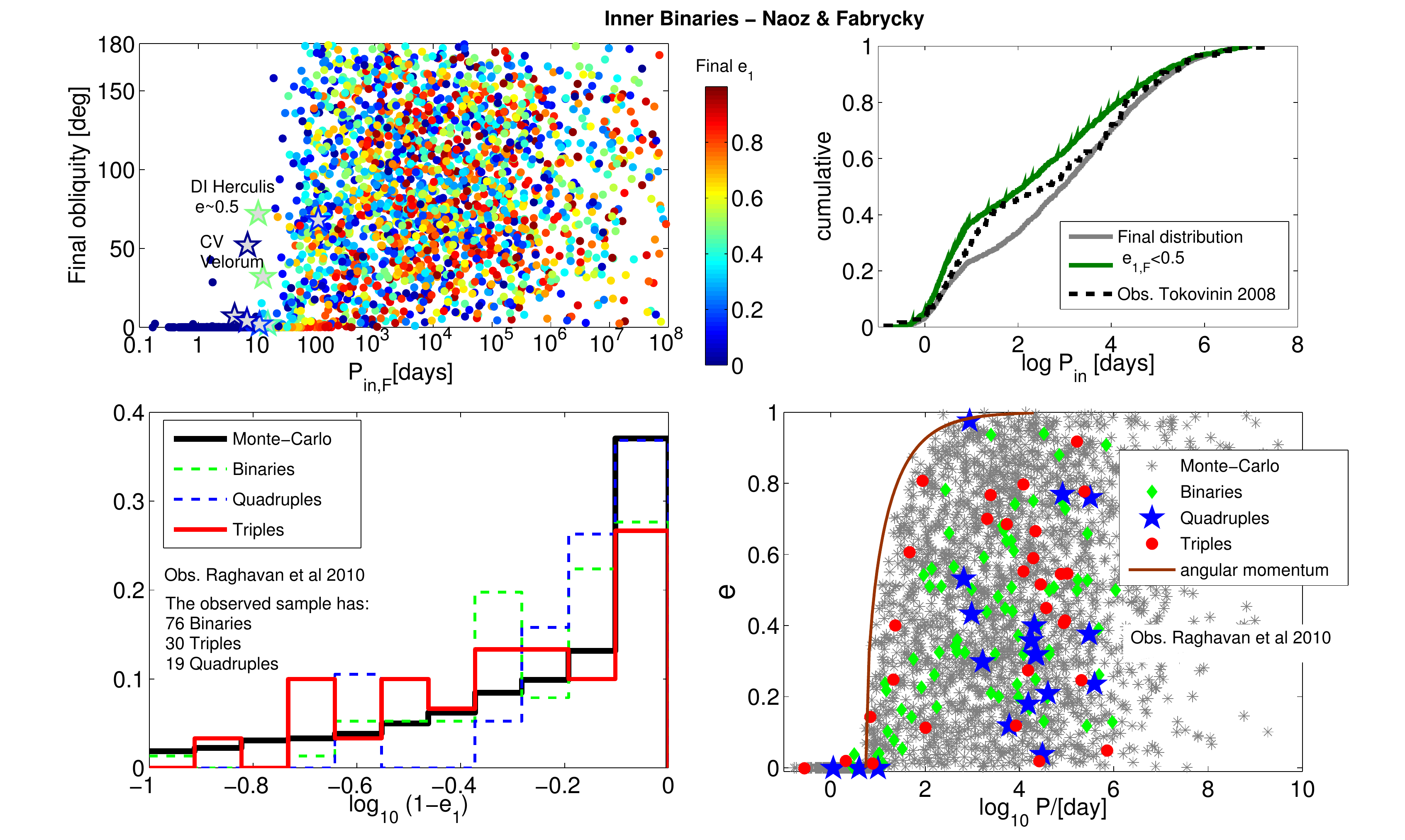}
\caption{\upshape {\bf Simulated inner binary orbital configuration compared to observations}. {\bf Top right panel}  the cumulative distribution of the observations distribution  taken from \citet{Tok08} public catalog (black dashed line), compared to the final distribution (grey solid line). Since the public catalog has typical inner orbital eccentricity of $0.5$, the final distribution is also shown for system with $e_{1,F}<0.5$ (green solid line). 
{\bf Top left panel:} Final distribution of the spin orbit angle (i.e., the obliquity) of the primary Vs the final period of the inner orbit, the color code is the final eccentricity of the inner binary. We also plot the observations \citep{Albrecht09,Albrecht11,Albrecht13,Albrecht+14,Triaud+13,Harding+13,Zhou+13}. {\bf Bottom left panel:} shows the inner orbit final eccentricity as a function of the final period. Over-plotted are observations adopted from  \citet{Raghavan+10} public catalog. The solid line represents a constant angular momentum curve with a final binary period of $5.5$~days. 
{\bf Bottom right panel:}  reproduction of  the inner orbit specific angular  momentum  distribution considered first by \citet{Har68}, compared to  \citet{Raghavan+10} observations. 
The top two panels and the Monte-Carlo simulations are adopted  from  \citet{NF}. 
} \label{fig:ObsPe}
  \end{center}
\end{figure}

Application of the secular hierarchical triple body system to triple stellar system was first considered by 
\citet{Har68,Har69}. His work was motivated by \citet{Heintz67} that observed triple stellar systems with possible perturbations  form the outer orbit. In this early work he already recognize the importance of the octupole level of approximation and expanded the Hamiltonian up to the octupole level of approximation. From the equations of motion he estimated a distribution for the inner orbit specific angular momentum $\sqrt{1-e_1^2}$ to match the observed distribution of triples. Later, \citet{Mazeh+79} showed that tidal effects during eccentricity excitations of the Kozai-Lidov cycle can circularize and shrink the orbit.

During the system evolution,  the their star can  cause large  eccentricity excitations for the inner orbit. Therefore, the nearly radial motion of the  binary  drives the two stars to merge, however, tidal forces tend to shrink and circularize  the orbit. If during the evolution the quadrupole--level of approximation  precession timescale is longer than the precession timescale associated with  short range forces (such as  tides, e.g., Eq.~(\ref{eq:tTF}) or GR, e.g., Eq.~(\ref{eq:tGR})) further eccentricity excitations are suppressed. In this channel, tidal forces can shrink  and circularize the inner orbit, forming a tight inner stellar binary decoupled from the tertiary.  
This process  was studied in great length in the literature as a promising channel to explain triples and close binaries observations  \citep[e.g.,][]{Sod75,Soderhjelm82,Sod84,1998KEM,1998EKH,Egg+01,Ford00Pls,Dan,PF09,Tho10,Shappee+13,NF}. We show here the updated inner orbit specific angular momentum simulated distribution compared to observations in Figure \ref{fig:ObsPe} bottom left panel, reproducing \citet{Har68,Har69} figure. Observations are taken from \citet{Raghavan+10} and Monte-Carlo simulations are adopted from \citet{NF} EKL triple star simulations.

\begin{figure}
\begin{center}
\includegraphics[width=\linewidth]{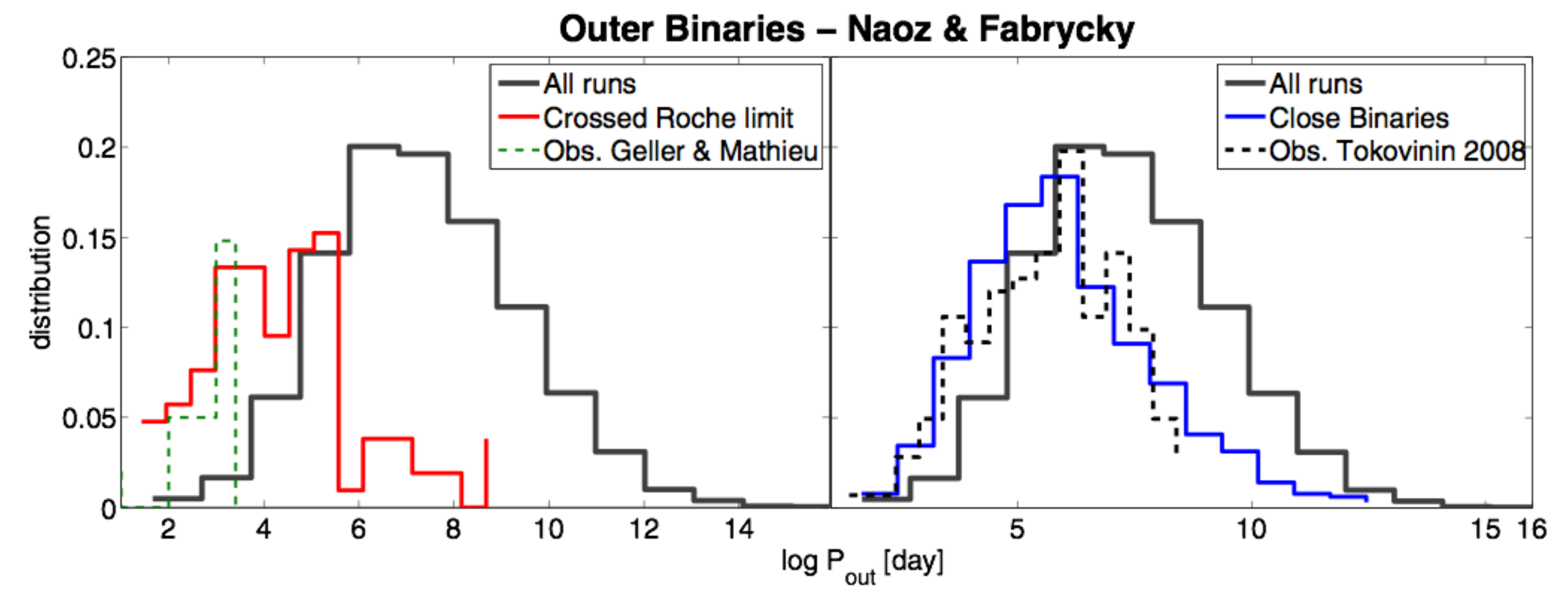}
\caption{\upshape {\bf Simulated outer binary period compared to observations}.  {\it Right  panel}:  shows the period distribution of the companion of the close binaries (blue line), the latter defined to have periods shorter than $\sim 16$~days. Over-plotted is     the observed distribution, scaled to match the theory lines, adopted from \citep{Tok08} public catalog. {\it Left  panel}:  shows the period of companions for the merged stellar population (red line) and the observed blue stragglers binary distribution of NGC 188 \citep{GM12}, also scaled to match the theory lines.  In both panels, grey lines represent the period distribution at the final stage of all of the outer companions in the Monte-Carlo runs.  
Figure adopted  from  \citet{NF}. 
} \label{fig:outer}
  \end{center}
\end{figure}

 \citet{NF}  ran a large Monte-Carlo simulations, including the EKL mechanism,  tides (as described in section \ref{sec:tides}) and general relativity, for $10$Gyr of evolution producing the  distribution   for  semi-major axis, eccentricity, inclination, and obliquity. The observed bimodal distribution of the inner orbit reported by  \citet{Tok08} public catalog (see Figure \ref{fig:ObsPe}, top left panel) is reproduced by  \citet{NF}  simulations. Their  Kolmogorov-Smirnov test 
does not reject the null hypothesis that the observed inner orbit period's distribution and the simulated one are from the same continuous distribution. Furthermore, they found that the simulated outer orbit distribution of the close binaries is consistent  with the one from 
 \citet{Tok08} catalog of observed triples (e.g., Figure \ref{fig:outer}).
Thus, they concluded that   secular evolution in triple's plays an important role in shaping the distribution of these systems.

\citet{TS02} reported that wide binaries are more likely to have non-negligible eccentricity (see also Tokovinin \& Kiyaeva 2015). For wide binaries in  triple systems this seems to be in agreement with the dynamical eccentricity excitation from an outer perturber where tidal shrinking and circularization are less efficient \citep[as can be seen in Figure \ref{fig:ObsPe} bottom left, adopted from][]{NF}.  The systems near the constant angular momentum line (solid line in the Figure), may represent 
a population of migrating binaries due to tidal dissipation \citep[as also seen in the {\it Kepler} binary stellar population, e.g.,][]{Dong+13}. Furthermore, The  formation channel  of close stellar binaries via EKL and tides was suggested to somewhat suppress the likelihood of finding aligned circumbinary planets around tight stellar binaries  \citep[e.g.][]{Hamers+15c,Martin+15cp,Munoz+15}.

An interesting and promising observable for triple stellar dynamics may be the obliquity angle. As more binary stars obliquities are being observed, e.g., the BANANA survey \citep{Banana}, and by other individual endeavors, the obliquity distribution may provide a key observable.  During the tidal evolution the obliquity of the tight binaries will 
most likely decay to zero faster than the eccentricity. This  results in  systems that are close to the angular momentum line to have typically low obliquities \citep{NF}. This behavior  is  depicted  in Figure \ref{fig:ObsPe} top left panel that shows that the  final obliquity distribution of close binaries with moderate eccentricities (blue to yellow color) have  moderate obliquities. Close  circular inner binaries with non-negligible  obliquities ($>10$
deg) are found to have smaller spin periods \citep[see also][]{Levrard+07,Fabrycky+07}.  The simulated stellar obliquities shown in Figure \ref{fig:ObsPe} are  consistent with the current available observations.

\begin{figure}
\begin{center}
\includegraphics[width=0.8\linewidth]{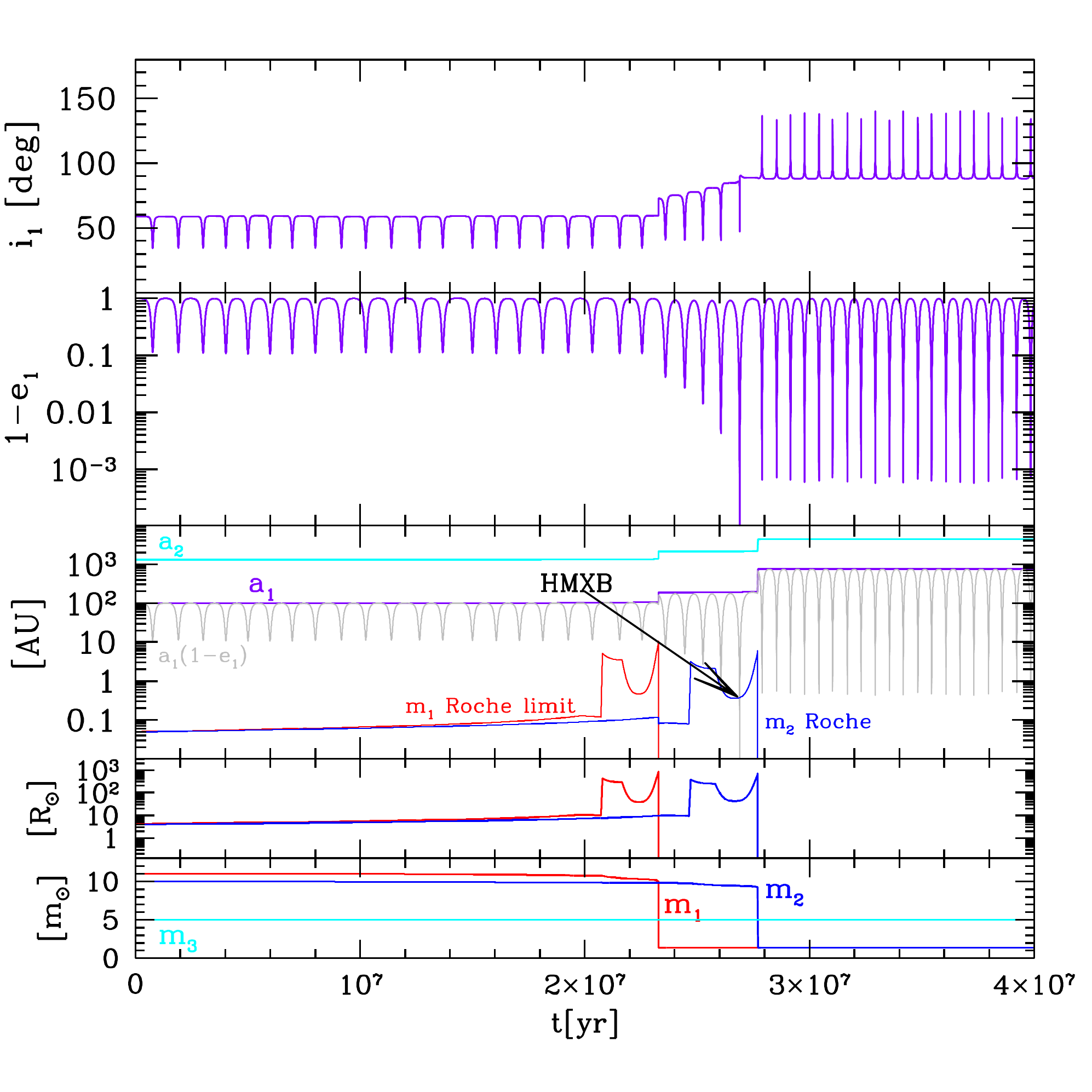}
\caption{\upshape {\bf Re-triggering EKL by mass loss}. This example produces a High Mass X-ray Binary (HMXB) or a supernova impostor.  We show (top to bottom) the inner orbit inclination $i_1$, the inner orbit eccentricity (depicted as $1-e_1$), the semi-major axis of the inner (purple) and outer  (cyan) orbit as well as the inner orbit peri-center and the two masses disruption distances (see Eq.~(\ref{eq:RLoeb}) for a popular  definition), the stellar radii and finally the bottom panel shows the masses of all three stars. This calculation includes solving the equations of the octupole-level of approximation, GR for both the inner and outer orbit, and stellar evolution according to SSE \citep{Hurley+00}, which includes mass loss and staler inflation. For simplicity the supernova was modeled  here as a simple mass loss and assuming no kicks. 
 Tidal evolutions were turned off for illustration purposes. 
A consequence of  the first mass loss episode and the formation of a neutron star is that the initially small $\epsilon_M$ increased. This yields eccentricity excitations leading to Roche limit crossing as the   $m_2$ star's radius  inflates . This may result in a     HMXB, or even a supernova impostor.
The system parameters are set initially: $m_1=11$~M$_\odot$, $m_2=10$~M$_\odot$, $m_3=5$~M$_\odot$, $a_1=100$~AU, $a_2=1300$~AU, $e_1=0.001$, $e_2=0.6$, $\omega_1=\omega_2=0$ and $i_\tot=79^\circ$. These parameters yield initial  $\epsilon_M=0.0034$. \citet{Naoz+15} discussed the formation scenario of low mass X-ray binaries via triple body evolution similarly to the example illustrated  here. 
} \label{fig:StEv}
  \end{center}
\end{figure}

Strong gravitational perturbations can lead to mergers of the inner members, if the tidal forces cannot react fast enough to stabilize the system (see for example Figure \ref{fig:MdwarfH} right panel).  In the previous  section we discussed tidal disruptions of Hot Jupiters due to large eccentricity excitations. In the context of triple stellar systems, extreme values of the eccentricity which take place on shorter timescales than the short range forces (such as GR and tides, but still long to allow the system to  remain secular), may lead to 
 the merger of stellar binary. If sufficient time has past from the merger time (perhaps at the order of Kelvin-Helmholtz timescale) this merger product may be identified as a blue  straggler. 
  \citet{PF09} envisioned a two-step process for which triple body interactions can form blue  stragglers. In their study, three-body dynamics plus tidal dissipation created a close binary, and that binary subsequently merged by magnetic breaking or had unstable or efficient mass transfer. \citet{NF} suggested that large eccentricity excitation during  the EKL evolution can lead to mergers. They found that their simulated  outer orbital period distribution is consistent with  observations for the companion of the merged population, adopted from \citet{GM12}, as depicted in Figure \ref{fig:outer}. 
  This further  emphasizes the notion that three body secular interactions may be the main channel for merged systems like blue stragglers. 
  
  Another interesting evidence for a merged system via perturbations from a distant perturber was recently found in the Galactic Center. 
Specifically, it seems that the object known as G2   \citep{Gillessen+12} is a binary star in disguise  \citep{Witzel+14}. Therefore, a similar mechanism to that of the formation of blue stragglers  may  operates in the Galactic center, where the massive black hole in the center of the galaxy causes large eccentricity excitations on a stellar binary in its vicinity  \citep[e.g.,][Stephan et al.~in Prep.]{AP12,Prodan+15}.

The secular approximation allows for very long integration times where stellar evolution may play an important role \citep[e.g.][]{PK12,Shappee+13}. In particular, systems that have inner binary members  with close  mass values (i.e., $m_1\approx m_2$), the octupole level is suppressed (recall the definition of $\epsilon_M$, Eq.~(\ref{eq:epslionM}), and  Figure \ref{fig:e1_excite}).   However, stellar  mass loss during post main sequence evolution can dramatically change the mass balance and re-trigger the EKL behavior \citep{Shappee+13}.  This is because the semi-major axis changes proportional to the mass loss ratio, i.e., $a_f/a_i=m_f/m_i$, where the subscripts ``$f$" and ``$i$" refer to the final and initial values. Note that adiabatic mass loss conserves the value of the orbital eccentricity.
Thus, the $\epsilon_M$ due to mass loss can change compared to the initial value  \citep[e.g.,][]{Shappee+13,Michaely+14,Naoz+15}
\begin{equation}\label{eq:EKM}
\frac{\epsilon_{M,f}}{\epsilon_{M,i}}= \frac{m_{1,f}+m_2+m_3}{m_{1,i}+m_2+m_3}\left( \frac{m_{1,f}-m_2}{m_{1,i}-m_2} \right)  \left(\frac{m_{1,i}+m_2}{m_{1,f}+m_2}  \right)^2 \ ,
\end{equation}
where for simplicity, for this equation, we assumed that only one mass will undergo mass loss ($m_{1,i}\to m_{1,f}$). Overall the absolute value of $\epsilon_{M}$ via this process increases. An example of this evolution is shown in Figure \ref{fig:StEv}. The system is set initially with an inner binary composed with two similar mass stars. As the more massive star losses mass the new  $\epsilon_{M}$ increased, according to Eq.~(\ref{eq:EKM}) allowing for larger eccentricity excitations. When the stars inflate in radius as they leave the main sequence, the disruption   distance associated  with their Roche limit increases as well (e.g., Eq.~(\ref{eq:RLoeb})). The eccentricity excitations were too small to affect the orbit before the first neutron   star was born. However, during the  large eccentricities excitation  after $\epsilon_{M}$ increased, the inflation in radius of the less massive star resulted in having this star crossing its Roche limit. 
 This may form a high mass X-ray binary which may be associated with  a supernova impostor \citep[as suggested for a binary interaction for supernova 2010d a, e.g.,][]{Binder+11}.  Another possible outcome for this system is a Thorne-$\dot{\rm Z}$ytkow object \citep[e.g.,][]{Thorne+75}, which has distinct observational signatures \citep[e.g.,][]{Levesque+14}. Recently, \citet{Naoz+15} showed that triple dynamics can offer a possible formation channel to low mass X-ray binaries, while skipping the common envelope phase, and by that overcoming the challenges that arise with the standard formation scenario   \citep[for more details about the challenges in the standard formation see][]{Podsiadlowski+03}. 

 \citet{Shappee+13} suggested that re-triggering the EKL  behavior  via mass loss may facilitate the formation of close neutron star (NS)--white dwarf (WD) binaries (or other combination such as NS-NS, or WD-WD)  without an initial common envelope phase. If compact objects such as double white dwarfs in triples find themselves in the right part of the parameter space, the above process may trigger large eccentricities, that can lead to grazing interactions or even collisions (recall that the approximation may break, yielding even larger eccentricities),  which may promote Type Ia supernovae \citep[e.g.,][]{Tho10,Hamers+13,Prodan+13,Katz+12,Kushnir+13,Dong+14}. 

\subsection{Compact objects}

\begin{figure}
\begin{center}
\includegraphics[width=0.9\linewidth]{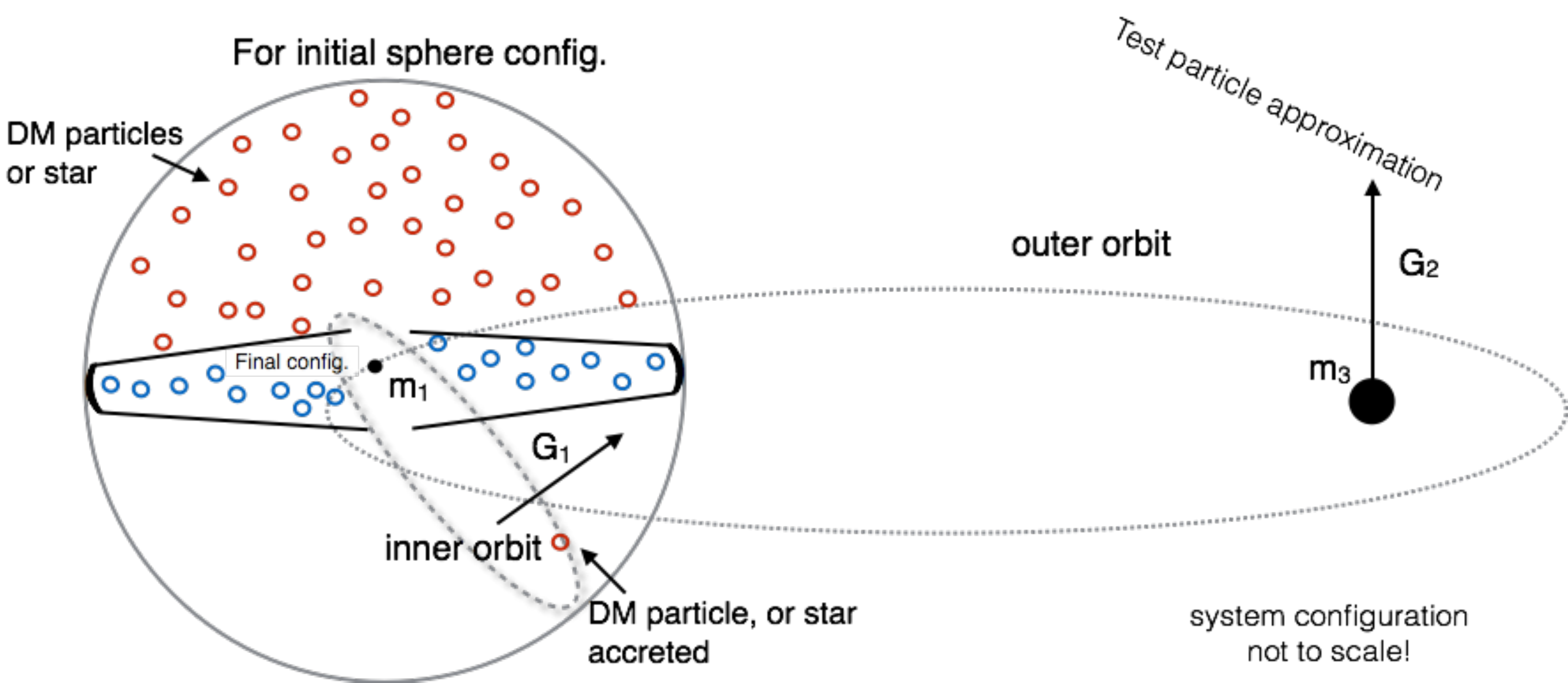}
\caption{\upshape {\bf Cartoon description of the resulted torus-like configuration from EKL in supermassive black hole binaries}.  The particles in a near-polar orbit, with respect to the black hole binary orbit, will undergo large eccentricity and inclination oscillations. This  leads to such large eccentricities that will result in either tidal disruption events for stars \citep[][]{Li+15}, or accretion of dark matter particles which may orbit the black hole \citep{NaozSilk}. 
} \label{fig:torus}
  \end{center}
\end{figure}

With in the hierarchical galaxy formation paradigm, and the strong observational evidence that a high abundance of the local galaxies host supermassive black holes, one expects that major galaxy mergers should inevitably result in the formation of supermassive black hole binaries or multiples  \citep[e.g.,][]{Valtonen+96,DiMatteo+05,Hoffman+07,Callegari+09,Dotti+12,Khan+12,Kulkarni+12}. The evolution of these binaries  highly depends on the conditions of the host galaxy. Numerical studies of spheroidal gas-poor galaxies suggest that these binaries can reach about a parsec separation and may stall there \citep[e.g.,][]{Begelman+80,Milosavljevic+01,Yu02}.
The effect of gravitational perturbations of supermassive black hole binaries on  an ambient star cluster has been discussed in length in the literature  
 \citep[e.g.,][]{Wen,MH02,Bla+02,Ivanov+05,Chen+09,Chen+11,Gualandris+09,Iwasawa+11,Sesana+11,Gualandris+12,Madigan+12,Meiron+13,Antonini+14,Bode+14,Wang+14,NaozSilk,Li+15}. In particular, it was suggested that the three body interactions may  play an important
role in both the growth of black holes at the centers of dense star 
clusters by increasing the tidal directions event rate of stars. It  was  also shown that interactions with the surrounding stars can either increase or decrease the eccentricity of the supermassive black hole binaries depending on the fraction of counter-rotating to co-rotating stars. Furthermore, the presence of supermassive black holes may increase the stellar tidal disruption event rate and even lead to a torus-like configuration of stars (or dark matter particles) around one of the black holes (see Figure \ref{fig:torus}). The supermassive black hole binary can also lead to an eccentric or ejected population of stars from the cluster.

For a supermassive black hole binaries embedded in a dense stellar environment, such as the one in the Galactic Center, other physical processes may affect the precession of a star around the primary black hole. Similarly to the short range forces discussed in Section \ref{sec:short}, if the extra precession takes place in an opposite direction to that induced due to the EKL mechanism, and it takes place on shorter timescale than $t_{\rm quad}$ eccentricity excitations may be suppressed.   These physical processes may include (but not limited to) precession caused by the stellar potential, scalar resonant relaxation or reorientation of the orbital plane due to vector resonant relaxation \citep{Kocsis+11,Kocsis+15} or Lense-Thirring precession \citep{Merritt+10,Merritt+12}. For the EKL mechanism of 
supermassive black hole binaries embedded in a dense stellar environment, \citet{Li+15} found that precession caused by the stellar potential and GR may have large effects on the dynamics while the others (such as tidal effects, scalar and vector resonant relaxation, and LenseÐThirring precession)  are typically less important. 

A dissipation mechanism which may play an important role when black holes (or other compact objects) are involved is  gravitational wave (GW) emission. In this scenario, black hole binary high orbital eccentricity induced by the outer perturber can lead to a more efficient merger rate, due to GW emission \citep[e.g.,][]{Bla+02}.  GW emission can also lead to the formation of extreme mass ratio binaries, such as supermassive black hole and a stellar mass black hole, or any other test particle, on a tight orbit \citep[e.g.,][]{Bode+14}. Considering the dynamical evolution of compact objects in the presence of an outer perturber, large eccentricities induced by the perturber can lead to a close approach between the two compact objects such that   GW emission will decay their orbital separation \citep[e.g.,][]{Wen,MH02,AP12,Seto13}. This perhaps can lead to  a detectable signal  using LIGO5 and VIRGO6 (e.g., \citet{Wen} and \citet{Naoz+12GR}, but see \citet{Mandel+08} and \citet{O'Leary+06}). Since GW emission associated with eccentric orbits is stronger  and have a very different spectrum relative to their circular counterparts,   it was suggested that using the GW information emitted by the close binary, it might be possible to constrain the mass or distance of the third body \citep[e.g.,][]{Yunes+11,Galaviz+11}.

Recently it was also suggested that black hole -low-mass X-ray binaries (BH-LMXBs) may  form via EKL mechanism \citep{Naoz+15}. During the dynamical evolution of the triple system, the EKL mechanism can cause large  eccentricity excitations on the LMXB progenitor, resulting BH-LMXB candidate, while skipping the common-envelope phase.  Interestingly, a substantial number of close binaries with an accreting compact object, e.g., LMXBs and their descendants (e.g., millisecond radio pulsars), are known or suspected triples \citep{1988ApJ...334L..25G,1999ApJ...523..763T,2001ASPC..229..117R,2003Sci...301..193S,2001ApJ...563..934C,2007MNRAS.377.1006Z,Prodan:2012ey,Prodan+15}.

\section{Beyond the three body secular approximation}

There are different channels to consider when going beyond the secular approximation. The first is to consider the validity of the approximation discussed in Section \ref{sec:valid}. In other words allowing for more compact systems (e.g., $\epsilon > 0.1$, Equation (\ref{eq:epsilon})) which means considering the  implications of having changes in the angular momentum on short timescale compared to the orbital timescale (e.g., Equation (\ref{eq:REO})). The second is to allow for  higher multiples. 

Considering compact systems, a popular application of the three body interaction is the merger of two white dwarfs to prompt the so called double degenerate type Ia supernova. It was suggested that double degenerate type Ia supernova may represents a  substantial fraction (if not all) of the type Ia supernovae. Observational evidences for this may lay in 
 distribution of times between star formation and the type Ia supernova explosion, usually called the delay-time distribution, that seems to favor the double degenerate scenario \citep[e.g.,][]{Maoz+14} or in the lack of hydrogen lines that are expected in the single degenerated (white dwarf with a stellar companion) scenario \citep[e.g.,][]{Shappee+13H}. There are  different theoretical models that address the double degenerate type Ia supernova formation. In the context of triple body interactions it was suggested that the large eccentricities associated with the EKL mechanism can lead to double degenerate type Ia supernova \citep[e.g.,][]{Tho10,Hamers+13,Prodan+13}.  Considering more compact  systems, the inner orbit specific angular momentum is  likely to reach almost zero (i.e., an almost radial motion) on timescales at the order of the inner orbit period (see Section   \ref{sec:valid})  
 causing the collision of two white dwarfs and resulting in type Ia supernova \citep[e.g.,][]{Katz+12,Kushnir+13,Dong+14}.

Another interesting astrophysical application for the insight gained in the triple study is by considering  higher multiples. There are of course many ways to address high multiple interaction. The first is to consider a scattering, short time scale, event, which has been discussed in length in the literature \citep[e.g.,][]{Hut+83,Rasio+96,Sourav+08,Nag+08,Antognini+15}. In a  stable system, which does not undergo  a scattering event, the additional fourth (or more) companion can have large effects on the eccentricity  and inclination evolution. In particular it can help tapping into large parts of the parameter space \citep[e.g.,][]{Takeda,Touma+09,Pejcha+13,Boue+14,Hamers+15}, and affect the spin orbit evolution \citep[e.g.,][]{Li+14Kepler56,Boue+14sp}. 
 A consequence of the latter effect is that circumbinary planets may be misaligned  \citep[e.g.][]{Hamers+15c,Martin+15cp,Munoz+15}. 
 In the context of the secular approximation, the Gauss averaging method can be utilized for N number of stable orbits \citep[e.g.][]{Touma+09}. This 
method is  a  phase-averaged calculation  for which the gravitational interactions between non-resonant orbits are equivalent in treating the  orbits as massive wires interacting with each other,  where the line-density is inversely proportional to the orbital velocity.  As explained above, a consequence of the secular approximation is that the  semi-major axes of the wires are constants of motion \citep[e.g.,][]{MD00}.   In general this method can be used to explore different many body secular effects, for example the evolution of a particle disk in the presence of a perturber \citep{Batygin12}.

\section{Summary}

The high abundance  of hierarchical triple systems in nature motivates the  investigation of their dynamics. Furthermore, this approximation seems to be very useful  in addressing a variety of puzzles and  systems that are observe, such as retrograde Hot Jupiters, blue stragglers, low and high mass X-ray binaries, compact object binaries, double degenerate type Ia supernova etc. Building on  the physical understandings gained in the past years in this subject, motivates us to go beyond the approximation for an even wider range of applications. 

The recent theoretical developments  can be summarized by the following:
\begin{itemize}
\item The z-component of the angular momentum of the inner and outer orbits (i.e., the nominal $\sqrt{1-e_{1,2}^2}\cos i_{1.2}$) are only conserved if one of the binary members is a test particle and the outer orbit is axisymmetric ($e_2=0$). 
\item Relaxing any of these assumptions may lead to high order resonances  characterized  by large eccentricity excitations and flips of the orbital oriention as well as  chaotic behavior. 
\item These high order resonances allow the system to tap into larger parts of the  initial  parameter space  for which the EKL mechanism is triggered.
\item Short range forces and other physical processes (such as GR and stellar mass loss) can also re-trigger the EKL mechanism for systems that did not exhibit these dynamics in the point mass approximation.    
\end{itemize}

The field continues to developed and to go beyond three body
systems and the secular or hierarchical approximations. These
improvements allow for the application to, and the understanding of, a larger
variety of systems. The intuition and insight that the
Eccentric Kozai-Lidov mechanism has provided is utilized for these approaches.

\section*{DISCLOSURE STATEMENT}
The author is not aware of any affiliations, memberships, funding, or financial holdings that might be perceived as affecting the objectivity of this review.

\section*{ACKNOWLEDGMENTS}
I greatly thank Fred Rasio, Brad Hansen, Gongjie Li and Diego Munoz for useful comments. I also thank the referee Ruth Murray-Clay, for reading the review thoroughly and providing useful suggestions.  Furthermore, I deeply  thank Alexander Stephan for reading the draft in great details.  I thank J. Antognini, B. Katz, G. Li, B. Liu, C. Petrovich, J. Teyssandier,  for sharing figures. 
I acknowledge the partial support from the Sloan Foundation Research Fellowships as well as the Annie Jump Cannon Prize. 

\section{Supplemental Material - The secular equations}\label{sec:eqs}

The full octupole-order  equations of motion for the most general case (i.e., relaxing the test particle and  axisymmetric potential approximations) presented in \citet{Naoz+11sec} are reiterated here for completeness.
We begin with reminding the reader of the definitions of a few useful parameters:
\begin{eqnarray}
\label{eq:C3_2}
C_3&=&-\frac{15}{16}\frac{k^4}{4}\frac{(m_1+m_2)^9}{(m_1+m_2+m_3)^4}\frac{m_3^9(m_1-m_2)}{(m_1m_2)^5}\frac{L_1^6}{L_2^3G_2^5} \nonumber \\
&=& -C_2\frac{15}{4}\frac{\epsilon_M}{e_2} \ 
\end{eqnarray}
where
\begin{equation}\label{eq:epslionM_2}
\epsilon_M=\frac{m_1-m_2}{m_1+m_2}\frac{a_1}{a_2}\frac{e_2}{1-e_2^2} \ 
\end{equation}
and  
\begin{equation}
\label{eq:A8}
A= 4+3e_1^2-\frac{5}{2}B\sin i_\tot^2 \ ,
\end{equation}
where
\begin{equation}
B=2+5e^2_1-7e_1^2\cos(2\omega_1) \ ,
\end{equation}
and
\begin{equation}
\cos \phi=-\cos \omega_1\cos \omega_2 -\cos i_\tot \sin \omega_1 \sin \omega_2 \ .
\end{equation}
As shown in \citet{Naoz+11sec} elimination of the nodes (i.e,. setting $\Omega_1 - \Omega_2 = \pi$) can be done as long as one does not conclude that the conjugate z-component of the angular momenta ($H_1$ and $H_2$) are constant of motion.  The
partial derivatives with respect to the other coordinates and momenta
are not affected by the substitution $\Omega_1 - \Omega_2 = \pi$. In that case, the time evolution of $H_1$ and
$H_2$ (and thus $i_1$ and $i_2$) can be derived from the total angular
momentum conservation. 
The
doubly averaged Hamiltonian after eliminating the nodes:
\begin{eqnarray}
  \Ham(\Delta h \to \pi)& =& C_2 \{ \left( 2 + 3 e_1^2 \right) \left( 3 \cos^2 i_\tot - 1 \right)  \\ & + &15 e_1^2 \sin^2 i_\tot \cos(2 \omega_1) \}  \nonumber \\ \nonumber
  &+& C_3 e_1 e_2 \{ A \cos \phi \nonumber \\ &+ & 10 \cos i_\tot \sin^2 i_\tot  (1-e_1^2) \sin \omega_1 \sin \omega_2 \} \nonumber \ .
\end{eqnarray} 

The time evolution of the argument of periapse for the inner and outer
orbits are given by:
\begin{eqnarray}
\label{eq:g1dot8} 
\dot{\omega}_1&=&6 C_2\bigg\{ \frac{1}{G_1}[4 \cos^2 i_\tot+(5\cos (2\omega_1)-1)  \\ \nonumber 
&\times& (1-e_1^2-\cos^2 i_\tot)] +\frac{\cos i_\tot}{G_2}[2 + e^2_1(3-5\cos (2\omega_1))]\bigg\} \\ \nonumber
&-&C_3e_2\bigg\{e_1 \left(\frac{1}{G_2}+\frac{\cos i_\tot}{G_1}\right) \\ \nonumber
&\times& [\sin \omega_1 \sin \omega_2(10(3\cos^2 i_\tot-1)(1-e_1^2)+A) \\ \nonumber
&-&5 B \cos i_\tot\cos \phi] -\frac{1-e_1^2}{e_1G_1}\times[\sin \omega_1\sin \omega_2 \\ \nonumber
&\times&10\cos i_\tot \sin i_\tot^2(1-3e_1^2) \\ \nonumber
&+&\cos\phi (3A-10\cos i_\tot^2+2)]\bigg\} \ ,
\end{eqnarray}
and 
\begin{eqnarray}
\label{eq:g2dot8}
\dot{\omega}_2&=&3 C_2\bigg\{ \frac{2\cos i_\tot}{G_1}[2+e_1^2(3-5\cos (2\omega_1))]  \\ \nonumber 
&+&\frac{1}{G_2}[4+6e_1^2+(5\cos^2 i_\tot -3)(2+ e_1^2[3-5\cos(2\omega_1)] )\bigg\} \\ \nonumber
&+&C_3e_1  \bigg\{\sin \omega_1\sin \omega_2 \bigg ( \frac{4e_2^2+1}{e_2G_2} 10\cos i_\tot\sin^2 i_\tot (1-e_1^2) \\ \nonumber
&-&e_2\left(\frac{1}{G_1}+\frac{\cos i_\tot}{G_2}\right) [A+10(3\cos^2 i_\tot-1)(1-e^2_1)]\bigg ) \\ \nonumber
&+&\cos \phi \bigg [ 5B\cos i_\tot e_2 \left(\frac{1}{G_1}+\frac{\cos i_\tot}{G_2}\right) +\frac{4e_2^2+1}{e_2G_2}A\bigg ]\bigg \}
\end{eqnarray}
The time evolution of the longitude of ascending nodes is given by:
\begin{eqnarray}
\label{eq:h1dot8}
\dot{\Omega}_1&=&-\frac{3 C_2}{G_1 \sin i_1} \left( 2 + 3 e_1^2 - 5 e_1^2 \cos\left( 2 \omega_1 \right) \right) \sin\left( 2 i_{\rm tot} \right) \\ \nonumber
&-&C_3e_1e_2[5B\cos i_\tot\cos\phi \nonumber \\ &-& A\sin \omega_1\sin \omega_2 +10(1-3\cos^2 i_\tot) \nonumber \\ \nonumber
&\times&(1-e_1^2)\sin \omega_1\sin \omega_2]\frac{\sin i_\tot}{G_1 \sin i_1} \ ,
\end{eqnarray}
where in the last part we have used again the law of sines for which
$\sin i_1= G_2 \sin i_\tot / G_\tot$.  The evolution of the longitude
of ascending nodes for the outer orbit can be easily obtained using:
\begin{equation}
\dot{\Omega}_2=\dot{\Omega}_1 \ .
\end{equation}

The evolution of the inner and outer eccentricities  is:
\begin{eqnarray}
\label{eq:dote1_8}
\dot{e}_1&=&C_2\frac{1-e_1^2}{G_1}[30e_1 \sin^2i_\tot\sin(2\omega_1)] \\ \nonumber
&+& C_3e_2\frac{1-e_1^2}{G_1}[35 \cos\phi\sin^2 i_\tot e_1^2 \sin (2\omega_1) \\ \nonumber
&-& 10\cos i_\tot \sin^2 i_\tot \cos \omega_1 \sin \omega_2 (1-e_1^2)\\ \nonumber
&-& A(\sin \omega_1\cos \omega_2  - \cos i_\tot \cos \omega_1 \sin \omega_2)] \ ,
\end{eqnarray}
and
\begin{eqnarray} 
\dot{e}_2 &=& -C_3 e_1\frac{1-e_2^2}{G_2}[10\cos \left(i_\tot\right) \sin^2 \left(i_\tot\right)(1-e_1^2)\sin \omega_1 \cos \omega_2 \nonumber \\ 
&+& A(\cos \omega_1 \sin \omega_2 - \cos(i_\tot)\sin \omega_1 \cos \omega_2)] \ .
\end{eqnarray}
The angular momenta derivatives of the inner and outer orbits as a function of time can be easily calculated, where for the inner orbit we write:
\begin{eqnarray} 
\label{eq:G1dot8}
\dot{G}_1 & =& - C_2 30 e_1^2  \sin (2\omega_1) \sin^2(i_\tot) + C_3 e_1 e_2 ( \\ \nonumber
&-&35 e_1^2 \sin^2(i_\tot)\sin (2 \omega_1) \cos\phi + A [\sin \omega_1 \cos \omega_2  \\ \nonumber
&-& \cos(i_\tot)\cos \omega_1 \sin \omega_2 ]  \\ \nonumber
&+& 10\cos(i_\tot)\sin^2(i_\tot)
[1-e_1^2]\cos \omega_1 \sin \omega_2  )\ ,
\end{eqnarray}
and for the outer orbit (where the quadrupole term is zero)
\begin{eqnarray} 
\label{eq:G2dot8}
\dot{G}_2 & =&  C_3 e_1 e_2 [A\{\cos \omega_1 \sin \omega_2  - \cos(i_\tot)\sin
\omega_1 \cos \omega_2 \}\nonumber \\  &+&10\cos(i_\tot)\sin^2(i_\tot)
[1-e_1^2]\sin \omega_1 \cos \omega_2  ]\ .
\end{eqnarray}
Also the z-component of the inner orbit angular momentum is
\begin{equation}
\label{eq:Hdot1_8}
  \dot{H}_1 = \frac{G_1}{G_{\tot}} \dot{G}_1 - \frac{G_2}{G_{\tot}} \dot{G}_2 \ ,
\end{equation} 
where using the law of sines we write:
\begin{equation}
\label{eq:Hdot1_8n}
  \dot{H}_1 = \frac{\sin i_2}{\sin i_{\tot}} \dot{G}_1 - \frac{\sin i_1}{\sin i_{\tot}} \dot{G}_2 \ .
\end{equation} 
Because the total angular momentum is conserved $G_{\tot}={\rm Const.}=H_1+H_2$ the 
 outer orbit z-component time evolution is simply $\dot{H}_2 = -\dot{H}_1$.
The inclinations equation of motion is
\begin{equation}
\dot{(\cos i_1)}= \frac{\dot{H}_1}{G_1}-\frac{\dot{G}_1}{G_1} \cos i_1 \ ,
\end{equation}
and
\begin{equation}\label{eq:cosi2}
\dot{(\cos i_2)}= \frac{\dot{H}_2}{G_2}-\frac{\dot{G}_2}{G_2} \cos i_2 \ .
\end{equation}

\section{Supplemental Material - Static tides equations}\label{sec:tides}
Tidal interaction considered in this review are limited to the inner orbit members equilibrium and static tides formalism   \citep[e.g.,][]{Hut,1998EKH,1998KEM,Egg+01}. 
A compact representation of the tidal interactions equation can be found when using the Laplace-
Runge-Lenz vector system. In this system the three vector base is composed from the inner orbit eccentricity vector ${\bf e}_1$ the specific angular momentum vector \begin{equation} {\bf J}_1=\sqrt{K^2 (m_1+m_2)a_1(1-e_1^2)} \hat{\bf J}_1= {\bf G}_1 (m_1+m_2)/(m_1 m_2) \ .\end{equation}  The vector $\hat{\bf q}=\hat{\bf J}_1\times \hat{\bf e}_1$
completes the right-hand triad of unit vectors $(\hat{\bf q},\hat{\bf J}_1,\hat{\bf e}_1)$. Each of the inner member masses have a spin vector ${\bf \Omega}_{s1}$ and ${\bf \Omega}_{s2}$, respectively.  The time evolution equations are (where subscript $1$ and $2$ refer to masses $m_1$ and $m_2$):
\begin{eqnarray} 
\label{eq:TF1}
\frac{1}{e_1}\frac{d{\bf e}_1}{dt}&=&(Z_1+Z_2) \hat{\bf q} - (Y_1+Y_2) \hat{\bf J}_1 - (V_1+V_2)\hat{\bf e}_1 \ , \\
\frac{1}{J_1}\frac{d{\bf J}_1}{dt}&=& - (X_1+X_2)\hat{\bf q} - (W_1+W_2)\hat{\bf J}_1 +  (Y_1+Y_2)\hat{\bf e}_1 \ , \\
I_1\frac{d{\bf{\Omega}_{s1}}}{dt} &=& \mu J_1 ( X_1 \hat{\bf q}  + W_1 \hat{\bf J}_1 - Y_1  \hat{\bf e}_1  ) \ , \\ \label{eq:TF1f}
I_2\frac{d{\bf{\Omega}_{s2}}}{dt} &=& \mu J_1 ( X_2 \hat{\bf q}  + W_2 \hat{\bf J}_1 - Y_2  \hat{\bf e}_1  ) \ ,
\end{eqnarray}
where $\mu=m_1 m_2/(m_1+m_2)$ is the reduced mass, $I_1$ ($I_2$) is the moment of inertia of mass $m_1$ ($m_2$).  The vector $(X,Y,Z)$ is the angular velocity of the $(\hat{\bf q},\hat{\bf J}_1,\hat{\bf e}_1)$ frame and can be easily related to the Delaunay's elements in the invariable plan as \citep{1998EKH}:
\begin{eqnarray} 
\label{eq:XYZ}
X&=& \frac{di_1}{dt}\cos \omega_1 + \frac{d\Omega_1}{dt}\sin \omega_1 \sin i_1 \ , \\
Y&=& - \frac{di_1}{dt}\sin \omega_1 + \frac{d\Omega_1}{dt}\cos \omega_1 \sin i_1 \ , \\
Z&=& \frac{d\omega_1}{dt} + \frac{d\Omega _1}{dt} \cos i_1
\end{eqnarray}
This set of equations gives the precession rate due to tides ${d\omega_1}/{dt}$ as well as how the other Delaunay's elements vary with time.
We note that these equations (\ref{eq:TF1})-(\ref{eq:TF1f}) are identical to that of \citet{Egg+01} and \citet{Dan}, up to the gravitational influence of the third body which they described by the tensor $\bf{S}$. In our formalism its redundant. The functional form of $W,V,X,Y$ and $Z$ were given in  \citet{Egg+01} and are simply:
\begin{eqnarray} 
\label{eq:TF2}
V_1&=&\frac{9}{t_{F1}}\left(\frac{1+15e_1^2/4+15e_1^4/8+5e_1^6/64}{(1-e_1^2)^{13/2} } - \frac{11\Omega_{s1,J}}{18 n} \frac{1+3e_1^2/2+e_1^4/8 }{(1-e_1^2)^5 } \right) \ , \\
W_1&=&\frac{1}{t_{F1}}\left(\frac{1+15e_1^2/2+45e_1^4/8+5e_1^6/16}{(1-e_1^2)^{13/2} } - \frac{11\Omega_{s1,J}}{ n} \frac{1+3e_1^2+3e_1^4/8 }{(1-e_1^2)^5 } \right) \ , \\
X_1 &=&- \frac{m_2 k_{1} R_1^5}{\mu n a_1^5}  \frac{ \Omega_{s1,J}\Omega_{s1,e} }{(1-e_1^2)^2} - \frac{\Omega_{s1,q}}{2nt_{F1}}\frac{1+9e_1^2/2+5e_1^4/8}{(1-e_1^2)^5} \ , \\
Y_1 &=& -\frac{m_2 k_{1} R_1^5}{\mu n a_1^5}  \frac{ \Omega_{s1,J}\Omega_{s1,q} }{(1-e_1^2)^2} + \frac{\Omega_{s1,e}}{2nt_{F1}}\frac{1+3e_1^2/2+e_1^4/8}{(1-e_1^2)^5} \ , \\
Z_1 &=& \frac{m_2 k_{1} R_1^5}{\mu n a_1^5}\left( \frac{2\Omega_{s1,J}^2-\Omega_{s1,q^2}-\Omega_{1s,e}^2 }{2(1-e_1^2)^2} + \frac{15 k^2 m_2}{a_1^3}\frac{1+3e_1^2/2+e_1/8}{(1-e_1^2)^5} \right) \ ,
\end{eqnarray}
where the expression for mass $m_2$ can be easily found by replacing subscript  $1$ with $2$. 
The mean motion is 
\begin{equation}
n=\frac{2\pi}{P_1}=\sqrt{\frac{k^2(m_1+m_2)}{a_1^3}} \ .
\end{equation}
also, $k_{1}$  is  classical apsidal motion constant, which is a  measure of quadrupolar deformability, and related to the Love parameter of mass $m_1$ by $k_L=2k_1$. It also related to  \citet{Egg+01} coefficient $Q_E$ by 
\begin{equation}
k_1=\frac{1}{2}\frac{Q_E}{1-Q_E}\ .
\end{equation}
The tidal friction timescale can be expressed in terms of the viscous timescale $t_{V1}$ (which is assume dot be constant in the tides applications in this review):
\begin{equation}
t_{F1}=\frac{t_{V1}}{9}\left(\frac{a_1}{R_1}\right)^8\frac{m_1^2}{(m_1+m_2)m_2}\frac{1}{(1+2k_1)^2} \ ,
\end{equation}
and similar equation for $t_{F2}$ can be found by replacing $1$ with $2$. 
This formalism describes viscosity that causes the tidal bulge to lag the instantaneous direction of the companion by a constant angle $1/(2Q)$ at constant time interval. The quality factor $Q$  can be expressed as a function of viscous timescale as well by \citep[e.g.,][]{Dan,Hansen10}
\begin{equation}
Q=\frac{4}{3}\frac{k_1}{(1+2k_1)^2}\frac{k^2m_1}{R_1^3}\frac{t_{V1}}{n} \ .
\end{equation}


\pagestyle{empty}
\bibliographystyle{jponew}
\bibliography{Kozai}


\end{document}